\documentclass[aps,prd,preprint,floatfix,nofootinbib,longbibliography,a4paper]{revtex4-1}
\pdfoutput=1
\usepackage{color}
\usepackage{graphicx}
\usepackage{textcomp}
\usepackage{subfigure}
\usepackage{feynmp}
\usepackage{amsmath,amssymb,array}
\usepackage{enumerate}
\usepackage{hhline}
\newcommand{\mathsym}[1]{{}} 
\def\lsim{\:\raisebox{-1.1ex}{$\stackrel{\textstyle<}{\sim}$}\:}
\def\gsim{\:\raisebox{-1.1ex}{$\stackrel{\textstyle>}{\sim}$}\:}

\newcommand{\beqa}{\begin{eqnarray}}
\newcommand{\eeqa}{\end{eqnarray}}
\newcommand{\be}{\begin{equation}}
\newcommand{\ee}{\end{equation}}
\newcommand{\ba}{\begin{array}} 
\newcommand{\ea}{\end{array}}

\begin{document} 
\vspace*{1cm}
\title{Effects of heavy neutrinos on vacuum stability in two-Higgs-doublet model with GUT scale supersymmetry}
\bigskip
\author{V Suryanarayana Mummidi}
\email{suryam@iisermohali.ac.in}
\author{Vishnu P. K.}
\email{vishnupk@iisermohali.ac.in}
\author{Ketan M. Patel}
\email{ketan@iisermohali.ac.in}
\affiliation{Indian Institute of Science Education and Research Mohali, Knowledge City, Sector  81, S A S Nagar, Manauli 140306, India.}

\begin{abstract}

We analyse the implications of right-handed neutrinos on the stability of the electroweak vacuum in two-Higgs-doublet models with supersymmetry at high scale. It is assumed that supersymmetry is broken at  scale $M_S = 2 \times 10^{16}$ GeV and effective theory below $M_S$ is two-Higgs-doublet model of type II with three generations of singlet neutrinos which induce small masses for the standard model neutrinos  through type I seesaw mechanism. We study the high and low scale versions of seesaw mechanism. In both these cases, we show that the presence of right-handed neutrinos significantly improves the stability of electroweak vacuum if their Yukawa couplings with the SM leptons are of ${\cal O}(1)$ or greater. However, this possibility is severely constrained by the measured mass and couplings of Higgs and limits on the mass of the charged Higgs from the flavour physics data. It is shown that the stable or metastable electroweak vacuum and experimentally viable low energy scalar spectrum require  $\tan\beta \lsim 2.5$ and the magnitude of neutrino Yukawa couplings smaller than ${\cal O}(1)$. The results obtained in this case are qualitatively similar to those without right-handed neutrinos. 
\end{abstract} 

\maketitle

\section{Introduction}
\label{sec:introduction}
An embedding of the Standard Model (SM) into supersymmetric grand unified theories (GUTs) leads to an elegant and predictive framework which resolves several technical problems obstructing the extendability of the SM gauge theory to very short length scales. Supersymmetry (SUSY), close to TeV scale, stabilizes the electroweak scale against the large radiative corrections that arise due to higher fundamental scales present in the theory. TeV scale SUSY provides precision gauge coupling unification at the scale $M_{\rm GUT} \approx 2 \times 10^{16}$ GeV, although this can also be achieved if SUSY is broken at very high scale but some of the super-partners remain as light as few hundred GeV \cite{Giudice:2004tc,ArkaniHamed:2004fb}. SUSY as a local symmetry is an essential ingredient in the superstring theory which provides a potential framework for unification of all the fundamental forces \cite{Green:1987sp}. However, it is generically expected that the SUSY breaking scale in such a theory would be very close to the string scale.  A similar situation arises in supersymmetric GUTs constructed in five or six dimensional spacetime in which the mechanisms used for breaking the unified gauge symmetry often break also the supersymmetry \cite{Kitano:2003cn,Buchmuller:2015jna}\footnote{Note that in \cite{Kitano:2003cn}, breaking of supersymmetry at high scale was prevented by introducing brane localized $D$-terms. Supersymmetry is broken at the GUT scale in absence of such terms.}. In these  frameworks, the effective theory below the GUT scale is described by non-supersymmetric SM which in some cases augmented by other possible light states as remnants of an underlying ultraviolet complete theory.

The experimental data collected in the first and second runs of the Large Hadron Collider (LHC) has shown no evidence for TeV scale supersymmetry \cite{Aaboud:2017bac,Sirunyan:2017cwe}. If this trend continues in the future runs of LHC and other experiments then it would imply that supersymmetry is not an underlying mechanism for stabilization of electroweak scale. In this case, the assumption of  existence of a weak scale supersymmetry is no longer necessarily required. Following this, in this paper we assume that SUSY exists in an underlying theory but it is broken at the scale much above the electroweak scale. It is assumed that supersymmetry does not play any role in stabilizing the electroweak scale and the electroweak scale remains finely tuned or some new dynamics in the high energy theory take care of the gauge hierarchy problem.  Although an absence of SUSY at the low energy seems to make it less interesting from the phenomenological point of view, however its existence at high scale still leads to nontrivial consequences on the low energy theory. For example, the scalar potential of an effective theory below the SUSY breaking scale arises from the D-term potential of an underlying ultraviolet supersymmetric theory. Therefore, the electroweak symmetry breaking, stability of  electroweak vacuum and mass spectrum of scalars in the effective theory are constrained by the high scale SUSY. Such consequences are already studied for an effective theory being only the SM \cite{Giudice:2011cg,EliasMiro:2011aa,Draper:2013oza,Ellis:2017erg}, SM with additional Higgs doublet \cite{Gorbahn:2009pp,Lee:2015uza}, and SM with higgsinos and gauginos \cite{Bagnaschi:2014rsa}.

It is observed that the SM alone as an effective theory cannot be matched to its minimal supersymmetric version (MSSM) when SUSY breaking scale is higher than $10^{11}$ GeV because of the vacuum stability constraints \cite{Giudice:2011cg,EliasMiro:2011aa}\footnote{The quoted limit is obtained by considering the mean value of the measured top quark mass. The limit is very sensitive with respect to the choice made for the value of top quark mass \cite{Draper:2013oza,Ellis:2017erg}.}. The SM with an additional Higgs doublet close to the electroweak scale, known as two-Higgs-doublet model (see \cite{Branco:2011iw} for a review), is another possible effective theory with SUSY at high scale. The pair of scalar doublets in the two-Higgs-doublet model (THDM) can be identified as the two Higgs doublets of the MSSM. Such a matching between MSSM and THDM at high SUSY breaking scale has already been  considered in \cite{Gorbahn:2009pp,Lee:2015uza,Bagnaschi:2015pwa,Bhattacharyya:2017ksj}. It is found in \cite{Bagnaschi:2015pwa} that THDM can be consistently matched to MSSM with SUSY breaking scale as high as the reduced Planck scale. This improvement over the SM is due to the presence of an additional Higgs doublet which modifies the stability conditions allowing more freedom in the effective potential. The stability or metastability of the electroweak vacuum however puts stringent constraints on the allowed values of $\tan\beta$ parameter which is the ratio of the vacuum expectation value (VEV) of two Higgs doublets.

In this paper, we investigate the effects of the so-called right handed (RH) neutrinos on the stability of scalar potential in THDM with supersymmetry broken at GUT scale. The SM augmented with such singlet fermions provides natural explanation for non-vanishing and tiny neutrino masses through type I seesaw mechanism \cite{Minkowski:1977sc,Yanagida:1979as,GellMann:1980vs,Glashow:1979nm,Lazarides:1980nt,Schechter:1980gr,Mohapatra:1980yp}. In the GUTs based on $SO(10)$ gauge group the RH neutrinos reside, along with the SM fermions, in three copies of 16-dimensional irreducible representation of the gauge group. In these models, if SUSY and gauge symmetry are both broken at the GUT scale leaving a pair of MSSM Higgs doublets light then the effective theory below the GUT scale is described by THDM with three generations of RH neutrinos. For example, this possibility is naturally realized in the GUT models based on flux compactification \cite{Buchmuller:2015jna,Buchmuller:2017vho,Buchmuller:2017vut}.  Motivated by this, we assume that the MSSM is broken at the scale $M_S = 2 \times 10^{16}$ GeV leaving THDM augmented with three generations of RH neutrinos as an effective theory below $M_S$. The RH neutrinos obtain their masses through lepton number violating interactions which in turn induce tiny masses for the SM neutrinos through type I seesaw mechanism. We consider two distinct possibilities in which the mass scale of RH neutrinos is either close to $M_S$ or electroweak scale. 

We find that RH neutrinos have considerable effects on the stability of electroweak vacuum. In particular, if these neutrinos are strongly coupled with the SM leptons then they lead to significant improvements in  the stability of the scalar potential in THDM. It is shown that the stable vacuum can be achieved for almost any value of $\tan\beta$ if the magnitude of Yukawa couplings of RH neutrinos are larger than that of the top quark Yukawa coupling. However, the observed Higgs mass, measured couplings of Higgs with the gauge bosons and limits on the charged Higgs mass from flavour physics data severely constrain on this scenario.  

The paper is organized as follows. We discuss the THDM framework with type I seesaw mechanism in the next section. The procedure of renormalization group (RG) evolution and matching at different scales have been discussed in section \ref{sec:rge}. Numerical analysis and their results are discussed in section \ref{sec:numerical}. The conclusion is presented in section \ref{sec:concl}. Technical details related to renormalization group equations, threshold corrections at the high scale, extraction of the gauge and Yukawa couplings at top quark mass scale and dependency of results on the choice of top quark mass are elaborated in the Appendices.

\section{The Framework}
\label{sec:model}
The scalar potential of the most general THDM can be parametrized as
\beqa \label{THDM_V}
V &=& m_1^2\, H_1^\dagger H_1\, +\, m_2^2\, H_2^\dagger H_2\, - \, \left(m_{12}^2\, H_1^\dagger H_2\, +\, {\rm H.c.} \right) \nonumber \\
& + & \frac{\lambda_1}{2}\, (H_1^\dagger H_1)^2\, +\, \frac{\lambda_2}{2}\, (H_2^\dagger H_2)^2\, +\, \lambda_3\, (H_1^\dagger H_1)  (H_2^\dagger H_2)\, +\, \lambda_4\, (H_1^\dagger H_2)  (H_2^\dagger H_1) \nonumber \\
& + & \left(\frac{\lambda_5}{2}\, (H_1^\dagger H_2)^2\, +\, \lambda_6\, (H_1^\dagger H_1) (H_1^\dagger H_2)\, + \, \lambda_7\, (H_1^\dagger H_2) (H_2^\dagger H_2)\, +\, {\rm H.c.}   \right)\,,
\eeqa
where $H_1$ and $H_2$ are two complex Higgs fields, each of them is a doublet under $SU(2)_L$ and carries hypercharge $Y=1$. With an addition of three copies of fermion singlet to the SM fermion spectrum, the most general Yukawa Lagrangian of the model can be written as
\beqa \label{THDM_LY}
-{\cal L}_Y &=& \overline{Q}_L^i \left( Y_d^{ij} H_1+\tilde{Y}_d^{ij} H_2\right) d_R^j\,+\, \overline{Q}_L^i \left( \tilde{Y}_u^{ij} \tilde{H}_1+\, Y_u^{ij} \tilde{H}_2\right) u_R^j\nonumber \\
& + &\,  \overline{L}_L^i \left( Y_e^{ij} H_1+\tilde{Y}_e^{ij} H_2\right) e_R^j\,+\, \overline{L}_L^i \left( \tilde{Y}_\nu^{ij} \tilde{H}_1+\, Y_\nu^{ij} \tilde{H}_2\right) \nu_R^j\, +\, {\rm H.c.}\,,
\eeqa
where  $i,j=1,2,3$ stand for three generations of fermions and $\tilde{H}_{1,2} = i \sigma^2 H^*_{1,2}$.

If THDM is assumed to be an effective theory, obtained from the MSSM after the SUSY is broken at the scale $M_S$, then the scalar potential in Eq. (\ref{THDM_V}) is matched to the MSSM Higgs potential at $M_S$. The potential of the MSSM Higgs doublets $H_u$ and $H_d$ (with hypercharge $Y=1$ and $Y=-1$, respectively) contains the D-terms of superpotential and the soft supersymmetry breaking terms. Setting $\tilde{H}_1=H_d$ and $H_2 = H_u$, tree level matching between the potentials leads to the following conditions at $M_S$ \cite{Haber:1993an,Lee:2015uza}: 
\be \label{lm_MS}
\lambda_1 = \lambda_2 = \frac{1}{4}\left(g_2^2+  g_Y^2 \right)\,,~~\lambda_3=\frac{1}{4}\left(g_2^2 - g_Y^2 \right)\,,~~\lambda_4 = -\frac{1}{2} g_2^2\,, \ee
\be \label{lm567_MS}
\lambda_5 = \lambda_6 = \lambda_7 = 0\,, \ee
where $g_Y = \sqrt{3/5}\, g_1 $. Further, the terms involving $H_d^\dagger$ and $H_u^\dagger$ in the Yukawa Lagrangian are absent at the scale $M_S$. This implies 
\be \label{Yuk_MS}
\tilde{Y}_d^{ij}=\tilde{Y}_u^{ij}=\tilde{Y}_e^{ij}=\tilde{Y}_\nu^{ij} =0\, \ee
in Eq. (\ref{THDM_LY}) at the scale $M_S$. The conditions in Eqs. (\ref{lm567_MS},\ref{Yuk_MS}) imply that the theory at $M_S$ is essentially type II THDM \cite{Branco:2011iw} with additional  boundary conditions on the scalar quartic couplings as given in Eq. (\ref{lm_MS}).

We assume that the supersymmetry breaking scale is very close to the GUT scale. If the type I seesaw mechanism is considered as an underlying mechanism to generate tiny masses for the SM neutrinos, it introduces new scales in the theory, namely the mass thresholds of RH neutrinos. The Majorana masses for the RH neutrinos can be written as
\be \label{MR}
-{\cal L}_{\nu_R} = \frac{1}{2} M_R^{ij}\, \nu_R^{T i} {\cal C} \nu_R^j\, + \, {\rm H.c.}\,. \ee
We denote the physical masses of RH neutrinos by $M_{R_i}$ with $i=1,2,3$ and adopt a convention in which  $M_{R_1}  \le M_{R_2} \le M_{R_3}$. It is typically expected that $M_{R_i}$ lie in between the electroweak and GUT scale. For $M_{R_1} \gg \langle H_u \rangle$, the effective light neutrino mass matrix becomes
\be \label{seesaw}
{\cal M}_\nu = - \langle H_u \rangle^2\, Y_\nu\, M_R^{-1}\, Y_\nu^T\,. \ee
The $Y_\nu$ and $M_R$ cannot be fixed uniquely from the available experimental information of neutrino masses and mixing parameters. This lack of information is best parametrized by the Casas-Ibarra parametrization \cite{Casas:2001sr} in which the Dirac neutrino Yukawa coupling matrix in the diagonal basis of the charged lepton and RH neutrino mass matrices is expressed as
\be \label{CI}
Y_\nu = \frac{1}{\langle H_u \rangle}\, U_{\rm PMNS}\, {\cal D}_\nu\, {\cal R}\, {\cal D}_R\,. \ee
Here $U_{\rm PMNS}$ is the leptonic mixing matrix, ${\cal D}_R = {\rm Diag.}(\sqrt{M_{R_1}},\sqrt{M_{R_2}},\sqrt{M_{R_3}})$, ${\cal D}_\nu = {\rm Diag.}(\sqrt{m_{\nu_1}},\sqrt{m_{\nu_2}},\sqrt{m_{\nu_3}})$ and $m_{\nu_i}$ are the light neutrino masses.  ${\cal R}$ is an unknown complex orthogonal matrix which parametrize the freedom in choice of $Y_\nu$ allowed by the seesaw formula, Eq. (\ref{seesaw}). We however do not consider this general case but discuss two phenomenologically interesting limits as described in the following.

\subsection{High scale seesaw (HSS)}

In $SO(10)$ based GUTs, all the quarks and leptons of a given generation are embedded in a single ${\bf 16}$-dimensional irreducible spinorial representation of the gauge group. In most of the situations, such a unification leads to an approximate equality between $Y_\nu$ and $Y_u$. For example, the renormalizable versions of supersymmetric $SO(10) $ models with one or more ${\bf 10}$-plet Higgs in the Yukawa sector always imply $Y_\nu = Y_u$  at the GUT scale \cite{Fritzsch:1974nn,Babu:1992ia}. The exact equality between $Y_u$ and $Y_\nu$ at the GUT scale is broken if the underlying model contains higher dimensional Higgs representations, such as ${\bf \overline{126}}$ and/or ${\bf 120}$, or if the corrections from higher order non-renormalizable operators are taken into consideration. In many of these cases an approximate relation, $Y_\nu \approx Y_u$, still holds (see \cite{Joshipura:2011nn} for example). There also exists a possibility in which the hierarchy among the couplings in $Y_\nu$ is widely different from those in $Y_u$. This situation is known to arise from the orbifolded $SO(10)$ GUTs in five or six spacetime dimensions \cite{Feruglio:2014jla,Feruglio:2015iua,Buchmuller:2017vut}. If  fermions are kept in the bulk and Higgs is localized on the brane, then the effective Yukawa coupling matrix in four dimensional theory is given by, for example $Y_f = F_{f_L} {\cal Y}_f F_{f_R}$ where ${\cal Y}_f$ is typically a matrix with elements of order unity, and $F_{f_L}$, $F_{f_R}$ are diagonal matrices with elements representing the values of profile factors at the given four dimensional fixed point. The later decides the inter-generational mass hierarchies in a given fermion sector. In general, if the GUT symmetry is broken by orbifolding then $F_{Q_L} \neq F_{L_L}$ and $F_{u_R} \neq F_{\nu_R}$. Therefore, the resulting $Y_\nu$ and $Y_u$ can have very different hierarchical structure and/or relative strength of magnitude \cite{Feruglio:2014jla,Feruglio:2015iua}. 

To accommodate these possibilities, we generically parametrize $Y_\nu$ at $M_S$ as:
\be \label{YnuHS}
Y_\nu =\epsilon\, {\rm Diag.}\left( \xi^N,\, \xi,\, 1 \right)\, Y_u\,, \ee
where $\epsilon$, $\xi$ and $N$ are real numbers which determine the relative  strength of Dirac Yukawa couplings of neutrinos with respect those of up type quarks. Since the largest coupling in $Y_u$ is already of ${\cal O}(y_t)$, very large value of $\epsilon$ leads to non-perturbative $Y_\nu$. We therefore consider $\epsilon \in [0.1-10]$, i.e. at most an order of magnitude difference between $Y_u$ and $Y_\nu$. The parameters $\xi$ and $N$ determine the hierarchical structure of couplings in $Y_\nu$.  For $\xi=1$, one obtains the hierarchy in $Y_\nu$ same as that in $Y_u$. Different hierarchical structure for $Y_\nu$ can be obtained using suitably chosen values of $\xi$ and $N$. For the above values of $\epsilon$, Eq. (\ref{seesaw}) leads to the masses of RH neutrinos in the range $10^7$ - $10^{16}$ GeV. This case is therefore named as the high scale seesaw (HSS) case. Since some of the couplings in $Y_\nu$ are of ${\cal O}(y_t)$ or large, one expects considerable running effects from the RH neutrinos even though their masses are close to $M_S$.

\subsection{Low scale seesaw (LSS)}
The running effects due to RH neutrinos are enhanced if the seesaw scale is close to the electroweak scale and they are strongly coupled with the THDM. A usual way to accommodate low seesaw scale is to consider $\epsilon \ll 1$ in Eq. (\ref{YnuHS}) which in turn decreases the masses of RH neutrinos by a factor of  $\epsilon^2$, as it can be seen from Eq. (\ref{seesaw}). Small $\epsilon$ however makes RH neutrinos very weakly coupled with the THDM and their effects on the running of couplings become negligible. There exists an alternate approach in which the low seesaw scale can be realized with ${\cal O}(1)$ couplings in $Y_\nu$ \cite{Kersten:2007vk,Dev:2013oxa,Chattopadhyay:2017zvs}. In this case, the smallness of the SM neutrino masses is attributed to the flavour structure of $Y_\nu$ and $M_R$ instead of the scale of RH neutrino masses or strength of couplings in $Y_\nu$. Due to the matrix structure of the seesaw formula, it is possible to choose the form of $Y_\nu$ and $M_R$ such that the Eq. (\ref{seesaw}) leads to vanishing ${\cal M}_\nu$. In the diagonal basis of RH neutrinos, they can be written as
\be \label{YnuLS}
Y_\nu = \left( \ba{ccc} y_1 & iy_1 & 0\\
y_2 & iy_2 & 0\\
y_3 & iy_3 & 0 \ea\right)\,,~~~~M_R =\left( \ba{ccc} M & 0 & 0\\
0 & M & 0\\
0 & 0 & M_3 \ea \right)\,. \ee
The above structures can be obtained from a global $U(1)$ symmetry \cite{Kersten:2007vk} or from a class of discrete symmetries \cite{Chattopadhyay:2017zvs}. One obtains ${\cal M}_\nu=0$ in this case  irrespective of the values of $y_i$, $M$ and $M_3$. It can be seen that the third RH neutrino does not couple with the SM leptons. The first two generations of RH neutrinos are strongly coupled if $y_i$s are chosen to be of order unity. Viable neutrino masses can be generated by introducing perturbations to the above structure. The strength of these perturbations are found to be very small \cite{Chattopadhyay:2017zvs} and therefore their contribution to RG effects are negligible. The form of $Y_\nu$ and $M_R$ given in Eq. (\ref{YnuLS}) therefore provides a good description of low scale seesaw (LSS) with strongly coupled RH neutrinos.

\section{RG evolution of the couplings and Constraints}
\label{sec:rge}
The framework under consideration involves many hierarchically separated scales. We perform renormalization group evolution of the couplings of effective field theory between different scales and match their values at the boundaries. The couplings are evolved using 2-loop RG equations and  matching at the thresholds are performed including 1-loop threshold corrections. The 2-loop RG equations are computed using a publicly available package SARAH \cite{Staub:2013tta} and they are listed in Appendix \ref{Appendix:RGE}. In the following subsections, we describe matching conditions, theoretical and phenomenological constraints on the couplings at various scales.

\subsection{Matching conditions at $M_S$}
We assume that supersymmetry is broken at the scale $M_S = 2 \times 10^{16}$ GeV and the theory below $M_S$ is an effective THDM with or without RH neutrinos at intermediate scales between $M_S$ and $M_t$. As it is discussed earlier, a tree-level matching between MSSM and THDM leads to relations given in Eqs. (\ref{lm_MS},\ref{lm567_MS}). The one-loop threshold corrections to these matching conditions and to the Yukawa couplings, in the absence of RH neutrinos, are given in \cite{Haber:1993an,Lee:2015uza}. These corrections depend on the sparticle spectrum at $M_S$ and also on the values of trilinear couplings and $\mu$ parameter. For simplicity, we assume that
\be \label{susy_spectrum}
m_{\tilde q} = m_{\tilde l} = M_1 = M_2 = M_3 = M_S\,, ~A_t = A_b = A_\tau =0\,,~\mu \approx {\cal O}(M_S)\,,  \ee
where $m_{\tilde q}$ and $m_{\tilde l}$ represent degenerate squark and slepton masses respectively, $\mu$ is higgsino mass parameter, $M_i$ are gaugino mass parameters and $A_{t,b,\tau}$ are the trilinear couplings of squarks and sleptons with relevant MSSM Higgs fields. The above assumption is realized in specific GUT based model \cite{Buchmuller:2015jna}. With these assumptions, the threshold corrections induced by squarks and sleptons are suppressed by the degeneracy of their masses and also by vanishing trilinear couplings. It can be seen from the expressions given in \cite{Lee:2015uza}, the one-loop threshold corrections to Yukawa couplings vanish entirely for the superpartner spectrum given in Eq. (\ref{susy_spectrum}). The  threshold corrections to the quartic couplings, induced through  one-loop box and triangle diagrams, depend only on the Yukawa couplings and $\mu$ parameter in the limit of vanishing trilinear couplings. We also estimate one-loop threshold corrections to quartic couplings which arise from the Dirac Yukawa couplings of RH neutrinos with the SM leptons. The expressions of  threshold corrections, used in our analysis, are listed in Appendix \ref{Appendix:threshold}. For the analysis presented in this paper, we have chosen $\mu = 0.1 M_S$ for definiteness.

\subsection{Constraints at intermediate scales}
While evolving gauge, Yukawa and quartic couplings from $M_S$ to $M_t$, we adopt the following procedure. If the mass scale $M_{R_j}$ of $j^{\rm th}$ RH neutrino $\nu_R^j$ appears below $M_S$, their running effects are taken into account by appropriate RG evolution. It is expected that for  renormalization scale $Q < M_{R_j}$, the $\nu_R^j$ should be integrated out from the spectrum and it should not contribute in the running of couplings. We implement this decoupling of heavy neutrinos by switching off the Yukawa couplings of $\nu_R^j$ with the SM leptons at the scale  $M_{R_j}$ and below. In other words, the values of elements of $j$th column in the matrix $Y_\nu$ is put to zero after the running scale $Q$ crosses the scale  $M_{R_j}$. This procedure is carried out sequentially for all RH neutrinos with masses between $M_S$ and $M_t$.

We also consider various theoretical constraints on the couplings which should be satisfied at every  scale. It is to be noted from Eq. (\ref{lm567_MS}) and RG equations that the couplings $\lambda_{5,6,7}$ vanish at all the scales. This happens because these couplings (as well as the Yukawa couplings in $\tilde{Y}_f$ for $f=u,d,e,\nu$) are protected by a softly broken $Z_2$ symmetry of an effective THDM theory and therefore if they are zero at one scale then they will not be generated by running\footnote{The threshold corrections at $M_S$ can generate non-zero values for $\lambda_{5,6,7}$ and $\tilde{Y}_f$. However, these threshold corrections are vanishing for the SUSY spectrum considered in Eq. (\ref{susy_spectrum}).}. A stability of scalar potential in Eq. (\ref{THDM_V}) would require the following conditions to be satisfied by the remaining couplings \cite{Gunion:2002zf}
\beqa \label{stability_cond}
\lambda_1(Q) & > & 0\,, \nonumber\\
\lambda_2(Q) & > & 0\,, \nonumber\\
\lambda_3^\prime(Q) \equiv  \lambda_3 (Q) + \sqrt{\lambda_1(Q) \lambda_2(Q)} & > & 0\,, \nonumber\\ 
\lambda_4^\prime(Q) \equiv  \lambda_4 (Q) + \lambda_3^\prime (Q)  & > & 0\,,
\eeqa
for  $M_t \le Q \le M_S$. The above conditions are sufficient to provide absolute stability for the electroweak vacuum. One may also consider a phenomenologically allowed and a more conservative possibility in which the electroweak vacuum is not completely stable but it is metastable with lifetime greater than the age of universe $\sim 10^{10}$ years. This replaces the last condition in Eq. (\ref{stability_cond}) by a weaker condition \cite{Bagnaschi:2015pwa}
\be \label{meta_cond}
\lambda(Q) \equiv \frac{4 \sqrt{\lambda_1(Q) \lambda_2(Q)}\, \lambda_4^\prime(Q)}{\lambda_1 (Q) + \lambda_2(Q) + 2 \sqrt{\lambda_1(Q) \lambda_2(Q)}} - \lambda_{\rm meta} \gsim 0\,,
\ee
where 
\be \label{lm_meta}
\lambda_{\rm meta} = -\frac{2.82}{41.1 + \log_{10}\left( \frac{Q}{\rm GeV}\right)}\,. \ee
The derivation of the above condition involves probability of tunnelling into the true vacuum which was estimated in case of single scalar field with $\phi^4$ potential in \cite{Isidori:2001bm} including the quantum effects. Following a similar approach, the metastability condition, Eq. (\ref{meta_cond}),  was derived in \cite{Bagnaschi:2015pwa} after mapping the THDM scalar potential into single field potential using the first three conditions in Eq. (\ref{stability_cond}) with convinient choice of gauge and field basis. More details about the derivation of Eq. (\ref{meta_cond}) can be found in an Appendix of \cite{Bagnaschi:2015pwa}. For most of the cases studied here, it is found that the first three conditions of Eq. (\ref{stability_cond}) are always satisfied as a consequence of the boundary values set by supersymmetry at $M_S$. Hence, the stability or metastability of the electroweak vacuum is solely decided by values of $\lambda_4^\prime$ and $\lambda$ at the intermediate scales.

\subsection{Matching conditions and constraints at $M_t$}
The RG equations determine the values of the couplings of effective THDM at the scale $M_t$. At $M_t$, the gauge and Yukawa couplings are matched with their experimentally measured values while the quartic couplings determine the Higgs potential which is subject to the constraints imposed by  consistent electroweak symmetry breaking and measurement of Higgs properties. 

The electroweak symmetry breaking is governed by the VEVs of two Higgs fields,
\be \label{Higgs_vev}
\langle H_i \rangle = \frac{1}{\sqrt{2}} \left( \ba{c} 0 \\ v_i \ea \right)\,
\ee
at the minimum of scalar potential. In our notation, $v_1 = v_d$ and $v_2 = v_u$ and they define electroweak VEV $v$ and a parameter $\tan\beta$ as the following
\be \label{tanbeta}
v \equiv \sqrt{v_u^2 + v_d^2} \approx 246\, {\rm GeV}\,,~~~\tan\beta \equiv \frac{v_u}{v_d}\,.
\ee
The breaking of electroweak symmetry gives rise to five physical Higgs bosons in the spectrum. These are two charged and CP-even ($H^\pm$), two neutral and CP-even ($h$ and $H$), and a neutral and  CP odd ($A$) scalars.  At the minimum of potential, the parameters $m_1^2$, $m_2^2$ and $m_{12}^2$ can be replaced by the following tree-level expressions \cite{Gunion:2002zf}:
\beqa \label{potential_replaced}
m_{12}^2 &=& M_A^2\, \sin\beta \cos\beta, \nonumber \\
m_1^2 &=& M_A^2 \sin^2\beta - \frac{1}{2} v^2 \left( \lambda_1 \cos^2 \beta + (\lambda_3 + \lambda_4) \sin^2\beta\right)\,, \nonumber \\
m_2^2 &=& M_A^2 \cos^2\beta -\frac{1}{2}  v^2 \left( \lambda_2 \sin^2 \beta + (\lambda_3 + \lambda_4) \cos^2\beta\right)\,,
\eeqa
where $M_A$ is the mass of pseudo-scalar Higgs in $\overline{\rm MS}$ renormalization scheme. With these replacements, the scalar potential given in Eq. (\ref{THDM_V}) is completely specified by $M_A$, $v$, $\tan\beta$ and the quartic couplings.

The mass of charged Higgs is given by
\be \label{M_Hp}
M_{H^\pm}^2 = M_A^2 - \frac{1}{2} \lambda_4\, v^2\,. \ee
The CP-even scalar states mix with each other and it is convenient to work in so-called Higgs basis in which only one of the combinations of $H_1$ and $H_2$, namely $h_1 \equiv \cos\beta H_1 + \sin\beta H_2$, acquires a non-trivial VEV. The combination orthogonal to $h_1$ is identified as $h_2$ such that $\langle h_2 \rangle = 0$.  In the basis $\{h_1, h_2\}$, the  squared mass matrix of CP-even neutral scalars is given as \cite{Gunion:2002zf}
\be \label{MH_squared}
{\cal M}^2 = M_A^2 \left( \ba{cc} 0 & 0\\ 0 & 1\ea\right) + v^2  \left( \ba{cc} g_{11} &  g_{12}\\  g_{12}& g_{22}\ea\right)\,,\ee
where
\beqa \label{g_ij}
g_{11} & = & \lambda_1 \cos^2 \beta + \lambda_2 \sin^2 \beta + 2(\lambda_3 + \lambda_4) \sin^2\beta \cos^2\beta\,, \nonumber \\
g_{12} & =& -\cos\beta \sin\beta \left( \lambda_1 \cos^2\beta - \lambda_2 \sin^2\beta - (\lambda_3 + \lambda_4) \cos2\beta \right)\,, \nonumber \\
g_{22} & = &  (\lambda_1+\lambda_2) \cos^2 \beta \sin^2\beta - 2(\lambda_3 + \lambda_4) \sin^2\beta \cos^2\beta\,. \eeqa
Performing another change of basis 
\be \label{rotation}
\left( \ba{c} H \\ h\ea\right) = U \left( \ba{c} h_1 \\ h_2\ea\right) \,,~~U \equiv \left( \ba{cc} \cos(\beta-\alpha) & -\sin(\beta-\alpha) \\
 \sin(\beta-\alpha) & \cos(\beta-\alpha) \ea \right)\,,\ee
 such that
 \be\label{}
 U\, {\cal M}^2\, U^\dagger = {\rm Diag.}(M_H^2, m_h^2)\,,\ee
where $m_h$ and $M_H$ are masses of physical CP-even neutral Higgs bosons. These masses and the mixing angle $\beta-\alpha$ are computed from the above diagonalization. We assume that $m_h \le M_H$ and identify the lighter state with the observed SM like Higgs. In order to make consistent matching between theory and data, we convert the running mass $m_h$ evaluated at the scale $M_t$ to the pole mass $M_h$ using the following formula
\be \label{Higgs_matching}
M_h^2 = m_h^2(M_t) + \delta m_h^2 (M_t)\,, \ee
where $\delta m_h^2$ is the SM one-loop self-energy correction and its expression in terms of $\overline{\rm MS}$ parameters is given in \cite{Draper:2013oza,Lee:2015uza}. We do not include the contributions from the other scalars in $\delta m_h^2$ and assume that they are sub-dominant compared to the SM contributions. Numerically we find that the correction to the Higgs mass induced by the second term in the above equation remains less than $0.6$ GeV.

The spectrum of physical scalars and the angle $\beta-\alpha$ are subject to several direct and indirect constraints. We consider the experimentally measured value from \cite{Aad:2015zhl} and allow a deviation of $\pm 3$ GeV from the central value to account for theoretical uncertainty in estimating the value of $M_h$. In THDM of type II, the charged Higgs with mass up to 580 GeV is disfavoured by $b \to s + \gamma$ measurements at $95\%$ confidence level \cite{Misiak:2017bgg} for almost any value of $\tan\beta$. Further, it can be seen from Eq. (\ref{rotation}) that the couplings of $h$ with the weak bosons is proportional to $\sin^2(\beta-\alpha)$.  These couplings are constrained by the signal strength of Higgs decaying into pair of vector bosons. The results of a recent global fit of THDM parameters indicate that the deviation from $\beta -\alpha = \pi/2$ cannot be larger than 0.055 in the case of type II THDM \cite{Chowdhury:2017aav}. The above constraints on the spectrum are summarized as
\beqa \label{constraints_lowscale}
M_h & = & (125 \pm 3)\, {\rm GeV}\, \nonumber \\
M_{H^\pm} & \gsim & 580\, {\rm GeV}\, \nonumber \\
|\cos(\beta - \alpha)| & \lsim & 0.055\,. \eeqa
We find that the limit on $M_{H^\pm}$ also puts a lower bound on $M_A$, from Eq. (\ref{M_Hp}), and hence a lower bound on $M_H$ as well, since all the quartic couplings are determined from the supersymmetry in this model. The bounds on the masses of THDM scalars obtained in this way are more stringent than the direct search bounds, see for example \cite{Chowdhury:2017aav} and references therein. We also investigate the effects of these scalars on electroweak precision observables. For this, we estimate corrections to $W$ boson mass $M_W$ and effective weak mixing angle $\theta_{\rm eff}^{\rm lept}$ in THDM and compare them with their measured values following the procedure adopted in \cite{Broggio:2014mna}. These constraints are found to be always satisfied for the values of $M_A$ allowed by the constraints listed in Eq. (\ref{constraints_lowscale}).

We use the experimental values of gauge couplings and fermion mass parameters measured at different scales and evolve them to the scale $M_t$. This is described in  Appendix \ref{Appendix:inputs}. These values are listed in Table \ref{tab:inputs}.
\begin{table}[!ht]
\begin{center}
\begin{tabular}{|cc|cc|cc|cc|}
	\hline
	Parameter & Value & Parameter & Value & Parameter & Value & Parameter & Value\\
	\hline
	$g_1$ & 0.46315 & $m_u$ & 1.21 MeV & $m_d$ & 2.58 MeV & $m_e$ & 0.499 MeV \\
	$g_2$ & 0.65403 & $m_c$ & 0.61 GeV & $m_s$ &  52.74  MeV & $m_{\mu}$&0.104 GeV\\
	$g_3$ & 1.1630 & $m_t$ & 163.74 GeV & $m_b$ & 2.72 GeV & $m_{\tau}$&1.759 GeV \\
	\hline
\end{tabular}
\caption{The values of the SM parameters, in ${\overline{\rm MS}}$ scheme, at renormalization scale $M_t$ used in our analysis. More details are given in Appendix \ref{Appendix:inputs}.}
\label{tab:inputs}
\end{center}
\end{table}
The Yukawa couplings are matched at $M_t$ using the following tree-level relations in a convenient basis:
\beqa \label{yuk}
Y_u (M_t) & =& \frac{\sqrt{2}}{v \sin\beta}\, {\rm Diag.}\left( m_u(M_t),\, m_c(M_t),\, m_t(M_t)\right)\, \nonumber \\
Y_d (M_t) & = &\frac{\sqrt{2}}{v \cos\beta}\,V_{\rm CKM}\,  {\rm Diag.} \left( m_d(M_t),\, m_s(M_t),\, m_b(M_t)\right)\, \nonumber \\
Y_e (M_t) & = &\frac{\sqrt{2}}{v \cos\beta}\, {\rm Diag.}\left( m_e(M_t),\, m_\mu(M_t),\, m_\tau(M_t)\right)\,,\eeqa
where $V_{\rm CKM}$ is quark mixing matrix. For its elements, we use the latest values from the PDG \cite{Patrignani:2016xqp}. In the standard parametrization, this matrix is given in terms of three mixing angles and a phase. We use their values: $\sin \theta_{12}^q =0.2251$, $\sin \theta_{23}^q=0.041$, $\sin\theta_{13}^q= 0.0036$ and $\delta_{\rm CKM} = 68.04^\circ$. In the next section, we discuss our procedure for solving RG equations and present the obtained results in details.

\section{Numerical results}
\label{sec:numerical}
We numerically solve the 2-loop RG equations for different cases and implement matching conditions and constraints as discussed in the previous sections. First, the gauge and Yukawa couplings are evolved from $M_t$ to $M_S$ using their values at $M_t$ and 1-loop RG equations. The running of these couplings do not depend on quartic couplings at 1-loop. The values of quartic couplings at $M_S$ are then obtained using conditions given in Eqs. (\ref{lm_MS}, \ref{lm567_MS}) and one-loop threshold corrections listed in Appendix \ref{Appendix:threshold}. We then perform full 2-loop evolution of the gauge, Yukawa and quartic couplings form $M_S$ to $M_t$ and compare the obtained values of gauge and Yukawa couplings with their input values at $M_t$. The couplings are then evolved again from $M_t$ to $M_S$ using 2-loop RG equations. This procedure is carried out iteratively until the values of couplings converge to their input values at $M_t$. During the running, we check that the stability or metastability conditions, given in Eqs. (\ref{stability_cond}, \ref{meta_cond}), are satisfied at all intermediate scales. Once the convergence is obtained, we calculate the masses of physical scalars using Eqs. (\ref{M_Hp},\ref{MH_squared}) as function of input parameter $M_A$ and $\tan\beta$.

Before we discuss the results of our numerical analysis, we outline our method of obtaining the RH neutrino mass spectrum in the case of high scale seesaw.  The RH neutrino masses are evaluated using the boundary condition given in Eq. (\ref{YnuHS}) and seesaw formula Eq. (\ref{seesaw}) with the following replacement for ${\cal M}_\nu$:
\be \label{neutrinos}
{\cal M}_\nu = U_{\rm PMNS}\, {\rm Diag.}(m_{\nu_1},m_{\nu_2},m_{\nu_3})\, U_{\rm PMNS}^T\,, \ee
where $U_{\rm PMNS}$ is leptonic mixing matrix and it is parametrized by three mixing angles and three CP phases in the standard parametrization \cite{Patrignani:2016xqp}. Since we consider a basis in which charged lepton Yukawa matrix is diagonal\footnote{The charged lepton Yukawa matrix is taken diagonal at $M_t$. Further, it remains almost diagonal at all the scales as the magnitude of off-diagonal elements in $Y_e$ induced by RG evolution is very small.}, the above form of ${\cal M}_\nu$ leads to realistic lepton mixing. We assume normal ordering in the masses of neutrinos, and obtain the values of $m_{\nu_2}$ and $m_{\nu_3}$ in terms of $m_{\nu_1}$ from the solar and atmospheric squared mass differences, $\Delta m_{21}^2 \equiv m_{\nu_2}^2-m_{\nu_1}^2$ and $\Delta m_{31}^2 \equiv m_{\nu_3}^2-m_{\nu_1}^2$, respectively. The values used in our analysis for leptonic mixing angles, CP phases and squared mass differences are as follows:
$$\sin^2\theta_{12} = 0.307,\, \sin^2\theta_{23} =  0.538,\, \sin^2\theta_{13} = 0.02206,\,$$
$$ \delta_{CP} = -\frac{\pi}{2},\, \alpha_{21} = \alpha_{31} = 0,\, $$
$$\Delta m_{21}^2 = 7.40 \times 10^{-5}\, {\rm eV}^2,\, \Delta m_{31}^2 = 2.494 \times 10^{-3}\, {\rm eV}^2.$$
The above values for mixing angles, $\delta_{\rm CP}$, $\Delta m_{21}^2$ and $\Delta m_{31}^2$ are taken from the latest global fit of neutrino oscillation data reported as NuFIT 3.2 (2018) \cite{Esteban:2016qun}. We also fix the lightest neutrino mass $m_{\nu_1} = 0.001$ eV for definiteness. Note that we use low energy values of neutrino masses and mixing parameters in seesaw formula to determine the RH neutrino mass spectrum. It is assumed that these parameters do not change significantly under RG evolution. Even if the running effects are taken into consideration for these parameters, the change in the masses of RH neutrinos obtained by Eq. (\ref{seesaw}) is small and therefore it has negligible effects on the results of our analysis.

\subsection{Without seesaw}
\begin{figure}[t]
\centering
\subfigure{\includegraphics[width=0.49\textwidth]{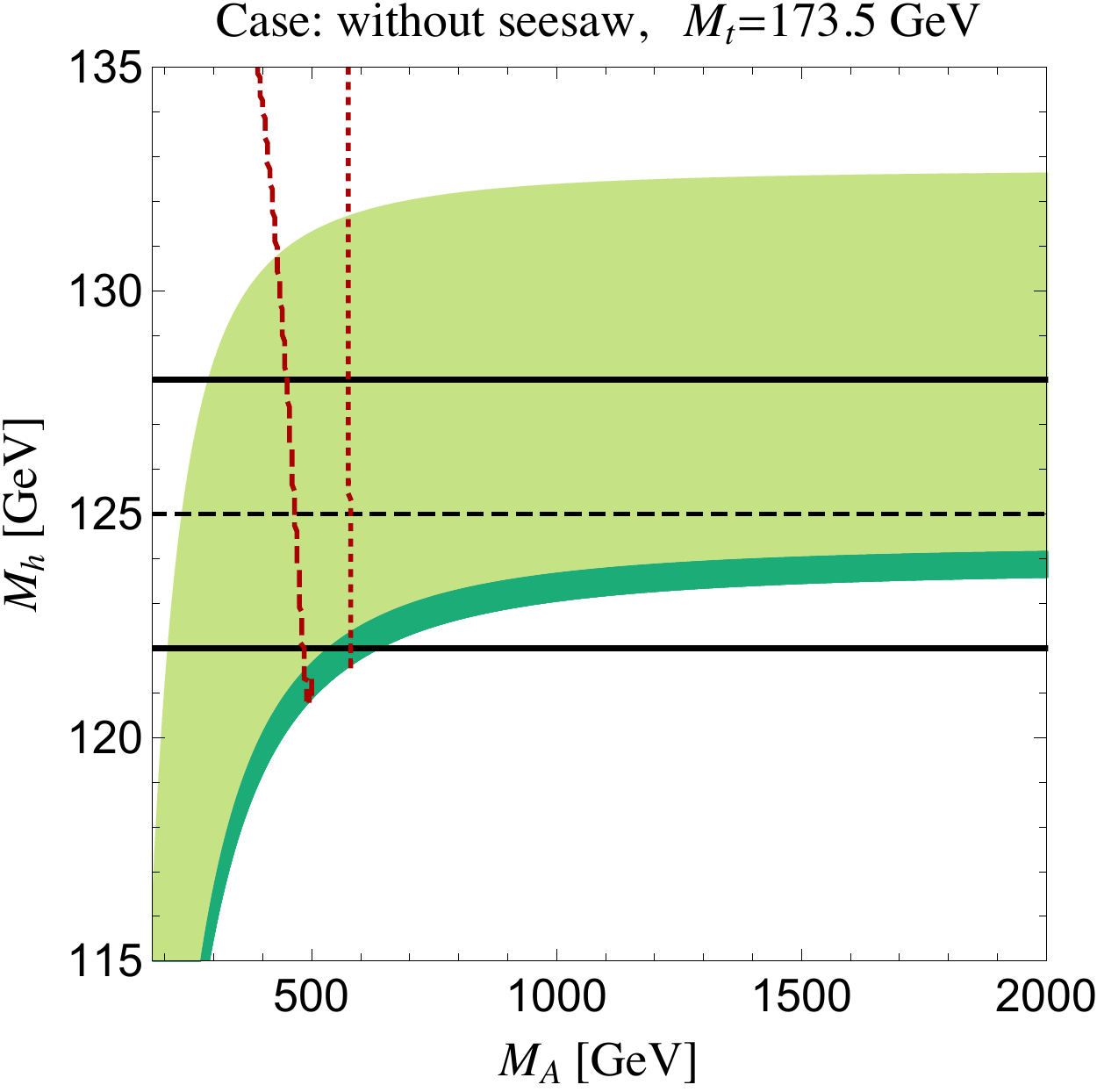}} \hspace*{0.1cm}
\subfigure{\includegraphics[width=0.48\textwidth]{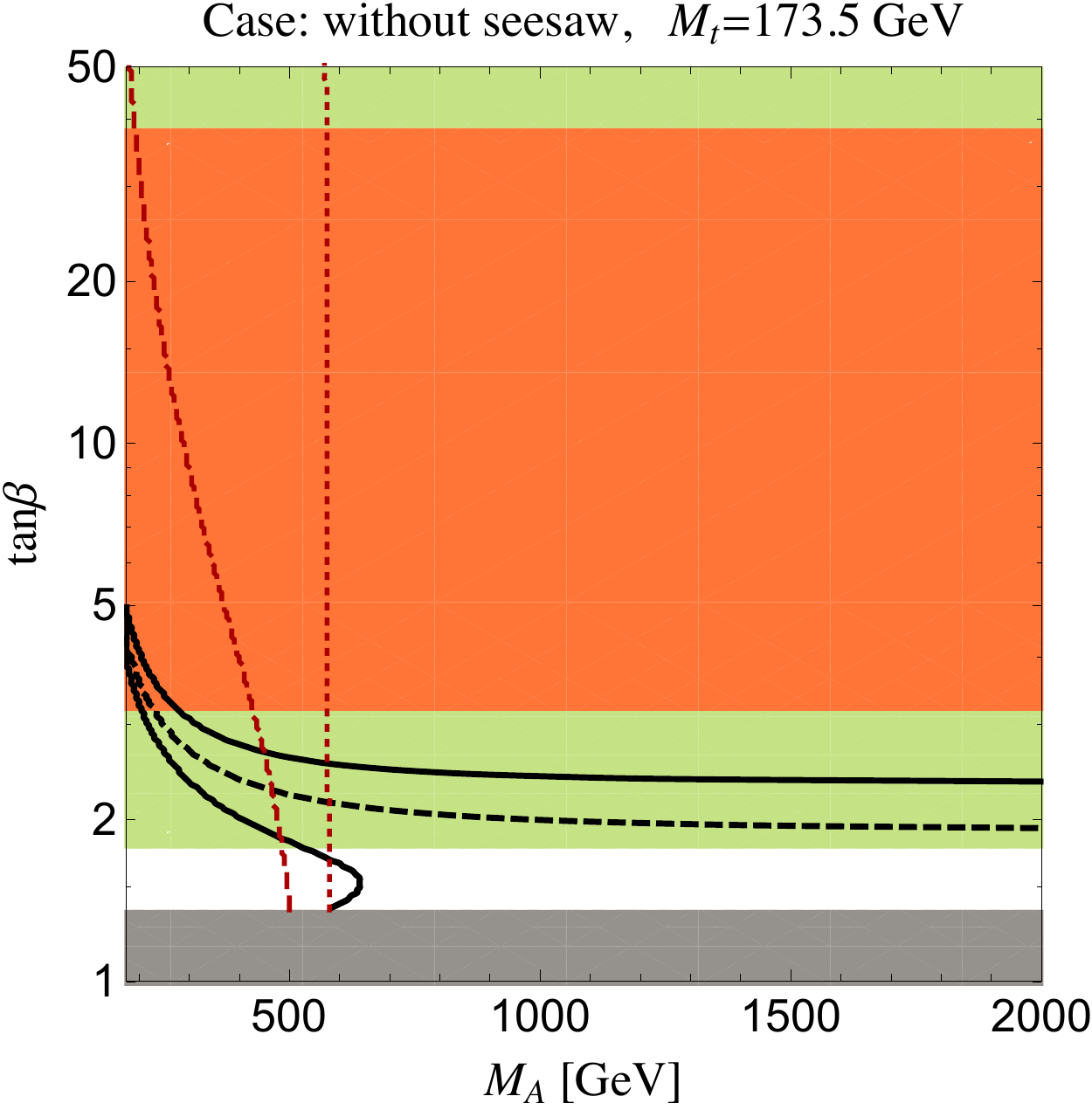}}\\
\caption{Left panel: Regions allowed in $M_A$-$M_h$ plane by absolute stability (darker green) and metastability (green) of the scalar potential. Right panel: Regions of absolute stability (unshaded), metastability (green) and unstability (orange) of the scalar potential in $M_A$-$\tan\beta$ plane. In the regions shaded by grey colour, the couplings become non-perturbative. In both the panels, the region between two black lines correspond to the values of $M_h$ in the range of 122-128 GeV while the dashed black line is for $M_h = 125$ GeV. The region on the left side of vertical dotted and dashed red lines are disfavoured by the constraints, $M_{H^\pm} > 580$ GeV and $|\cos(\beta-\alpha)| < 0.055$, respectively.}
\label{fig1}
\end{figure}

We first analyse a case without RH neutrinos. In this case, it is observed in \cite{Bagnaschi:2015pwa} that the quartic couplings $\lambda_1$, $\lambda_2$ and $\lambda_3$ remain positive at all the scales because of their boundary conditions at $M_S$ and hence the first three of the conditions of Eq. (\ref{stability_cond}) are always satisfied. $\lambda_4$ is negative at $M_S$ and it remains negative at the intermediate scales which in turn requires sufficiently large $\lambda_{1,2,3}$ in order to satisfy the last condition of Eq. (\ref{stability_cond}) or Eq. (\ref{meta_cond}) for stability or metastability respectively. For very small values of $\tan\beta$, it is observed that the couplings $\lambda_1$ and $\lambda_3$ evolve slowly. The magnitude of $\lambda_2$ however increases rapidly while running from $M_S$  to $M_t$ because of large and negative contribution to 1-loop beta function from the top quark loop. In this case, it is large and positive $\lambda_2$ which makes $\lambda^\prime_4(Q)>0$ and leads to a stable scalar potential. With increasing $\tan\beta$ the top quark Yukawa coupling $y_t$ decreases which in turn slows down the running of $\lambda_2$. In this case, $\lambda_2$ does not attain large enough value to make the potential stable. For very large values of $\tan\beta$, the bottom quark Yukawa coupling $y_b$ becomes as strong as $y_t$ and hence the running effect in $\lambda_1$ becomes as strong as that in $\lambda_2$ and both are driven to positive and large values at  scales below $M_S$. Their combined contributions increase the value of $\lambda^\prime_4(Q)$ which drive electroweak vacuum towards metastability or stability. We obtain the values of quartic couplings at $M_t$ and compute the scalar mass spectrum and impose the constraints given in Eq. (\ref{constraints_lowscale}). The results are displayed in Fig. \ref{fig1}.

We find that the top Yukawa coupling becomes non-perturbative for $\tan\beta \le 1.36$ and hence the perturbative approach of RG evolution breaks down. A stable scalar potential is achieved for $1.36 < \tan\beta \le 1.77$.  As it is explained earlier, the large positive value of $\lambda_2$ makes $\lambda_4^\prime (Q) > 0$ for all $Q$ between $M_S$ and $M_t$ in this case. For $1.77 < \tan\beta \le 3.18$, the electroweak vacuum becomes metastable while the region $3.18 < \tan\beta \le 38.5$ is disfavoured by an unstable vacuum. The potential again becomes metastable for $\tan\beta > 38.5$ because of large contribution of bottom quark in the running of $\lambda_1$. It can be seen from Fig. \ref{fig1} that the $b \to s+\gamma$ constraint on $M_{H^\pm}$ implies $M_A \gsim 580$ GeV. This together with constraint on Higgs mass restrict the values of $\tan\beta$ in a very narrow range, i.e. $1.36 < \tan\beta \le 2.5$. We find that our results are in qualitative agreement with the results obtained in \cite{Bagnaschi:2015pwa} but for $M_S = 2 \times 10^{17}$ GeV.

\subsection{Case: HSS}
\begin{figure}[t]
\centering
\subfigure{\includegraphics[width=0.48\textwidth]{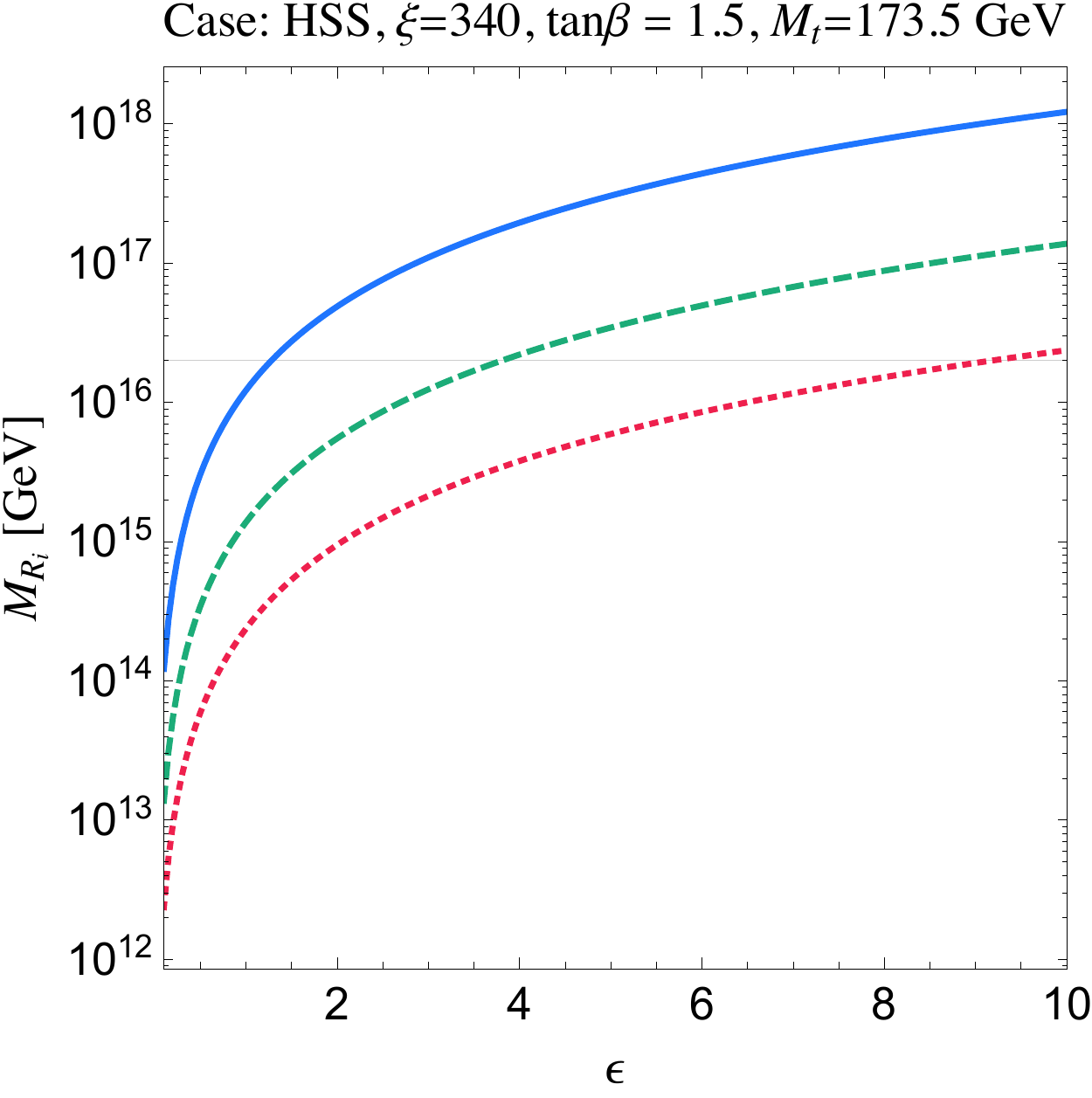}} \hspace*{0.1cm}
\subfigure{\includegraphics[width=0.48\textwidth]{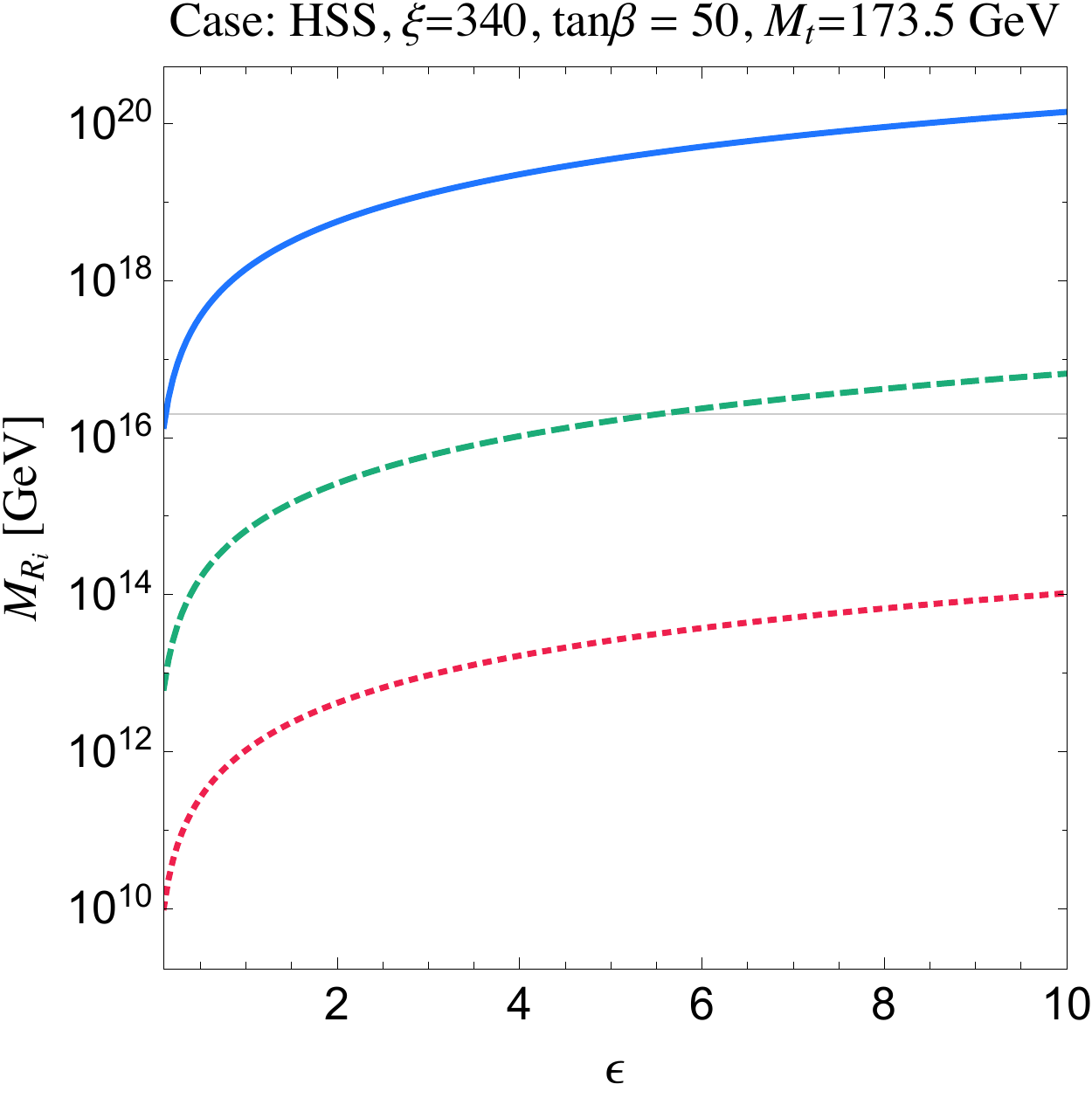}}\\
\caption{The mass spectrum of RH neutrinos with respect to the $\epsilon$ parameter, as obtained from the condition in Eq. (\ref{YnuHS}) with $\xi =340$ and $N=2.07$ and seesaw formula Eq. (\ref{seesaw}) and Eq. (\ref{neutrinos}) for $\tan\beta =1.5$ (left panel) and $\tan\beta = 50$ (right panel). The horizontal line corresponds to the value $M_S = 2 \times 10^{16}$ GeV.}
\label{fig2a}
\end{figure}

\begin{figure}[t]
\centering
\subfigure{\includegraphics[width=0.48\textwidth]{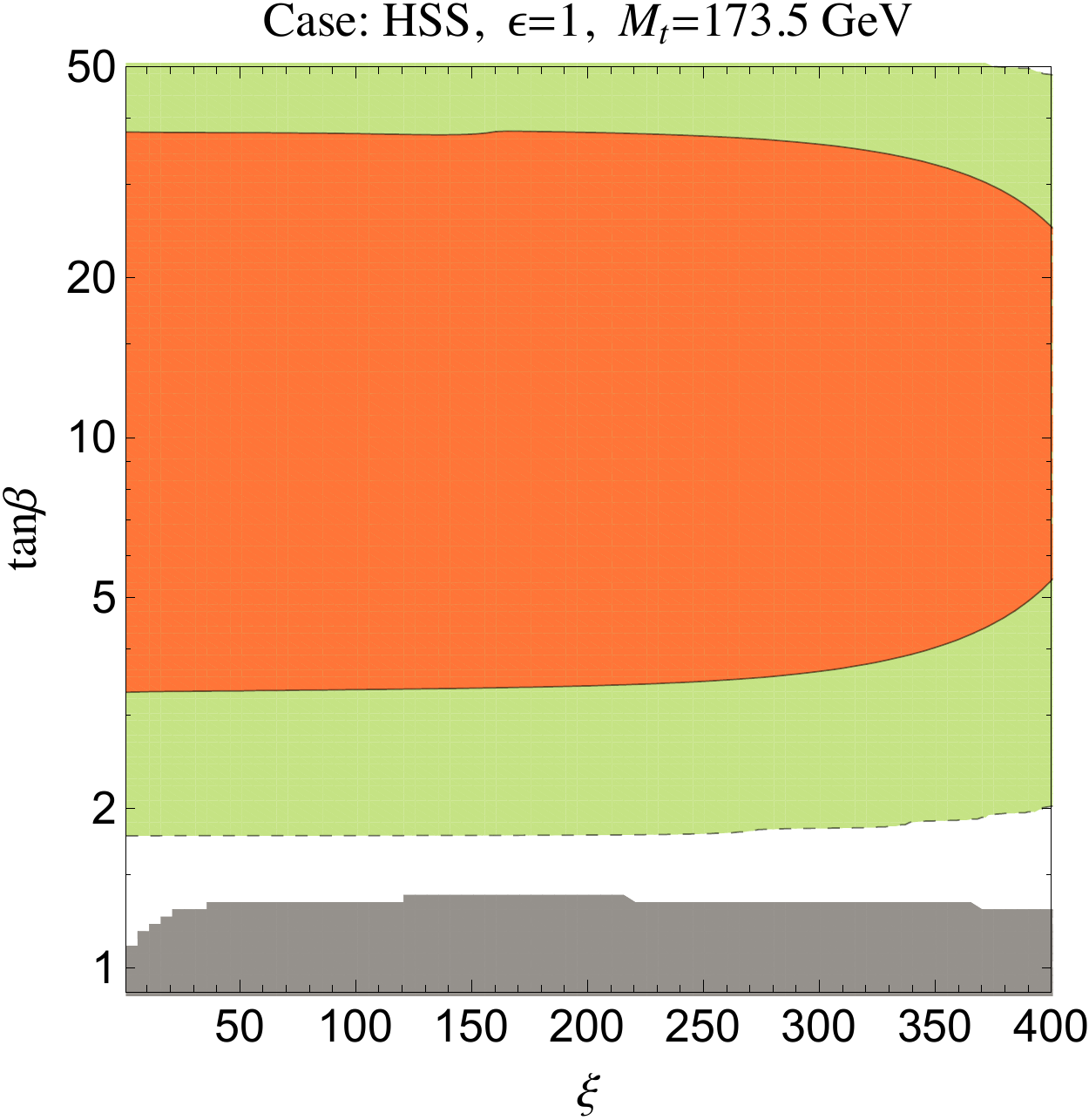}} \hspace*{0.1cm}
\subfigure{\includegraphics[width=0.48\textwidth]{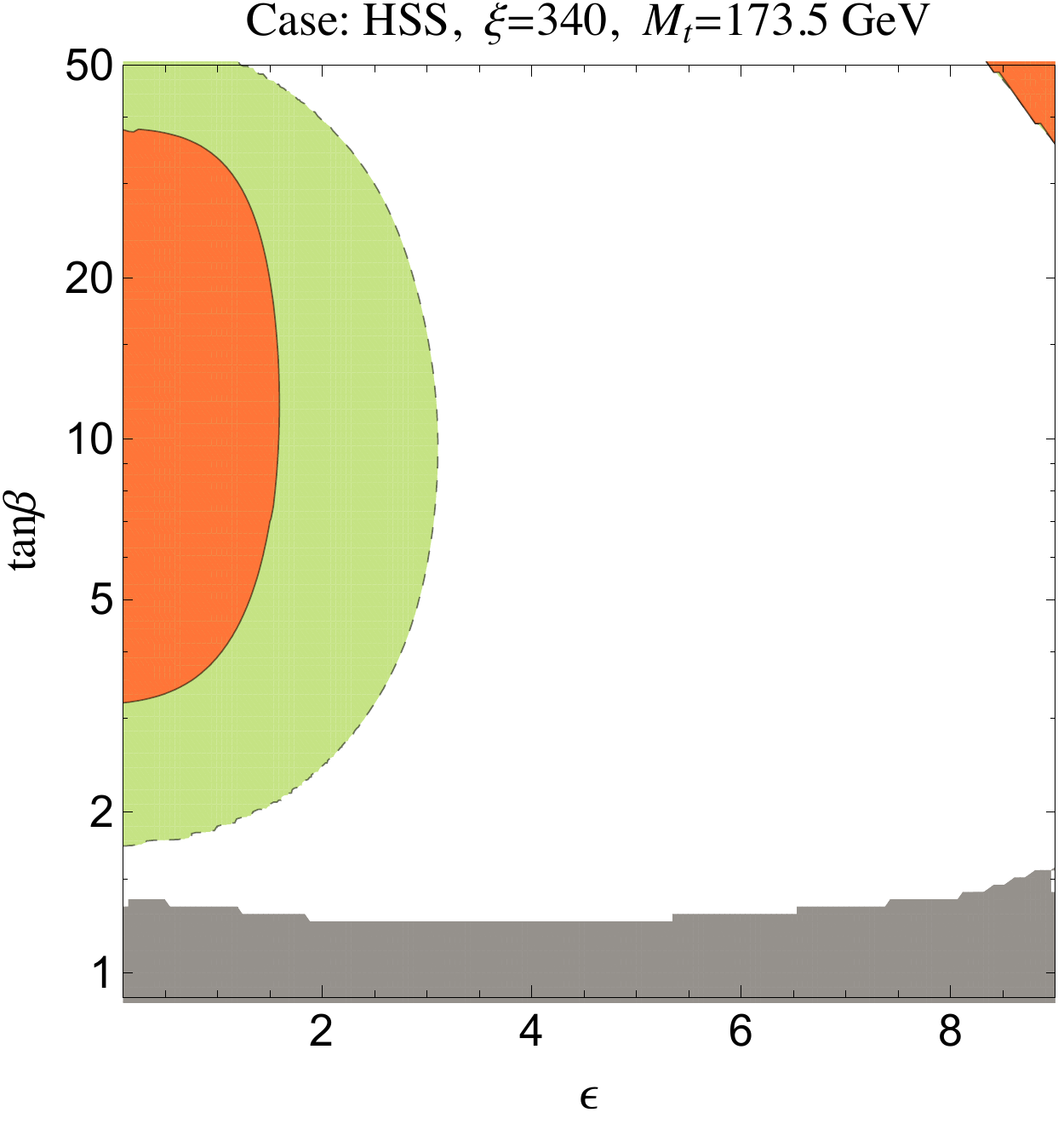}}\\
\caption{The allowed values of $\tan\beta$ by an absolute stability (unshaded) and metastability (green) of the scalar potential as function of $\xi$ (in the left panel) and $\epsilon$ (in the right panel) parameters as defined in Eq. (\ref{YnuHS}) with $N=2.07$. The orange region corresponds to unstable scalar potential while the grey region in the bottom is where the perturbativity of the couplings is lost.}
\label{fig2}
\end{figure}

The effects of high scale seesaw mechanism on vacuum stability is studied by incorporating RH neutrino thresholds in the RG evolution and by considering the $SO(10)$ GUT inspired boundary conditions, Eq. (\ref{YnuHS}), at the scale $M_S$. The parametrization is chosen in such a way that $\xi$ and $N$ control inter-generational hierarchy in $Y_\nu$. For example, the couplings in  $Y_\nu$ are as hierarchical as those in $Y_u$ for $\xi \approx 1$. Further, from the extrapolated values of couplings in $Y_u$ we find that all the three RH neutrino species couple with the SM leptons with equal strength for $N \approx 2.07$ and $\xi \approx 340$. The parameter $\epsilon$ sets overall scale of $Y_\nu$ and hence the scale of RH neutrinos. For example, the RH neutrino masses obtained using Eqs. (\ref{seesaw},\ref{YnuHS}), as explained in the beginning of this section, are displayed  Fig. \ref{fig2a}  for $\xi = 340$ and $N=2.07$. 
It can be seen that for small values of $\tan\beta$, all the neutrinos decouple from the effective theory for $\epsilon\ge 9$. For $\epsilon \in [0.1,9]$, at least one neutrino remains in the spectrum below $M_S$. We fix $N=2.07$ and evaluate the effect of neutrino(s) below $M_S$ on the vacuum stability in two different cases. First we fix $\epsilon =1$ and vary $\xi$ in the range from $1$ to $400$ to investigate the effects of different hierarchical structure of $Y_\nu$. In the second, we fix $\xi = 340$  which corresponds to all the three neutrinos coupled with approximately equal strength and vary $\epsilon$ in the range $0.1$ to $9$. The constraints on $\tan\beta$ obtained by the consideration of vacuum stability for these cases are displayed in Fig. \ref{fig2}.

The presence of RH neutrinos below $M_S$ modifies the running of quartic couplings in the following way.  As it can be seen from the expressions of beta functions given in Appendix \ref{Appendix:RGE},  RH neutrinos contribute to the running of $\lambda_2$ with a term proportional to $-{\rm Tr}(Y_\nu Y_\nu^\dagger Y_\nu Y_\nu^\dagger)$.  This contribution increases the value of $\lambda_2$ from its already positive value at $M_S$ while evolving from $M_S$ to $M_{R_i}$.  As a result of this, $\lambda_2$ takes relatively larger value at the scales between $M_{R_i}$ and $M_t$ in comparison to the case without RH neutrinos. This helps in obtaining increased $\lambda^\prime_4(Q)$ and hence this effect can drive the potential towards stability depending on the magnitude of couplings in $Y_\nu$. As it can be seen in the right panel in Fig. \ref{fig2}, this happens only when $\epsilon > 1.5$ for which RH neutrinos are very strongly coupled with  SM leptons. One finds that all the values of $\tan\beta$ between 1.2 and 50 are allowed by stability constraints for $\epsilon \in [1.5, 8]$.  For $\epsilon > 8$ and large $\tan\beta$, the quartic coupling $\lambda_2$ suddenly becomes negative resulting into an unstable region displayed in the upper right corner of $\tan\beta$-$\epsilon$ plane in Fig. \ref{fig2}. At two-loop, the contribution from RH neutrino Yukawa couplings in the running of $\lambda_2$ comes with opposite sign in comparison to that of one-loop as it can be seen from the expressions of beta functions given in Appendix \ref{Appendix:RGE}. For large $\epsilon$ and just  before the coupling $Y_\nu$ becomes non-perturbative, the two-loop beta function dominates over the one-loop. This effect is further enhanced by large $\tan\beta$ through large $Y_d$ and $Y_e$. This results into large positive value of the total beta function which rapidly drives $\lambda_2$ from its initial positive value at $M_S$ to negative value at the scale below $M_S$.

For $\epsilon \le 1.5$, the heavy neutrinos do not significantly change the stability constraints obtained in the case without RH neutrinos as it can be seen from both the panels in Fig. \ref{fig2}. Although all the neutrinos are not decoupled in this case, they do not have couplings strong enough to make any significant change in the running of quartic couplings. Even for $\xi = 340$, $N =2.07 $ which leads to all the three neutrinos with couplings of ${\cal O}(y_t)$, one obtains similar constraints on $\tan\beta$ as in the case without RH neutrinos.

\begin{figure}[t]
\centering
\subfigure{\includegraphics[width=0.45\textwidth]{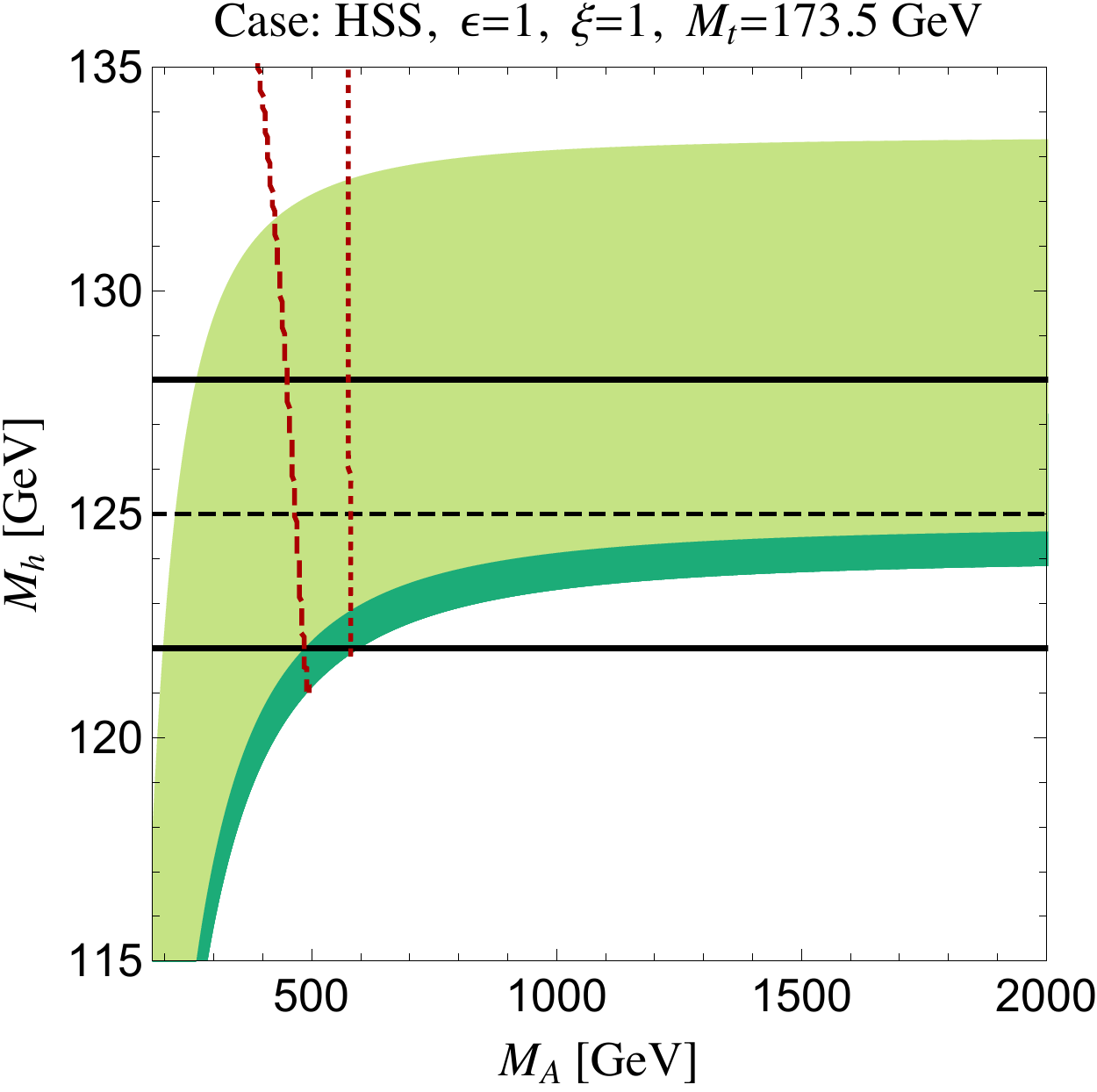}} \hspace*{0.1cm}
\subfigure{\includegraphics[width=0.44\textwidth]{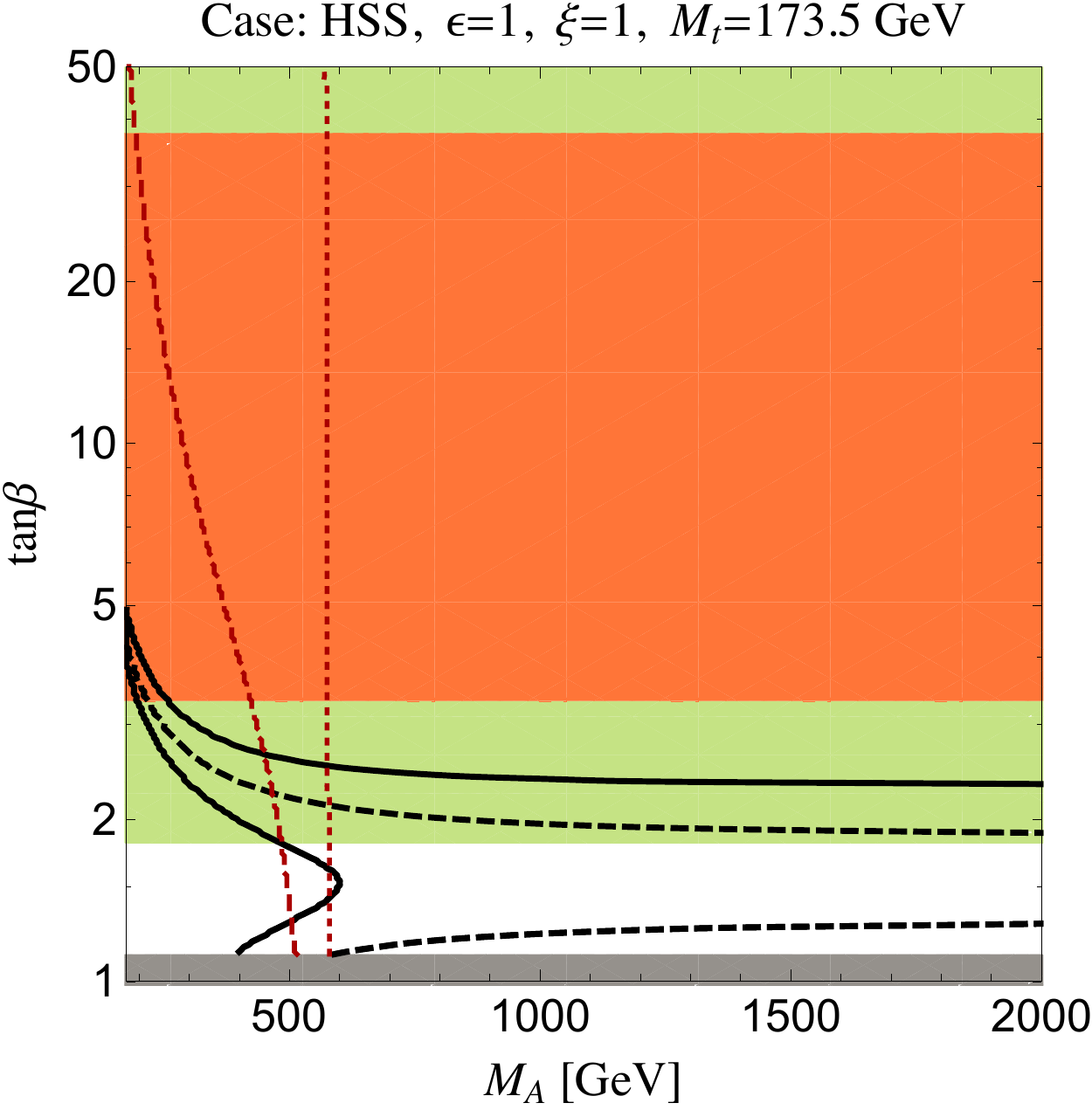}}
\subfigure{\includegraphics[width=0.45\textwidth]{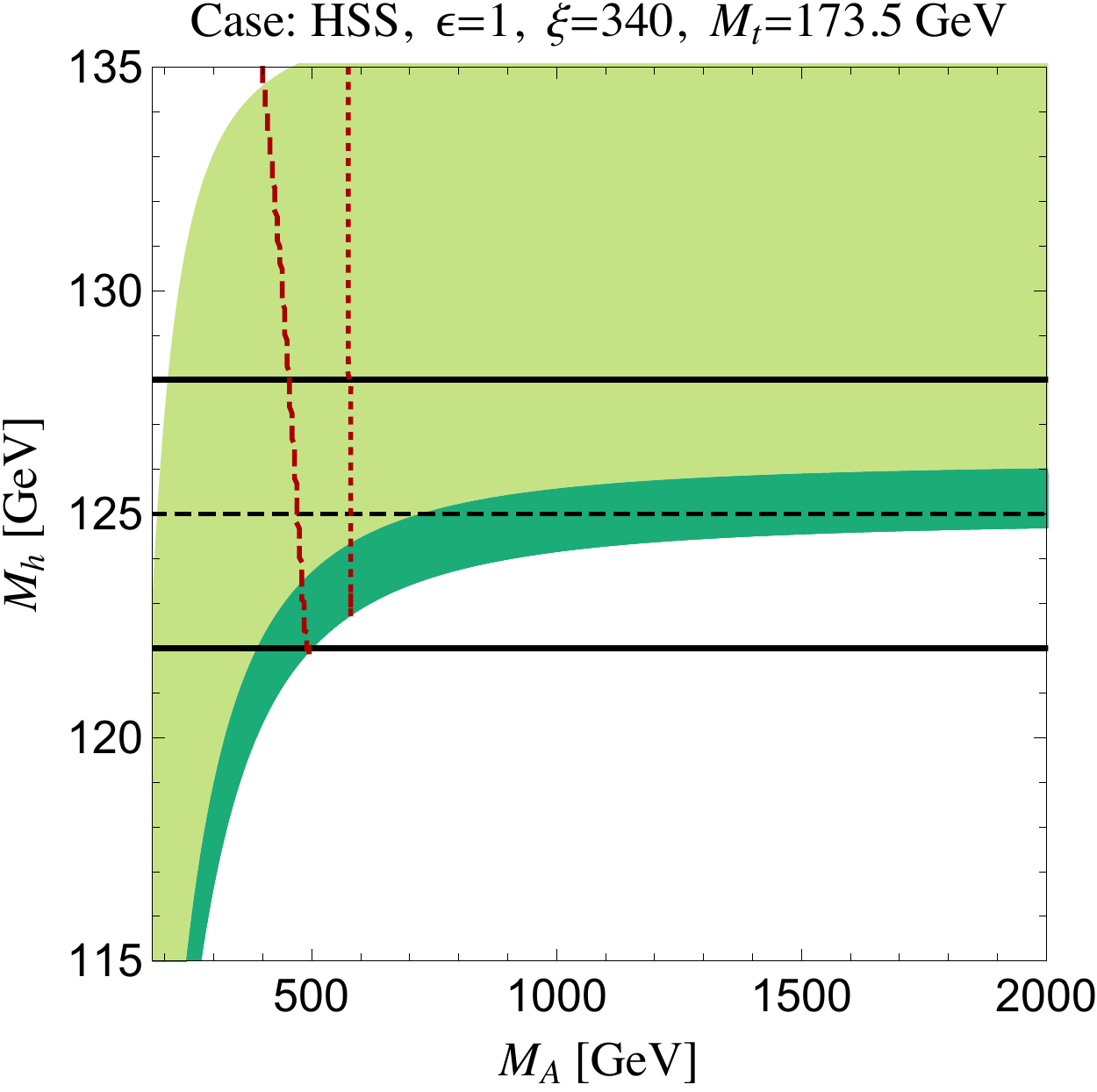}} \hspace*{0.1cm}
\subfigure{\includegraphics[width=0.44\textwidth]{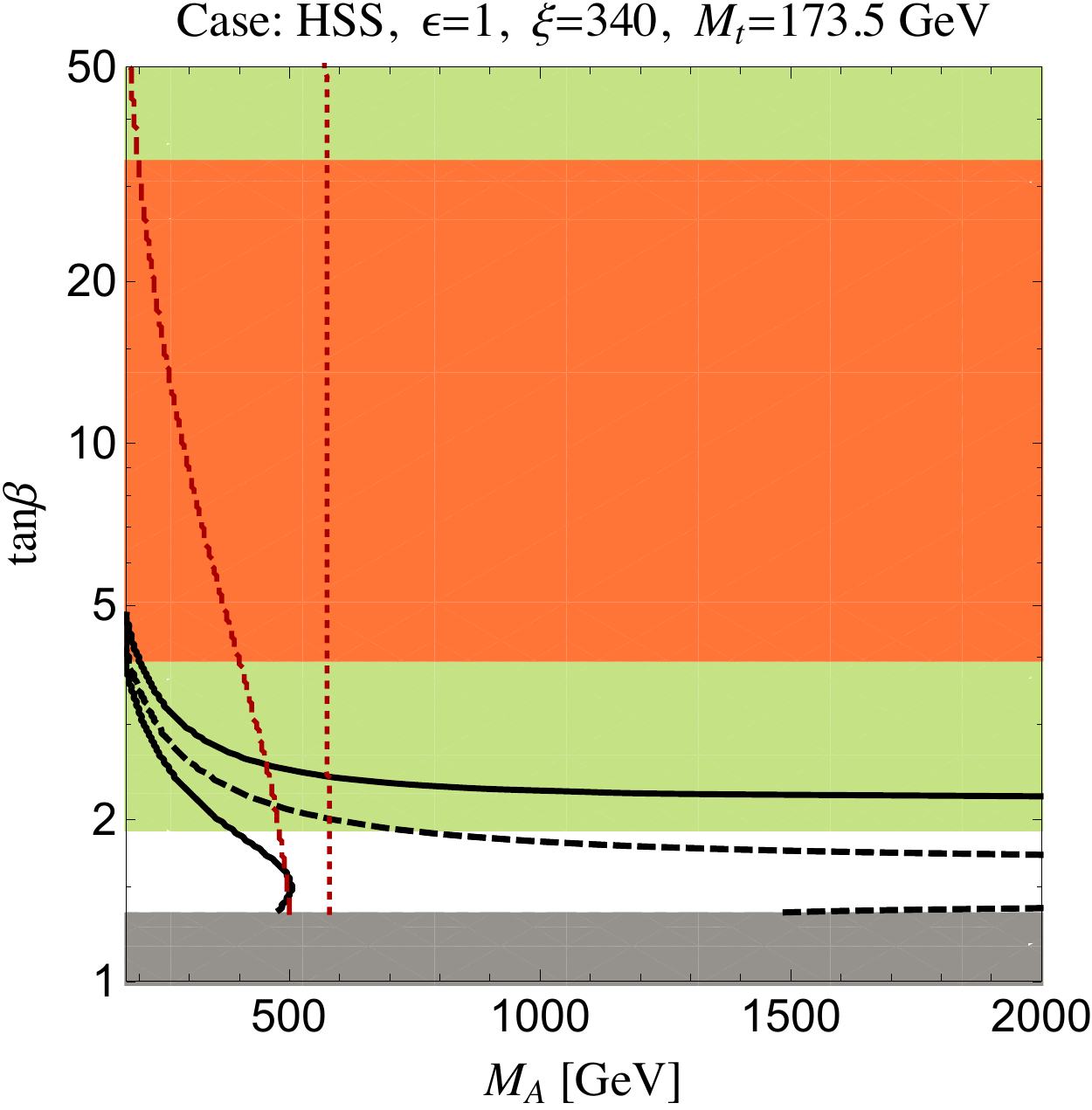}}
\subfigure{\includegraphics[width=0.45\textwidth]{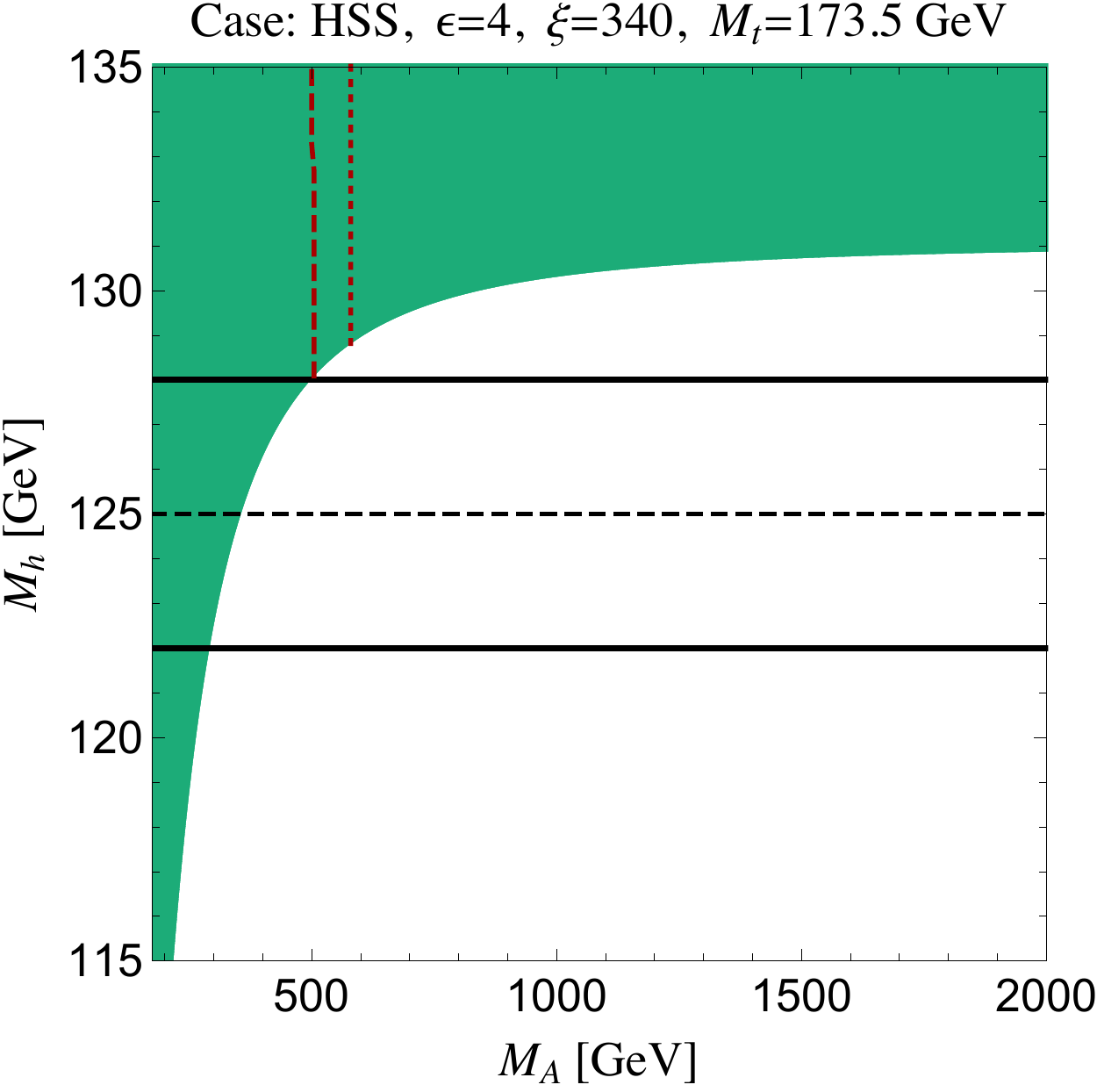}} \hspace*{0.1cm}
\subfigure{\includegraphics[width=0.44\textwidth]{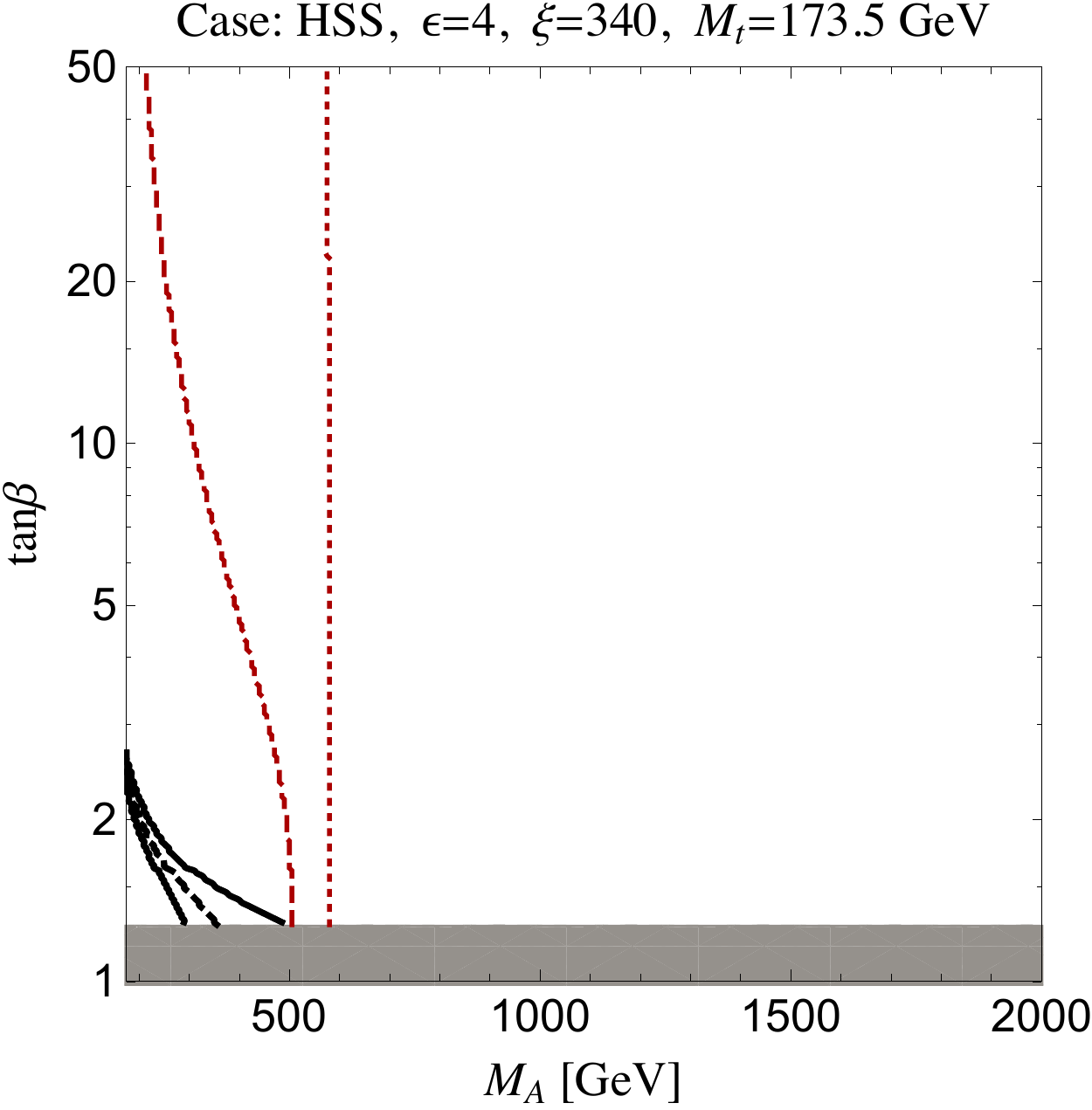}}
\caption{The details are same as given in the caption of Fig. \ref{fig1}. For all the cases, $N=2.07$.}
\label{fig3}
\end{figure}

In order to check the viability of above results with Higgs mass and other low energy constraints, we evaluate the scalar spectrum for three different benchmark points. These are: (i) $\epsilon = \xi = 1$, (ii) $\epsilon = 1$, $\xi = 340$ and (iii) $\epsilon = 4$, $\xi = 340$. The choices (i) and (ii) are motivated from a consideration that at least one or all the three neutrinos have couplings as large as ${\cal O}(y_t)$ while (iii) corresponds to very strongly coupled neutrinos which lead to significant improvements in stability constraints.  The results are displayed in Fig. \ref{fig3}.
For $\epsilon =\xi = 1$ (or equivalently $Y_\nu=Y_u$ at the GUT scale), the results are almost identical to the ones obtained in the case without RH neutrinos as it can be seen from Figs. \ref{fig1} and \ref{fig2}. The results do not change significantly also for $\epsilon=1$ and for any value of $\xi$ between 1 and $340$. For these benchmark points, the stability or metastability of electroweak vacuum and constraints on the masses of scalars imply  $1.2 \le \tan\beta \le 2.2$ and $M_A > 580$ GeV as viable ranges in which the effective THDM theory can be extrapolated to $M_S$. $\epsilon > 1.5$ leads to stability of the scalar potential for almost all the values of $\tan\beta$, however they face stringent constraints from $M_h$ and $M_{H^\pm}$ as it can be seen from the bottom panel in Fig. \ref{fig3}. Strongly coupled neutrinos in this case increase the magnitude of $\lambda_2$ at $M_t$ which results into relatively higher Higgs mass for a given value of $M_A$. Hence to obtain $M_h < 128$ GeV, one  needs lower $M_A$ which is already disfavoured by the constraints on the charged Higgs mass. This disfavours the values of $\epsilon > 1$. The above results imply that an existence of strongly coupled heavy neutrinos below $M_S$ improves the stability of THDM scalar potential, however this scenario is very strongly constrained by the observed Higgs mass and branching ratio of $b \to s + \gamma$.

\subsection{Case: LSS}
\begin{figure}[t]
\centering
\subfigure{\includegraphics[width=0.48\textwidth]{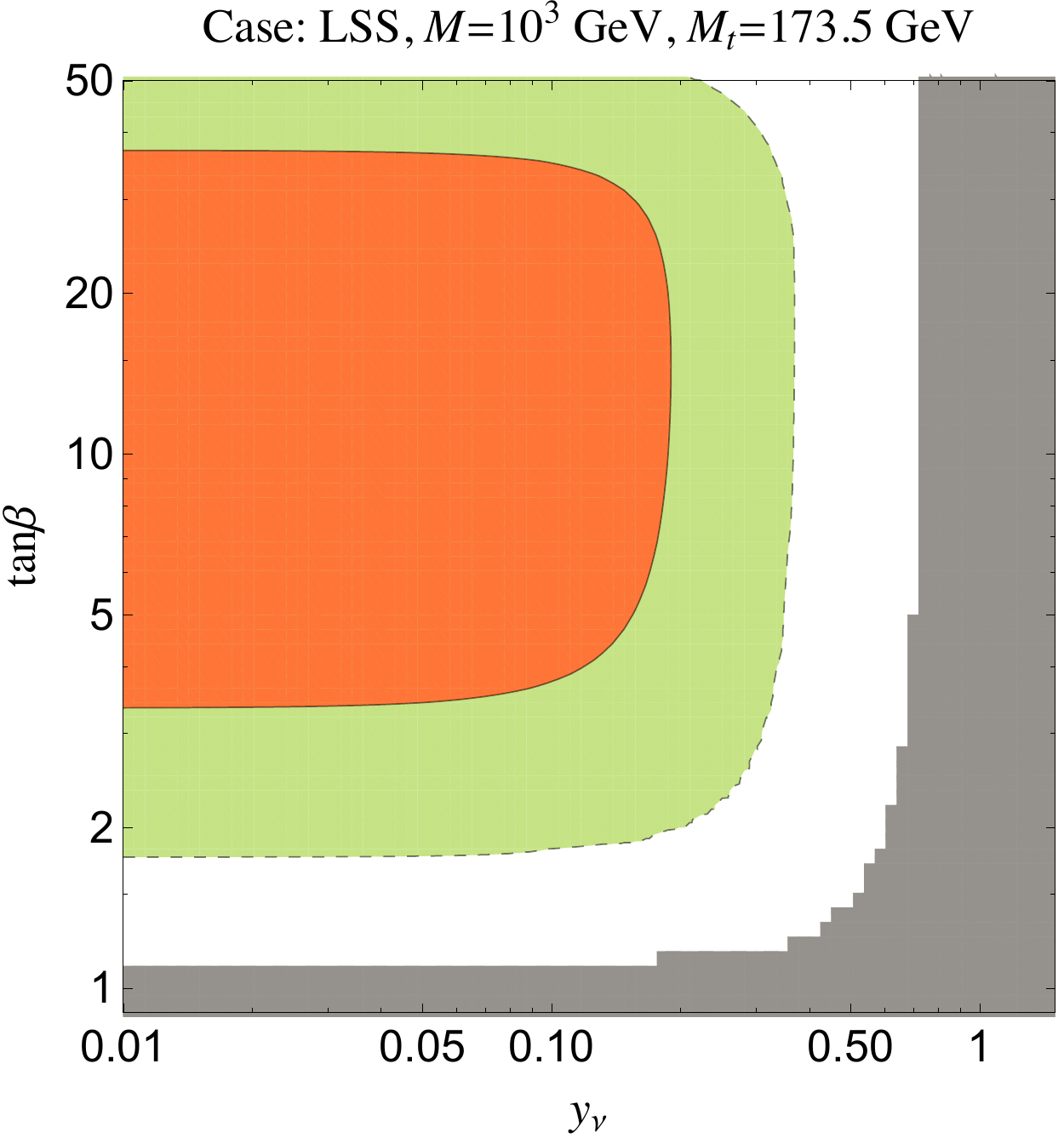}} \hspace*{0.1cm}
\subfigure{\includegraphics[width=0.48\textwidth]{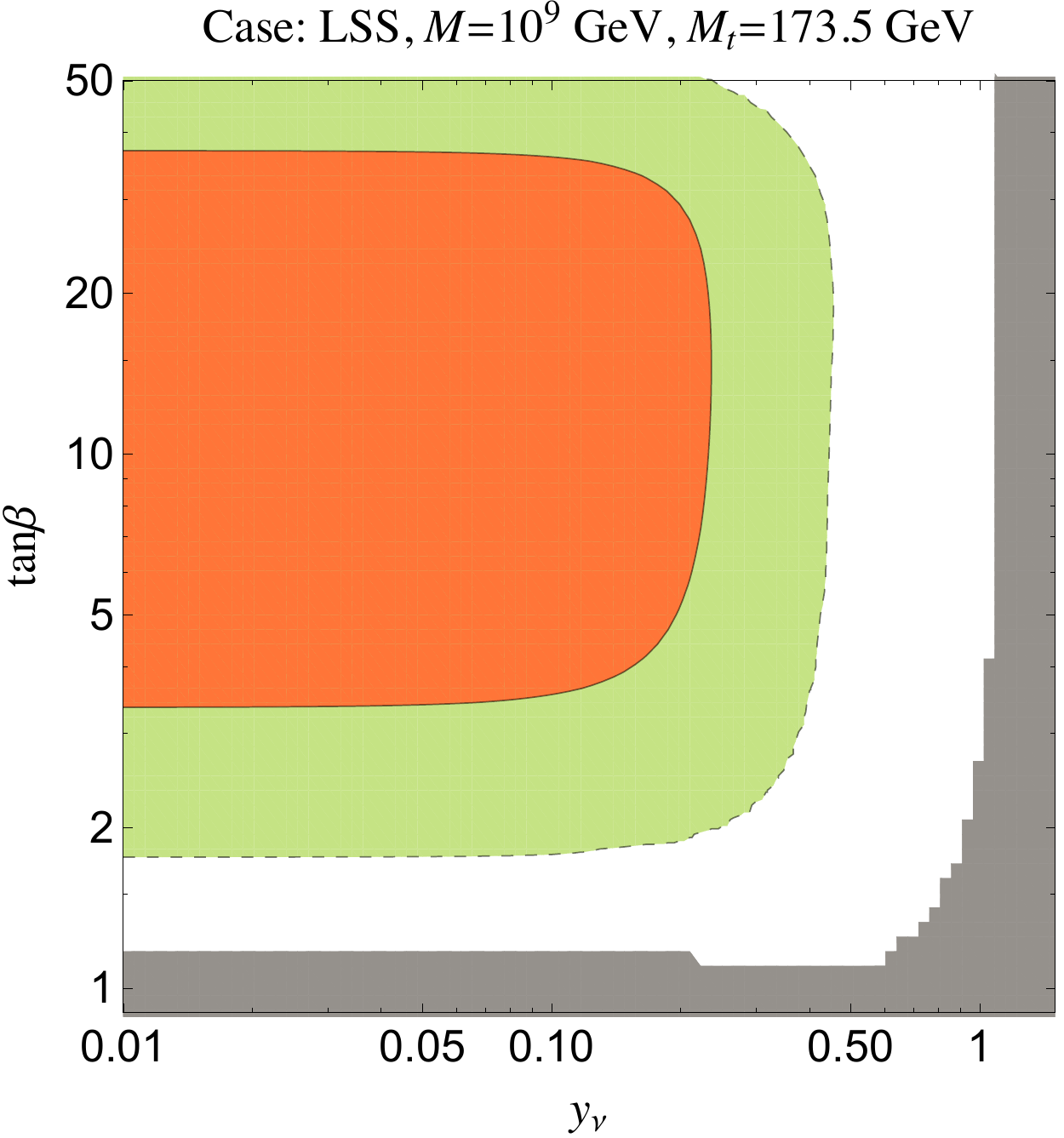}}\\
\caption{The allowed values of $\tan\beta$ by an absolute stability (unshaded) and metastability (green) of the scalar potential as function of $y_1=y_2=y_3 \equiv y_\nu$ as defined (and for $M = 10^3$ and $10^9$ GeV) in Eq. (\ref{YnuLS}). The orange region corresponds to unstable scalar potential while the grey region in the bottom is where the perturbativity of the couplings is lost.}
\label{fig4}
\end{figure}

We perform a similar analysis for a low scale seesaw scenario discussed in the section \ref{sec:model}. For simplicity, we assume that $y_1=y_2=y_3 \equiv y_\nu$ in Eq. (\ref{YnuLS}). The third RH neutrino does not couple to THDM as it can be seen from the flavour structure of $Y_\nu$ in Eq. (\ref{YnuLS}). We take two sample values for mass of the remaining degenerate heavy neutrinos, $M=10^3$ and $10^9$ GeV. The results are displayed in Fig. \ref{fig4}.
As it can be seen, strongly coupled neutrinos lead to significant changes in the stability of the scalar potential. The quartic couplings receive significant contributions from such neutrinos in this case which in turn helps in making the scalar potential more stable. The stability constraints on $y_\nu$ and $\tan\beta$ do not depend considerably on the mass scale of RH neutrinos.

\begin{figure}[t]
\centering
\subfigure{\includegraphics[width=0.44\textwidth]{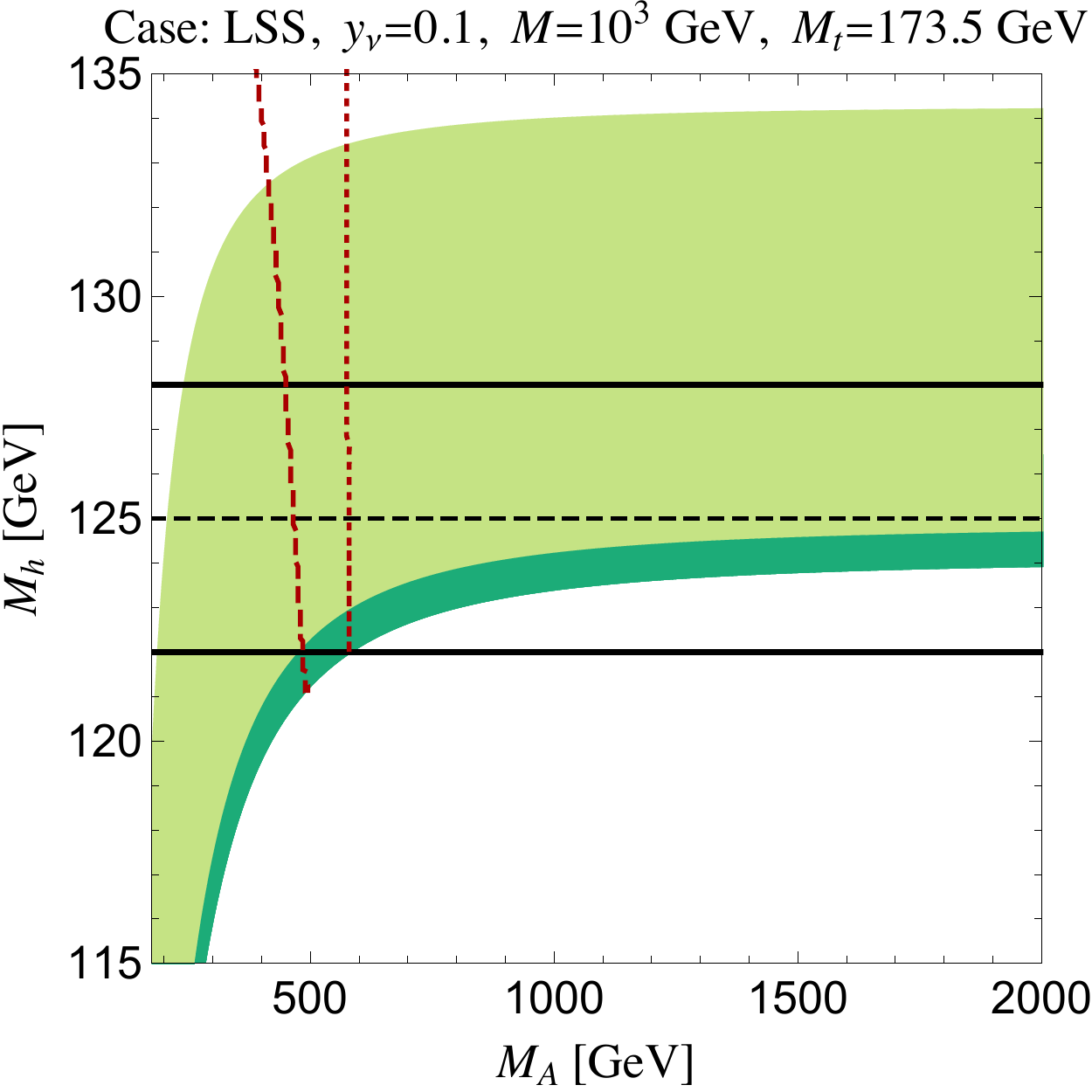}} \hspace*{0.1cm}
\subfigure{\includegraphics[width=0.43\textwidth]{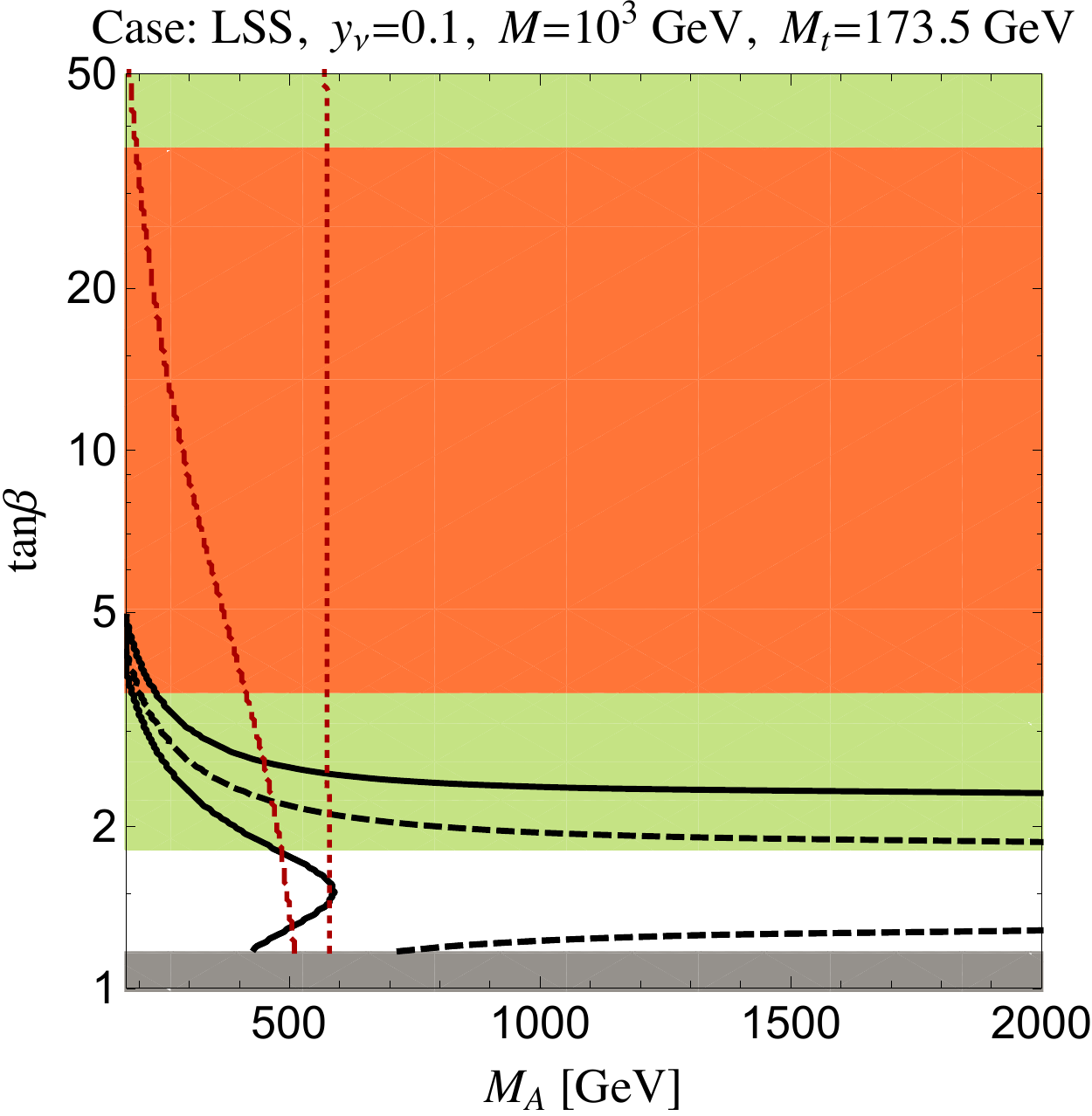}}
\subfigure{\includegraphics[width=0.44\textwidth]{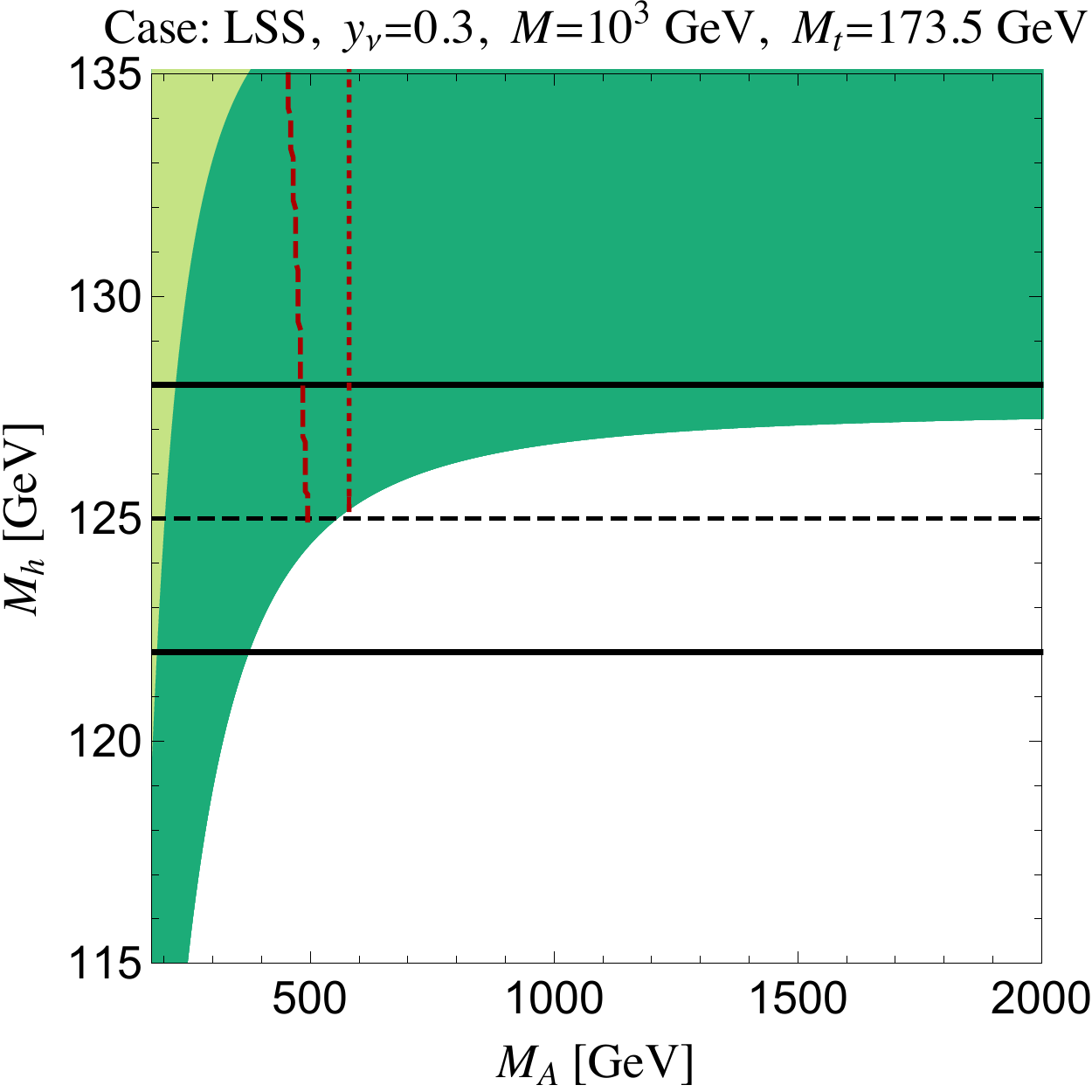}} \hspace*{0.1cm}
\subfigure{\includegraphics[width=0.43\textwidth]{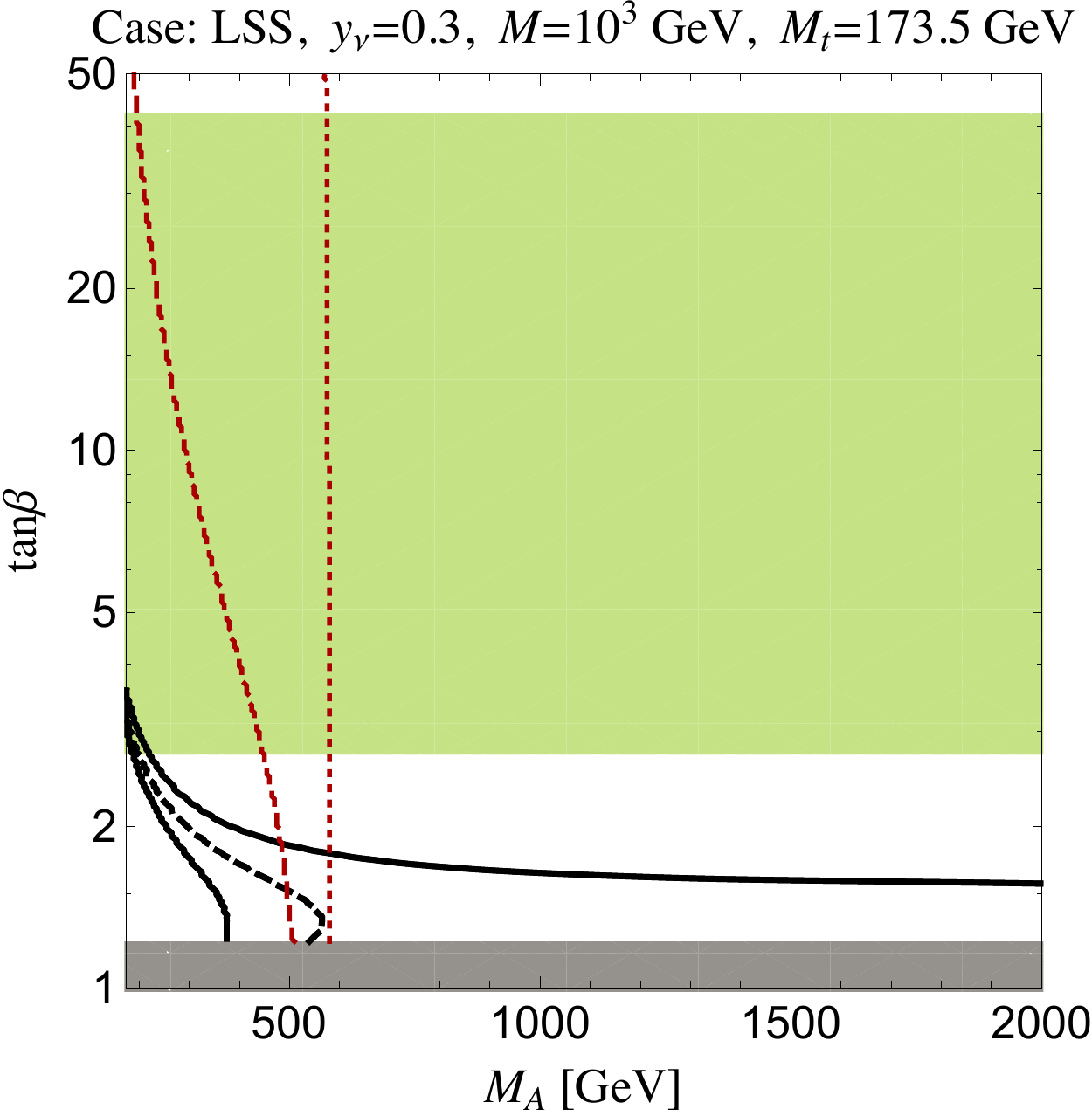}}
\subfigure{\includegraphics[width=0.44\textwidth]{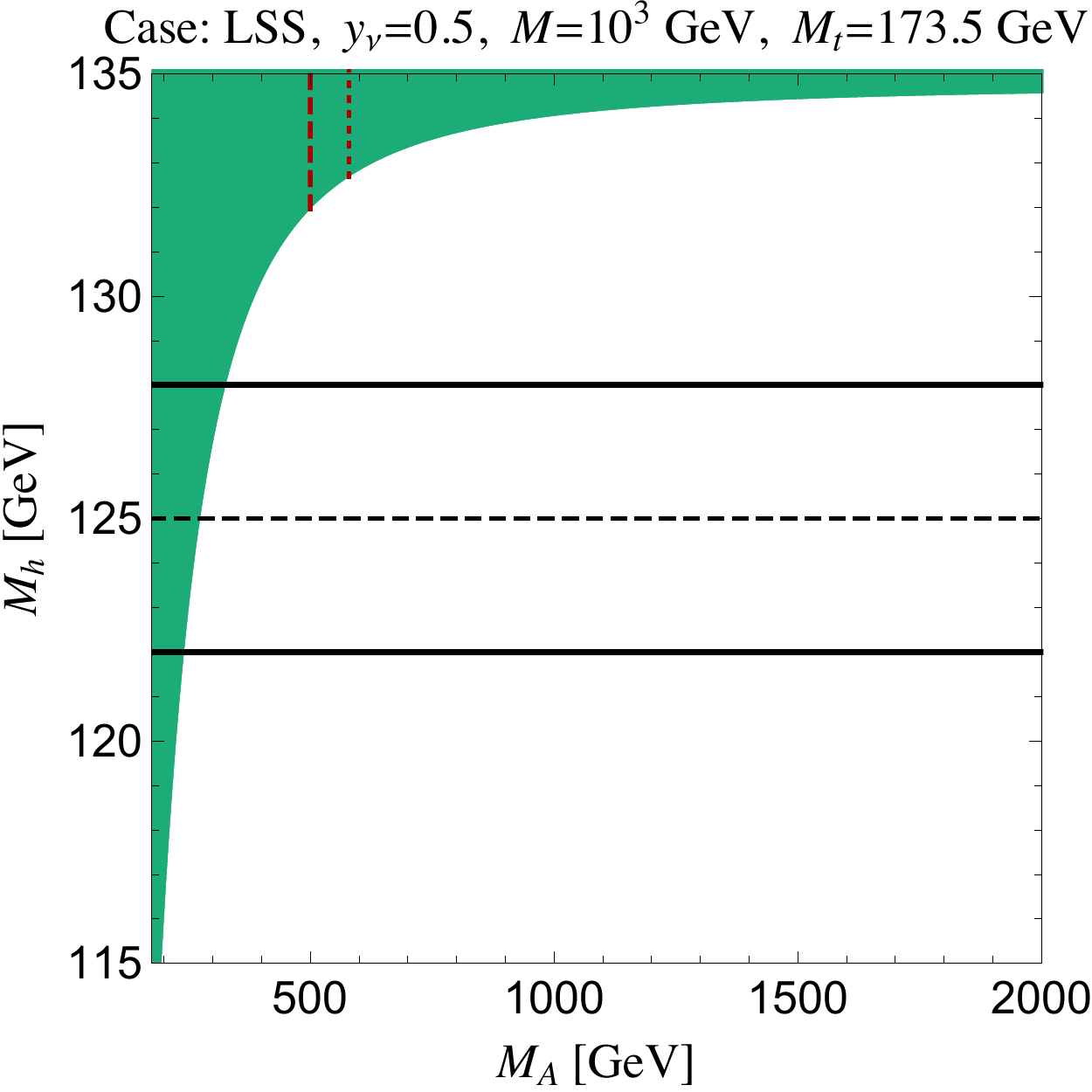}} \hspace*{0.1cm}
\subfigure{\includegraphics[width=0.43\textwidth]{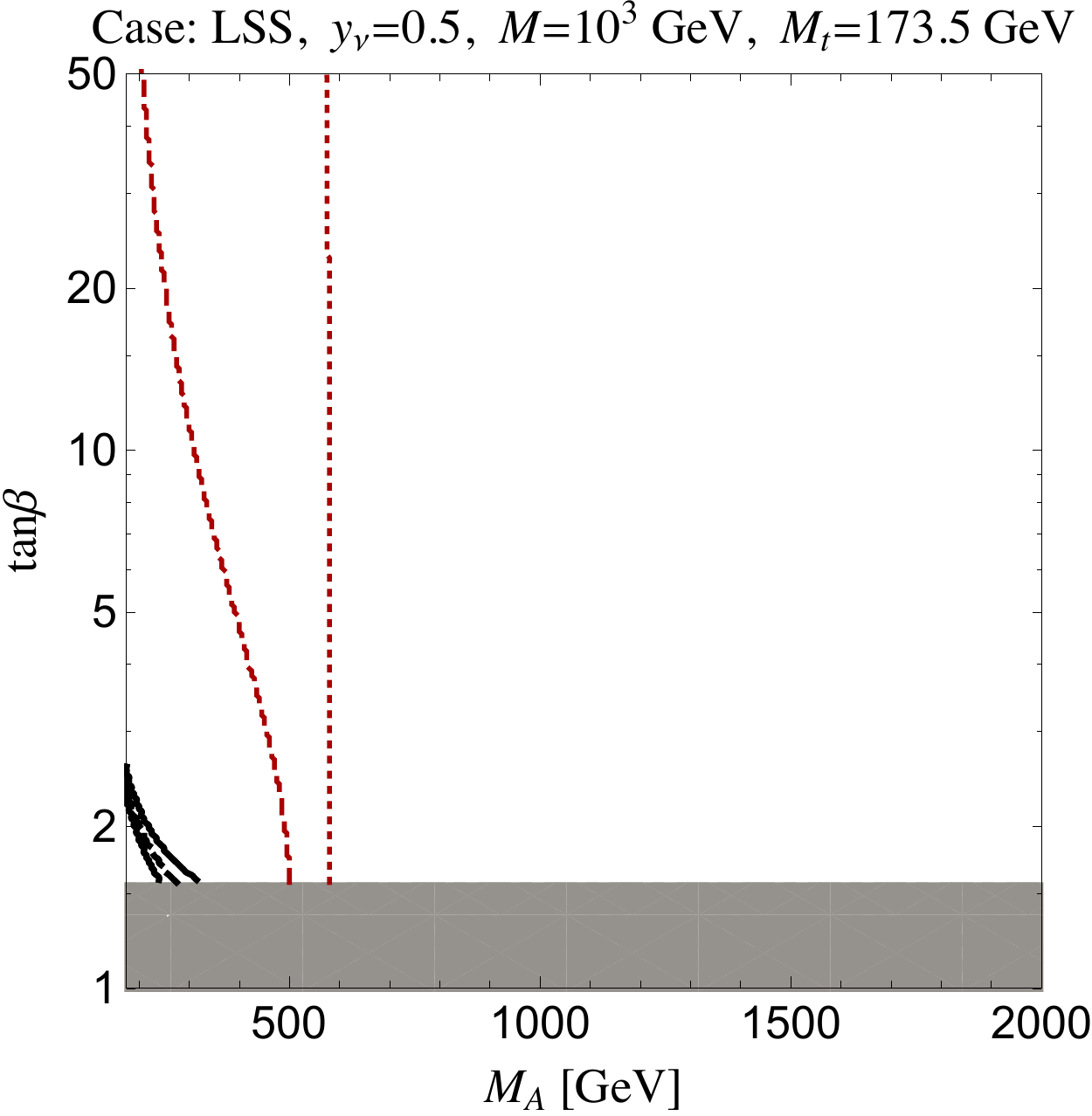}}
\caption{The details are same as given in the caption of Fig. \ref{fig1}.}
\label{fig5}
\end{figure}

We analyse the low energy scalar spectrum for three benchmark values of coupling: $y_\nu = 0.1$, $0.3$ and $0.5$. The results are displayed in Fig. \ref{fig5}. 
For $y_\nu \le 0.1$, the obtained constraints on $\tan\beta$ and $M_A$ are very similar to those obtained in the case without RH neutrinos. In the regime of strong $y_\nu$, these results change significantly. For $y_\nu = 0.3$, one always obtains $M_h > 125$ GeV if constraint on the charged Higgs mass is considered. Very strongly coupled neutrinos further push the value of $M_h$ on the higher side and such cases are disfavoured by limits on $M_{H^\pm}$ and $\beta-\alpha$.

In the analysis discussed above, we consider the top pole mas $M_t = 173.5$ GeV. The change in top mass is known to have considerable effects on the RG evolution. To evaluate such effects, we repeat the analysis carried out for the case presented in Figs. \ref{fig3} and \ref{fig5} for $M_t = 172.4$ GeV and $M_t=174.6$ GeV. These are values of top quark mass at $-1\sigma$ and $+1\sigma$ from the mean value respectively. The results of such analysis are displayed in Appendix \ref{Appendix:mt_var}.

\section{Conclusion}
\label{sec:concl}
We have investigated the viability of two-Higgs-doublet model, with type I seesaw mechanism incorporated in it, as an effective theory below the GUT scale in the presence of its minimal supersymmetric completion at the same scale. Supersymmetry is assumed to be broken at the scale $M_S = 2 \times 10^{16} $ GeV leaving only the SM particles, three copies of RH neutrinos and an additional Higgs doublet below $M_S$. It also leaves its imprints on the effective scalar potential of THDM by relating quartic couplings with gauge couplings at the SUSY breaking scale.  Because of these constraints provided by SUSY, the effective potential of THDM can be written in terms of only two free parameters: the ratio of vacuum expectation value of two Higgs doublets $\tan\beta$, and the mass of pseudo-scalar Higgs $M_A$. The same potential governs the electroweak symmetry breaking and determines the masses of physical scalars in the theory. The requirement that the electroweak vacuum is stable or metastable (i.e. its lifetime is greater than the age of universe) and constraints on the scalar spectrum from various direct and indirect experimental searches restrict the allowed values of $\tan\beta$ and $M_A$. In the absence of RH neutrinos, these constraints favour the values for $\tan\beta$ in the range from $1.4$ to $2.5$ and $M_A \gsim 580$ GeV.

Existence of RH neutrinos below the scale $M_S$ causes significant changes in the above results if they are very strongly coupled with the SM leptons. We have studied this possibility in the context of high and low scale type I seesaw mechanism. In the case of high scale seesaw, we find that if one or more neutrinos have Dirac type Yukawa couplings greater than $y_t$ at the GUT scale, then they considerably modify the running of quartic couplings and allow stable or metastable electroweak vacuum for almost all the values of $\tan\beta$ for which the effective theory remains perturbative. The RH neutrinos improve the stability of scalar potential in this case unlike in the case of the SM without high scale SUSY in which the strongly coupled RH neutrinos are known \cite{EliasMiro:2011aa,Rodejohann:2012px,Khan:2012zw,Chakrabortty:2012np,Rose:2015fua,Bambhaniya:2016rbb,Coriano:2014mpa,Coriano:2015sea} to destabilize the scalar potential. The difference between the two cases arise mainly because of the presence of SUSY at the high scale which ensures the stability of scalar potential in ultraviolet completion. The difference is also attributed to the presence of additional Higgs doublet in THDM which modifies the stability conditions allowing more freedom in the effective potential. We find that the observed Higgs mass and limit on the charged Higgs mass from the flavour physics data put stringent  constraints on $\tan\beta$ and $M_A$ when RH neutrinos are strongly coupled. Similar results are obtained in case of low scale seesaw when neutrino Yukawa couplings are ${\cal O}(1)$.

It is observed that if Yukawa couplings of RH neutrinos are of  ${\cal O}(y_t)$ or smaller at the GUT scale then they do not have significant impact on vacuum stability constraints in THDM. In the case of high scale seesaw, particular cases have been explored in which it is assumed that the neutrino Yukawa coupling matrix $Y_\nu$ is equal to up-type quark Yukawa coupling matrix $Y_u$ or all the couplings in $Y_\nu$ are ${\cal O}(y_t)$ at the GUT scale. This class of boundary conditions are often realized in $SO(10)$ based GUTs. It is found in these cases that RH neutrinos have negligible effect on stability of the scalar potential and scalar spectrum at low energy.  A stable or metastable scalar potential consistent with low energy constraints is obtained for $\tan\beta \in [1.2 - 2.5]$ in these cases. These results are qualitatively similar to those obtained in case without right-handed neutrinos in \cite{Bagnaschi:2015pwa}.

The results obtained in this paper are useful in constraining a class of models in which supersymmetry is broken at the GUT scale and an effective theory below this scale is THDM of type II with type I seesaw mechanism. It is shown here that the stability or metastability of electroweak vacuum and a consistent low energy scalar spectrum are achieved for only small $\tan\beta$ and for neutrino Yukawa couplings of ${\cal O}(y_t)$ or smaller. In concrete ultraviolet models, see for example the ones given in \cite{Buchmuller:2017vho,Buchmuller:2017vut}, the neutrino Yukawa couplings and/or $\tan\beta$ are often determined from the enhanced symmetry structure of underlying theory. Therefore, the results obtained in this paper can be used to constrain such frameworks.

 Our analysis provides a generic understanding of the effects of RH neutrinos. The numerical results obtained in this paper are subject to change for different choice of values of $M_S$ and/or GUT scale threshold corrections. We have assumed a particular value of SUSY breaking scale, $M_S = 2 \times 10^{16}$ GeV, throughout our analysis. It is known that stability of vacuum potential is considerably sensitive to the choice of such a scale \cite{Bagnaschi:2015pwa}. In particular, the constraints on the values of $\tan\beta$ imposed by unstability of potential become feeble (stronger) for relatively smaller (larger) values of $M_S$. Another important issue is that the precise unification of gauge couplings does not occur at the scale $M_S$ as the effective theory below this scale is pure THDM with singlet neutrinos. Exact gauge coupling unification may require either new fields at intermediate scales (for example, a pair of TeV scale Higgsinos \cite{Bagnaschi:2015pwa}) or sizeable threshold corrections at the GUT scale \cite{Ellis:2017erg}. For all these cases, a dedicated analysis would be required in order to derive quantitative constraints on $\tan\beta$ and $M_A$. However, we anticipate that the qualitative effects of RH neutrinos will remain the same. The presence of one or more strongly coupled RH neutrinos below $M_S$ improves the stability of vacuum in an underlying framework.

\section*{Acknowledgements}
This work was supported by Early Career Research Award (ECR/2017/000353) and by a research grant under INSPIRE Faculty Award (DST/INSPIRE /04/2015/000508) from the Department of Science and Technology, Government of India. KMP thanks DESY Theory Group for the kind hospitality during the final stage of this work.

\appendix
\section{Renormalization group equations for THDM of type II with right handed neutrinos}
\label{Appendix:RGE}
In this appendix, we provide 2-loop renormalization group equations for THDM of type II. They are obtained using publicly available package SARAH \cite{Staub:2013tta}. Note that we use different convention for the quartic couplings $\lambda_1$ and $\lambda_2$ in Eq. (\ref{THDM_V}) in comparison to the one used in SARAH. The RG equations listed below are therefore modified accordingly. The same equations are also listed in \cite{Lee:2015uza} but with considering only the third generation of fermions.

The couplings evolve according to the following equation:
\be \label{}
\mu\,\frac{d C}{d\mu} = \frac{1}{16\,\pi^2}\,\beta_{C}^{(1)}+\Big(\frac{1}{16\,\pi^2}\Big)^2\,\beta_{C}^{(2)}\,, \ee
where $C$ represents gauge, Yukawa and quartic couplings and $\mu$ is the renormalization scale. The one and two-loop beta functions for the different couplings are as the following.

\subsection{Gauge couplings}
{\allowdisplaybreaks  \begin{align} 
	\beta_{g_1}^{(1)} & =  
	\frac{21}{5}\, g_{1}^{3} \\ 
	\beta_{g_1}^{(2)} & =  
	\frac{1}{50} \,g_{1}^{3}\, \Big(180\, g_{2}^{2}  + 208\, g_{1}^{2}  + 440\, g_{3}^{2}  -75 \,\mbox{Tr}\Big({Y_{e}^{\dagger}Y_e  }\Big) -15\, \mbox{Tr}\Big({ Y_{\nu}^{\dagger}Y_\nu }\Big) \nonumber\\
	& -25 \,\mbox{Tr}\Big({ Y_{d}^{\dagger}Y_d }\Big)  -85\, \mbox{Tr}\Big({Y_{u}^{\dagger} Y_u  }\Big) \Big)\\ 
	\beta_{g_2}^{(1)} & =  
	-3\, g_{2}^{3} \\ 
	\beta_{g_2}^{(2)} & =  
	\frac{1}{10}\, g_{2}^{3}\, \Big(80\, g_{2}^{2}+120\, g_{3}^{2}  + 12\, g_{1}^{2}  -5\, \mbox{Tr}\Big({Y_e  Y_{e}^{\dagger}}\Big)  -5\, \mbox{Tr}\Big({Y_\nu  Y_{\nu}^{\dagger}}\Big)\nonumber\\
	&-15 \,\mbox{Tr}\Big({Y_d  Y_{d}^{\dagger}}\Big) -15\, \mbox{Tr}\Big({Y_u  Y_{u}^{\dagger}}\Big)     \Big)\\ 
	\beta_{g_3}^{(1)} & =  
	-7\, g_{3}^{3}\\ 
	\beta_{g_3}^{(2)} & =  
	-\frac{1}{10}\, g_{3}^{3}\, \Big(-11\, g_{1}^{2}  + 260\, g_{3}^{2}  -45\, g_{2}^{2}+ 20\, \mbox{Tr}\Big({Y_d  Y_{d}^{\dagger}}\Big)  + 20\, \mbox{Tr}\Big({Y_u  Y_{u}^{\dagger}}\Big)   \Big)
	\end{align}} 
\subsection{Yukawa couplings}
{\allowdisplaybreaks  \begin{align} 
	\beta_{Y_u}^{(1)} & =  
	 Y_u\, \Big(-8\, g_{3}^{2}  -\frac{17}{20}\, g_{1}^{2}  -\frac{9}{4}\, g_{2}^{2} + 3\, \mbox{Tr}\Big({Y_{u}^{\dagger}  Y_{u}}\Big)  + \mbox{Tr}\Big({Y_{\nu}^{\dagger}Y_\nu  }\Big)\Big)\nonumber\\
	 &+\frac{1}{2} \,\Big(3\,{Y_u  Y_{u}^{\dagger}  Y_u}  + {Y_d  Y_{d}^{\dagger}  Y_u  }\Big) \\ 
	\beta_{Y_u}^{(2)} & =  Y_u \,\Big(\frac{1267}{600} \,g_{1}^{4} -\frac{9}{20}\, g_{1}^{2} g_{2}^{2} -\frac{21}{4}\, g_{2}^{4} +\frac{19}{15}\, g_{1}^{2} g_{3}^{2} +9\, g_{2}^{2} g_{3}^{2} -108\, g_{3}^{4} +\frac{3}{2}\, \lambda_{2}^{2} +\lambda_{3}^{2}\nonumber\\
	&+\lambda_3 \lambda_4 +\lambda_{4}^{2}+\frac{1}{8}\, \Big(160\, g_{3}^{2}  + 17\, g_{1}^{2}  + 45 \,g_{2}^{2} \Big)\,\mbox{Tr}\Big({Y_u  Y_{u}^{\dagger}}\Big) +\frac{3}{8} \,\Big(5\, g_{2}^{2}  + g_{1}^{2}\Big)\,\mbox{Tr}\Big({Y_\nu  Y_{\nu}^{\dagger}}\Big)\nonumber\\
	&-\frac{9}{4} \,\mbox{Tr}\Big({Y_d  Y_{d}^{\dagger}  Y_u  Y_{u}^{\dagger}}\Big)-\frac{3}{4}\, \mbox{Tr}\Big({Y_e  Y_{e}^{\dagger}  Y_\nu  Y_{\nu}^{\dagger}}\Big) -\frac{27}{4}\, \mbox{Tr}\Big({Y_u  Y_{u}^{\dagger}  Y_u  Y_{u}^{\dagger}}\Big) -\frac{9}{4}\, \mbox{Tr}\Big({Y_\nu  Y_{\nu}^{\dagger}  Y_\nu  Y_{\nu}^{\dagger}}\Big) \Big)\nonumber\\ 
	&+{Y_d  Y_{d}^{\dagger}  Y_u}\, \Big(-2\, \lambda_3  + 2\, \lambda_4  + \frac{16}{3} \,g_{3}^{2}  + \frac{33}{16}\, g_{2}^{2}  -\frac{3}{4}\, \mbox{Tr}\Big({Y_e  Y_{e}^{\dagger}}\Big)  -\frac{41}{240}\, g_{1}^{2}  -\frac{9}{4}\, \mbox{Tr}\Big({Y_d  Y_{d}^{\dagger}}\Big) \Big)\nonumber \\ 
	&+\frac{1}{80}\,{Y_u  Y_{u}^{\dagger}  Y_u} \Big(1280\, g_{3}^{2}  -180\, \mbox{Tr}\Big({Y_\nu  Y_{\nu}^{\dagger}}\Big)  + 223\, g_{1}^{2}  -540\, \mbox{Tr}\Big({Y_u  Y_{u}^{\dagger}}\Big)  + 675\, g_{2}^{2}  -480\, \lambda_2 \Big)\nonumber\\
	&-\frac{1}{4} \, \Big(-6\, {Y_u  Y_{u}^{\dagger}  Y_u  Y_{u}^{\dagger}  Y_u}  + {Y_d  Y_{d}^{\dagger}  Y_d  Y_{d}^{\dagger}  Y_u} + {Y_u  Y_{u}^{\dagger}  Y_d  Y_{d}^{\dagger}  Y_u}\Big) \\ 
	\beta_{Y_d}^{(1)} & =  Y_d\,\big(-8\, g_{3}^{2}  -\frac{1}{4}\, g_{1}^{2}  -\frac{9}{4}\, g_{2}^{2} +3\, \mbox{Tr}\Big({Y_d  Y_{d}^{\dagger}}\Big)  + \mbox{Tr}\Big({Y_e  Y_{e}^{\dagger}}\Big)\Big)\nonumber\\
	&+\frac{1}{2} \,\Big(3\,{Y_d  Y_{d}^{\dagger}  Y_d}  + {Y_u  Y_{u}^{\dagger}  Y_d  }\Big) 
	\\ 
   \beta_{Y_d}^{(2)} & =  
   Y_d \,\Big(-\frac{113}{600}\, g_{1}^{4} -\frac{27}{20}\, g_{1}^{2} g_{2}^{2} -\frac{21}{4}\, g_{2}^{4} +\frac{31}{15}\, g_{1}^{2} g_{3}^{2} +9\, g_{2}^{2} g_{3}^{2} -108\, g_{3}^{4} +\frac{3}{2}\, \lambda_{1}^{2} +\lambda_{3}^{2}\nonumber\\
   &+\lambda_3 \lambda_4 +\lambda_{4}^{2}+\frac{5}{8}\, \Big(32\, g_{3}^{2}  + 9\, g_{2}^{2}  + g_{1}^{2}\Big)\mbox{Tr}\Big({Y_d  Y_{d}^{\dagger}}\Big) +\frac{15}{8}\, \Big(g_{1}^{2} + g_{2}^{2}\Big)\mbox{Tr}\Big({Y_e  Y_{e}^{\dagger}}\Big)\nonumber\\
   &-\frac{27}{4}\, \mbox{Tr}\Big({Y_d  Y_{d}^{\dagger}  Y_d  Y_{d}^{\dagger}}\Big)-\frac{9}{4}\, \mbox{Tr}\Big({Y_{d}^{\dagger} Y_u Y_{u}^{\dagger} Y_d }\Big) -\frac{9}{4}\, \mbox{Tr}\Big({Y_e  Y_{e}^{\dagger}  Y_e Y_{e}^{\dagger} }\Big) -\frac{3}{4}\, \mbox{Tr}\Big({Y_{e}^{\dagger}Y_\nu Y_{\nu}^{\dagger} Y_e  }\Big) \Big)\nonumber\\
   &+\frac{1}{80}\, {Y_d  Y_{d}^{\dagger}  Y_d}\, \Big(1280\, g_{3}^{2}  -180\, \mbox{Tr}\Big({Y_e  Y_{e}^{\dagger}}\Big)  + 187\, g_{1}^{2}  -540\, \mbox{Tr}\Big({Y_d  Y_{d}^{\dagger}}\Big)  + 675\, g_{2}^{2}  -480\, \lambda_1 \Big)\nonumber \\ 
   &+\frac{1}{240}\,{Y_u  Y_{u}^{\dagger} Y_d  } \,\Big(1280\, g_{3}^{2}  -180\, \mbox{Tr}\Big({Y_\nu  Y_{\nu}^{\dagger}}\Big)  -480\, \lambda_3  + 480\, \lambda_4  + 495\, g_{2}^{2}  -53 \,g_{1}^{2} \nonumber\\ &-540\, \mbox{Tr}\Big({Y_u  Y_{u}^{\dagger}}\Big) \Big)
   +\frac{1}{4} \, \Big(6\, {Y_d  Y_{d}^{\dagger}  Y_d  Y_{d}^{\dagger}  Y_d}  - {Y_d Y_{d}^{\dagger} Y_u  Y_{u}^{\dagger} Y_d  }  - { Y_u Y_{u}^{\dagger} Y_u Y_{u}^{\dagger} Y_d } \Big)\\ 
	\beta_{Y_\nu}^{(1)} & =  
	\frac{1}{2}\, \Big(3\, {Y_\nu  Y_{\nu}^{\dagger}  Y_\nu}  + {Y_e Y_{e}^{\dagger} Y_\nu }\Big) + Y_\nu\, \Big(3\, \mbox{Tr}\Big({Y_u  Y_{u}^{\dagger}}\Big)  -\frac{9}{20}\, \Big(5\, g_{2}^{2}  + g_{1}^{2}\Big) + \mbox{Tr}\Big({Y_\nu  Y_{\nu}^{\dagger}}\Big)\Big)\\ 
		\beta_{Y_\nu}^{(2)} & =  
	Y_\nu \,\Big(\frac{117}{200} \,g_{1}^{4} -\frac{27}{20}\, g_{1}^{2} g_{2}^{2} -\frac{21}{4} \,g_{2}^{4} +\frac{3}{2} \,\lambda_{2}^{2} +\lambda_{3}^{2}+\lambda_3 \lambda_4 +\lambda_{4}^{2} +\frac{1}{8}\, \Big(160\, g_{3}^{2}  + 17\, g_{1}^{2}  \nonumber\\
	&+ 45\, g_{2}^{2} \Big)\mbox{Tr}\Big({Y_u  Y_{u}^{\dagger}}\Big)+\frac{3}{8}\, \Big(5\, g_{2}^{2}  + g_{1}^{2}\Big)\mbox{Tr}\Big({Y_\nu  Y_{\nu}^{\dagger}}\Big) -\frac{9}{4}\, \mbox{Tr}\Big({Y_{d}^{\dagger} Y_u  Y_{u}^{\dagger} Y_d   }\Big) \nonumber\\
	&-\frac{3}{4}\, \mbox{Tr}\Big({Y_{e}^{\dagger} Y_\nu Y_{\nu}^{\dagger}  Y_e  }\Big) -\frac{27}{4}\, \mbox{Tr}\Big({Y_{u}^{\dagger} Y_u  Y_{u}^{\dagger}  Y_u  }\Big) -\frac{9}{4} \mbox{Tr}\Big({Y_\nu  Y_{\nu}^{\dagger}  Y_\nu  Y_{\nu}^{\dagger}}\Big) \Big)\nonumber\\ 
	 &+\frac{1}{80}\,{Y_\nu  Y_{e}^{\dagger}  Y_e} \Big(-160\, \lambda_3  + 160\, \lambda_4  + 165\, g_{2}^{2}  -180\, \mbox{Tr}\Big({Y_d  Y_{d}^{\dagger}}\Big)  + 21\, g_{1}^{2} \nonumber\\
	 & -60\, \mbox{Tr}\Big({Y_e  Y_{e}^{\dagger}}\Big) \Big)+\frac{3}{80}\, {Y_\nu  Y_{\nu}^{\dagger}  Y_\nu} \Big(-180\, \mbox{Tr}\Big({Y_u  Y_{u}^{\dagger}}\Big)  + 225\, g_{2}^{2}  -160\, \lambda_2\nonumber\\
	 &  -60\, \mbox{Tr}\Big({Y_\nu  Y_{\nu}^{\dagger}}\Big)  + 93\, g_{1}^{2} \Big)-\frac{1}{4}\, \Big(-6\, {Y_\nu  Y_{\nu}^{\dagger}  Y_\nu  Y_{\nu}^{\dagger}  Y_\nu}  + {Y_e Y_{e}^{\dagger} Y_e  Y_{e}^{\dagger}  Y_\nu   } + { Y_\nu Y_{\nu}^{\dagger}  Y_e Y_{e}^{\dagger}  Y_\nu    }\Big) \\
	\beta_{Y_e}^{(1)} & =  
	\frac{1}{4} \, \Big(6\, {Y_e  Y_{e}^{\dagger}  Y_e}  +2\, {Y_e  Y_{\nu}^{\dagger}  Y_\nu}\Big) + Y_e\, \Big(3\, \mbox{Tr}\Big({Y_d  Y_{d}^{\dagger}}\Big)  + \mbox{Tr}\Big({Y_e  Y_{e}^{\dagger}}\Big)  -\frac{9}{4} \Big(g_{1}^{2} + g_{2}^{2}\Big)\Big)\\ 
	\beta_{Y_e}^{(2)} & =  Y_e\, \Big(\frac{1449}{200}\, g_{1}^{4} +\frac{27}{20}\, g_{1}^{2} g_{2}^{2} -\frac{21}{4}\, g_{2}^{4} +\frac{3}{2}\, \lambda_{1}^{2} +\lambda_{3}^{2}+\lambda_3 \lambda_4 +\lambda_{4}^{2}\nonumber \\ 
	&+\frac{5}{8} \,\Big(32\, g_{3}^{2}  + 9 \,g_{2}^{2}  + g_{1}^{2}\Big)\mbox{Tr}\Big({Y_d  Y_{d}^{\dagger}}\Big) 
	+\frac{15}{8} \,\Big(g_{1}^{2} + g_{2}^{2}\Big)\,\mbox{Tr}\Big({Y_e  Y_{e}^{\dagger}}\Big) -\frac{27}{4}\, \mbox{Tr}\Big({Y_d  Y_{d}^{\dagger}  Y_d  Y_{d}^{\dagger}}\Big) \nonumber\\
	&-\frac{9}{4}\, \mbox{Tr}\Big({Y_{d}^{\dagger} Y_u Y_{u}^{\dagger}  Y_d  }\Big) -\frac{9}{4}\, \mbox{Tr}\Big({Y_e  Y_{e}^{\dagger}  Y_e  Y_{e}^{\dagger}}\Big) -\frac{3}{4}\, \mbox{Tr}\Big({Y_{e}^{\dagger} Y_\nu Y_{\nu}^{\dagger}  Y_e  }\Big) \Big)\nonumber\\
	&+\frac{3}{80} \,{Y_e  Y_{e}^{\dagger}  Y_e} \Big(129\, g_{1}^{2}  -180\, \mbox{Tr}\Big({Y_d  Y_{d}^{\dagger}}\Big)  + 225\, g_{2}^{2}  -160\,\lambda_1  -60\, \mbox{Tr}\Big({Y_e  Y_{e}^{\dagger}}\Big) \Big)\nonumber \\ 
	&+\frac{1}{80}\,{Y_\nu Y_{\nu}^{\dagger} Y_e  } \Big(-160\, \lambda_3  + 160\, \lambda_4  + 165\, g_{2}^{2}  -180\, \mbox{Tr}\Big({Y_u  Y_{u}^{\dagger}}\Big)  + 33\, g_{1}^{2}\nonumber\\
	&  -60\, \mbox{Tr}\Big({Y_\nu  Y_{\nu}^{\dagger}}\Big) \Big)+\frac{1}{4}\,\Big(6\, {Y_e  Y_{e}^{\dagger}  Y_e  Y_{e}^{\dagger}  Y_e}  - {Y_e  Y_{e}^{\dagger} Y_\nu Y_{\nu}^{\dagger}  Y_e    }  - {Y_\nu  Y_{\nu}^{\dagger}  Y_\nu Y_{\nu}^{\dagger} Y_e  } \Big)
	\end{align}} 
\subsection{Quartic couplings of scalars}
{\allowdisplaybreaks  \begin{align} 
	\beta_{\lambda_1}^{(1)} & =  
	+\frac{27}{100} \,g_{1}^{4} +\frac{9}{10}\, g_{1}^{2} g_{2}^{2} +\frac{9}{4}\, g_{2}^{4} -\frac{9}{5} \,g_{1}^{2} \lambda_1 -9 \,g_{2}^{2} \lambda_1 +12\, \lambda_{1}^{2} +4\, \lambda_{3}^{2} +4\, \lambda_3 \lambda_4 +2\,\lambda_{4}^{2}\nonumber\\
	&+12\, \lambda_1 \mbox{Tr}\Big({Y_d  Y_{d}^{\dagger}}\Big)+4 \,\lambda_1 \mbox{Tr}\Big({Y_e  Y_{e}^{\dagger}}\Big) -12 \,\mbox{Tr}\Big({Y_d  Y_{d}^{\dagger}  Y_d  Y_{d}^{\dagger}}\Big) -4\, \mbox{Tr}\Big({Y_e  Y_{e}^{\dagger}  Y_e  Y_{e}^{\dagger}}\Big) \\ 
	\beta_{\lambda_1}^{(2)} & =  
	-\frac{3537}{1000}\, g_{1}^{6} -\frac{1719}{200} \,g_{1}^{4} g_{2}^{2} -\frac{303}{40} \,g_{1}^{2} g_{2}^{4} +\frac{291}{8}\, g_{2}^{6} +\frac{1953}{200} \,g_{1}^{4} \lambda_1 +\frac{117}{20} \,g_{1}^{2} g_{2}^{2} \lambda_1 -\frac{51}{8}\, g_{2}^{4} \lambda_1\nonumber \\  &+\frac{108}{10}\, g_{1}^{2} \lambda_{1}^{2} 
	+\frac{108}{2}, g_{2}^{2} \lambda_{1}^{2} -78\, \lambda_{1}^{3} +\frac{9}{5}\, g_{1}^{4} \lambda_3 +15\, g_{2}^{4} \lambda_3 +\frac{24}{5}\, g_{1}^{2} \lambda_{3}^{2} +24\, g_{2}^{2} \lambda_{3}^{2} -20\, \lambda_1 \lambda_{3}^{2} \nonumber \\ 
	&-16\, \lambda_{3}^{3} +\frac{9}{10}\, g_{1}^{4} \lambda_4 +3 \,g_{1}^{2} g_{2}^{2} \lambda_4 +\frac{15}{2}\, g_{2}^{4} \lambda_4 +\frac{24}{5}\, g_{1}^{2} \lambda_3 \lambda_4 +24\, g_{2}^{2} \lambda_3 \lambda_4 -20\, \lambda_1 \lambda_3 \lambda_4 -24\, \lambda_{3}^{2} \lambda_4 \nonumber \\ 
	&+\frac{12}{5}\, g_{1}^{2} \lambda_{4}^{2} +6\, g_{2}^{2} \lambda_{4}^{2} -12\, \lambda_1 \lambda_{4}^{2} -32\, \lambda_3 \lambda_{4}^{2} -12\, \lambda_{4}^{3} \nonumber \\ 
	&+\Big( 80\, \lambda_1  \,g_{3}^{2}  + \frac{9}{2} \,\lambda_1^2 +\frac{45}{2}\, g_{2}^{2} \lambda_1   -\frac{9}{2}\, g_{2}^{4} + \frac{9}{10}\, g_{1}^{4}  + \frac{1}{20}\,g_{1}^{2} \Big(50\, \lambda_1  + 108\, g_{2}^{2} \Big)\Big)\mbox{Tr}\Big({Y_d  Y_{d}^{\dagger}}\Big) \nonumber \\ 
	&-\frac{3}{20} \,\Big(30\, g_{1}^{4}  -2\, g_{1}^{2} \Big(22 \,g_{2}^{2}  + 25\, \lambda_1 \Big) -20 \,g_{2}^{2} \lambda_1  + 160 \,\lambda_{1}^{2}  +10\, g_{2}^{4}\Big)\mbox{Tr}\Big({Y_e  Y_{e}^{\dagger}}\Big)-6\,\Big(4\, \lambda_{3}^{2}  \nonumber\\
	&+4\, \lambda_3 \lambda_4 +2\,\lambda_{4}^{2}\Big) \mbox{Tr}\Big({Y_u  Y_{u}^{\dagger}}\Big) -2\,\Big(4\, \lambda_{3}^{2} +4\,\lambda_3 \lambda_4 +2\, \lambda_{4}^{2} \Big) \mbox{Tr}\Big({Y_\nu  Y_{\nu}^{\dagger}}\Big)+\Big(\frac{8}{5}\, g_{1}^{2}\nonumber\\
	&-64\, g_{3}^{2} -3\, \lambda_1 \Big) \mbox{Tr}\Big({Y_d  Y_{d}^{\dagger}  Y_d  Y_{d}^{\dagger}}\Big) -9\, \lambda_1 \mbox{Tr}\Big({Y_{d}^{\dagger} Y_u   Y_{u}^{\dagger} Y_d   }\Big) -\Big(\frac{25}{5}\, g_{1}^{2} +\lambda_1\Big)\, \mbox{Tr}\Big({Y_e  Y_{e}^{\dagger}  Y_e  Y_{e}^{\dagger}}\Big) \nonumber\\
	&-3 \,\lambda_1 \mbox{Tr}\Big({Y_{e}^{\dagger} Y_\nu Y_{\nu}^{\dagger}  Y_e  }\Big) +60\, \mbox{Tr}\Big({Y_d  Y_{d}^{\dagger}  Y_d  Y_{d}^{\dagger}  Y_d  Y_{d}^{\dagger}}\Big) +12\, \mbox{Tr}\Big({Y_{d}^{\dagger} Y_d Y_{d}^{\dagger} Y_u Y_{u}^{\dagger} Y_d  }\Big) \nonumber \\ 
	&+20\, \mbox{Tr}\Big({Y_e  Y_{e}^{\dagger}  Y_e  Y_{e}^{\dagger}  Y_e  Y_{e}^{\dagger}}\Big) +4\, \mbox{Tr}\Big({Y_{e}^{\dagger} Y_e Y_e  Y_{\nu}^{\dagger}  Y_\nu  Y_{e}^{\dagger}    }\Big) \\ 
	\beta_{\lambda_2}^{(1)} & =  
	+\frac{27}{100} \,g_{1}^{4} +\frac{9}{10}\, g_{1}^{2} g_{2}^{2} +\frac{9}{4}\, g_{2}^{4} -\frac{9}{5} \,g_{1}^{2} \lambda_2 -9 \,g_{2}^{2} \lambda_2 +12\, \lambda_{2}^{2} +4\, \lambda_{3}^{2} +4\, \lambda_3 \lambda_4 +2\,\lambda_{4}^{2}\nonumber\\
	&+12\, \lambda_2 \mbox{Tr}\Big({Y_u  Y_{u}^{\dagger}}\Big)+4 \,\lambda_2 \mbox{Tr}\Big({Y_\nu  Y_{\nu}^{\dagger}}\Big) -12 \,\mbox{Tr}\Big({Y_u  Y_{u}^{\dagger}  Y_u  Y_{u}^{\dagger}}\Big) -4\, \mbox{Tr}\Big({Y_\nu  Y_{\nu}^{\dagger}  Y_\nu  Y_{\nu}^{\dagger}}\Big) \\ 
	\beta_{\lambda_2}^{(2)} & =  
	-\frac{3537}{1000}\, g_{1}^{6} -\frac{1719}{200}\, g_{1}^{4} g_{2}^{2} -\frac{303}{40}\, g_{1}^{2} g_{2}^{4} +\frac{291}{8} \,g_{2}^{6} +\frac{1953}{200}\, g_{1}^{4} \lambda_2 +\frac{117}{20} \,g_{1}^{2} g_{2}^{2} \lambda_2 -\frac{51}{8}\, g_{2}^{4} \lambda_2 \nonumber\\
	&+\frac{108}{10} g_{1}^{2} \lambda_{2}^{2} +54\, g_{2}^{2} \lambda_{2}^{2} -78\, \lambda_{2}^{3} +\frac{9}{5}\, g_{1}^{4} \lambda_3 +15 \,g_{2}^{4} \lambda_3 +\frac{24}{5} \,g_{1}^{2} \lambda_{3}^{2} +24\, g_{2}^{2} \lambda_{3}^{2} -20\,\lambda_2 \lambda_{3}^{2} \nonumber\\
	&-16\, \lambda_{3}^{3} +\frac{9}{10} g_{1}^{4} \lambda_4 +3 \,g_{1}^{2} g_{2}^{2} \lambda_4 +\frac{15}{2}\, g_{2}^{4} \lambda_4 +\frac{24}{5}\, g_{1}^{2} \lambda_3 \lambda_4 +24\, g_{2}^{2} \lambda_3 \lambda_4 -20\, \lambda_2 \lambda_3 \lambda_4 -24\, \lambda_{3}^{2} \lambda_4 \nonumber\\
	&+\frac{12}{5} \,g_{1}^{2} \lambda_{4}^{2} +6\, g_{2}^{2} \lambda_{4}^{2} -12\, \lambda_2 \lambda_{4}^{2} -32\, \lambda_3 \lambda_{4}^{2} -12 \,\lambda_{4}^{3}   \nonumber\\
	&-6\, \Big(4\, \lambda_{3}^{2}  + 4\, \lambda_3 \lambda_4  +2\, \lambda_{4}^{2} \Big)\mbox{Tr}\Big({Y_d  Y_{d}^{\dagger}}\Big) -2\, \Big(4\, \lambda_{3}^{2}  + 4\, \lambda_3 \lambda_4  +2\, \lambda_{4}^{2} \Big)\mbox{Tr}\Big({Y_e  Y_{e}^{\dagger}}\Big)\nonumber \\  &+\Big(-\frac{171}{50} \,g_{1}^{4} +\frac{63}{5}\, g_{1}^{2} g_{2}^{2} -\frac{9}{2}\, g_{2}^{4}  +\frac{17}{2}\, g_{1}^{2} \lambda_2  +\frac{45}{2}\, g_{2}^{2} \lambda_2 +80\, g_{3}^{2} \lambda_2 -72\, \lambda_{2}^{2}\Big) \mbox{Tr}\Big({Y_u  Y_{u}^{\dagger}}\Big) \nonumber\\
	&+\Big(-\frac{9}{50}\, g_{1}^{4}-\frac{3}{5}\, g_{1}^{2} g_{2}^{2} -\frac{3}{2}\, g_{2}^{4}  +\frac{3}{2}\, g_{1}^{2} \lambda_2 +\frac{15}{2}\, g_{2}^{2} \lambda_2 -24\, \lambda_{2}^{2}\Big) \mbox{Tr}\Big({Y_\nu  Y_{\nu}^{\dagger}}\Big)\nonumber\\
	&-9 \,\lambda_2 \mbox{Tr}\Big({Y_{d}^{\dagger} Y_u Y_{u}^{\dagger} Y_d   }\Big) -3 \,\lambda_2 \mbox{Tr}\Big({ Y_{e}^{\dagger} Y_\nu Y_{\nu}^{\dagger} Y_e      }\Big) -\frac{16}{5} \,g_{1}^{2} \mbox{Tr}\Big({Y_u  Y_{u}^{\dagger}  Y_u  Y_{u}^{\dagger}}\Big)\nonumber\\
	& -64\, g_{3}^{2} \mbox{Tr}\Big({Y_u  Y_{u}^{\dagger}  Y_u  Y_{u}^{\dagger}}\Big)-3\, \lambda_2 \mbox{Tr}\Big({Y_u  Y_{u}^{\dagger}  Y_u  Y_{u}^{\dagger}}\Big) - \lambda_2 \mbox{Tr}\Big({Y_\nu  Y_{\nu}^{\dagger}  Y_\nu  Y_{\nu}^{\dagger}}\Big) \nonumber\\
	&+12\, \mbox{Tr}\Big({Y_{d}^{\dagger} Y_u Y_{u}^{\dagger}  Y_u Y_d}\Big) +4\, \mbox{Tr}\Big({ Y_{e}^{\dagger} Y_\nu Y_{\nu}^{\dagger} Y_\nu   Y_{\nu}^{\dagger}  Y_e      }\Big) +60\, \mbox{Tr}\Big({Y_u  Y_{u}^{\dagger}  Y_u  Y_{u}^{\dagger}  Y_u  Y_{u}^{\dagger}}\Big) \nonumber\\
	&+20 \,\mbox{Tr}\Big({Y_\nu  Y_{\nu}^{\dagger}  Y_\nu  Y_{\nu}^{\dagger}  Y_\nu  Y_{\nu}^{\dagger}}\Big) \\
	\beta_{\lambda_3}^{(1)} & =  
	+\frac{27}{100}\, g_{1}^{4} -\frac{9}{10}\, g_{1}^{2} g_{2}^{2} +\frac{9}{4} \,g_{2}^{4} -\frac{9}{5}\, g_{1}^{2} \lambda_3 -9 \,g_{2}^{2} \lambda_3 +6\, \lambda_1 \lambda_3 +6\, \lambda_2 \lambda_3 +4\, \lambda_{3}^{2} +2 \,\lambda_1 \lambda_4\nonumber\\
	& +2 \,\lambda_2 \lambda_4 +2 \,\lambda_{4}^{2} +6\, \lambda_3 \mbox{Tr}\Big({Y_d  Y_{d}^{\dagger}}\Big) +2\, \lambda_3 \mbox{Tr}\Big({Y_e  Y_{e}^{\dagger}}\Big) +6\, \lambda_3 \mbox{Tr}\Big({Y_u  Y_{u}^{\dagger}}\Big) \nonumber\\
	&+2 \,\lambda_3 \mbox{Tr}\Big({Y_\nu  Y_{\nu}^{\dagger}}\Big) -12\, \mbox{Tr}\Big({ Y_{d}^{\dagger} Y_u Y_{u}^{\dagger} Y_d  }\Big) -4 \,\mbox{Tr}\Big({ Y_{e}^{\dagger} Y_\nu Y_{\nu}^{\dagger} Y_e    }\Big) \\ 
	\beta_{\lambda_3}^{(2)} & =  
	-\frac{3537}{1000}\, g_{1}^{6} +\frac{909}{200}\, g_{1}^{4} g_{2}^{2} +\frac{33}{40} \,g_{1}^{2} g_{2}^{4} +\frac{291}{8}\, g_{2}^{6} +\frac{27}{20}\, g_{1}^{4} \lambda_1 -\frac{3}{2}\, g_{1}^{2} g_{2}^{2} \lambda_1 +\frac{45}{4}\, g_{2}^{4} \lambda_1 \nonumber\\
	&+\frac{27}{20} \,g_{1}^{4} \lambda_2 -\frac{3}{2}\,g_{1}^{2} g_{2}^{2} \lambda_2 +\frac{45}{4}\, g_{2}^{4} \lambda_2 +\frac{1773}{200} \,g_{1}^{4} \lambda_3 +\frac{33}{20} \,g_{1}^{2} g_{2}^{2} \lambda_3 -\frac{111}{8}\, g_{2}^{4} \lambda_3 +\frac{36}{5}\, g_{1}^{2} \lambda_1 \lambda_3 \nonumber\\
	&+36\, g_{2}^{2} \lambda_1 \lambda_3 -60\, \lambda_{1}^{2} \lambda_3 +\frac{36}{5}\, g_{1}^{2} \lambda_2 \lambda_3 +36 \,g_{2}^{2} \lambda_2 \lambda_3 -15\, \lambda_{2}^{2} \lambda_3 +\frac{6}{5} \,g_{1}^{2} \lambda_{3}^{2} +6\, g_{2}^{2} \lambda_{3}^{2} -36 \,\lambda_1 \lambda_{3}^{2} \nonumber\\
	&-36\, \lambda_2 \lambda_{3}^{2}-12 \,\lambda_{3}^{3} +\frac{9}{10}\, g_{1}^{4} \lambda_4 -\frac{9}{5}\, g_{1}^{2} g_{2}^{2} \lambda_4 +\frac{15}{2} \,g_{2}^{4} \lambda_4 +\frac{12}{5}\, g_{1}^{2} \lambda_1 \lambda_4 +18 \,g_{2}^{2} \lambda_1 \lambda_4 -4 \,\lambda_{1}^{2} \lambda_4 \nonumber\\
	&+\frac{12}{5}\, g_{1}^{2} \lambda_2 \lambda_4 +18\, g_{2}^{2} \lambda_2 \lambda_4 -4 \,\lambda_{2}^{2} \lambda_4 -12\, g_{2}^{2} \lambda_3 \lambda_4 -16\, \lambda_1 \lambda_3 \lambda_4 -16\, \lambda_2 \lambda_3 \lambda_4 -4\, \lambda_{3}^{2} \lambda_4 \nonumber\\
	&-\frac{6}{5}\, g_{1}^{2} \lambda_{4}^{2} +6\, g_{2}^{2} \lambda_{4}^{2} -14\, \lambda_1 \lambda_{4}^{2} -14\, \lambda_2 \lambda_{4}^{2} -16\, \lambda_3 \lambda_{4}^{2} -12\, \lambda_{4}^{3} \nonumber \\ 
	&+\frac{1}{20}\, \Big(-120\, \Big(2 \lambda_{3}^{2} + 2 \,\lambda_1 \Big(3\, \lambda_3  + \lambda_4\Big) + \lambda_{4}^{2} \Big) + 225\, g_{2}^{2} \lambda_3 \nonumber\\
	& -45\, g_{2}^{4}  + 800\, g_{3}^{2} \lambda_3  + 9\, g_{1}^{4}  + g_{1}^{2} \Big(25 \lambda_3  -54\, g_{2}^{2} \Big)\Big)\mbox{Tr}\Big({Y_d  Y_{d}^{\dagger}}\Big) -\frac{1}{20}\, \Big(15 \,g_{2}^{4}  + 40\, \Big(2 \lambda_{3}^{2}  \nonumber\\
	&+ 2\, \lambda_1 \Big(3\, \lambda_3  + \lambda_4\Big) + \lambda_{4}^{2} \Big) + 45\, g_{1}^{4}  -75\, g_{2}^{2} \lambda_3  + g_{1}^{2} \Big(66\, g_{2}^{2}  -75\, \lambda_3 \Big)\Big)\mbox{Tr}\Big({Y_e  Y_{e}^{\dagger}}\Big) \nonumber \\ 
	&+\Big(-\frac{171}{100}\, g_{1}^{4}  -\frac{63}{10} \,g_{1}^{2} g_{2}^{2} -\frac{9}{4}\, g_{2}^{4}+\frac{17}{4}\, g_{1}^{2} \lambda_3 +\frac{45}{4} \,g_{2}^{2} \lambda_3 +40\, g_{3}^{2} \lambda_3 -36\, \lambda_2 \lambda_3  -12\, \lambda_{3}^{2} \nonumber\\
	&-12 \,\lambda_2 \lambda_4 -6\, \lambda_{4}^{2} \Big) \mbox{Tr}\Big({Y_u  Y_{u}^{\dagger}}\Big)+\Big( -\frac{9}{100}\, g_{1}^{4} +\frac{3}{10}\, g_{1}^{2} g_{2}^{2}  -\frac{3}{4}\, g_{2}^{4} +\frac{3}{4}\, g_{1}^{2} \lambda_3 \nonumber\\
	&+\frac{15}{4}\, g_{2}^{2} \lambda_3 -12\, \lambda_2 \lambda_3  -4 \,\lambda_{3}^{2}  -4\, \lambda_2 \lambda_4  -2\, \lambda_{4}^{2} \Big) \mbox{Tr}\Big({Y_\nu  Y_{\nu}^{\dagger}}\Big) -\frac{27}{2}\, \lambda_3 \mbox{Tr}\Big({Y_d  Y_{d}^{\dagger}  Y_d  Y_{d}^{\dagger}}\Big) \nonumber\\
	&+\Big(-\frac{4}{5}\, g_{1}^{2}  -64\, g_{3}^{2}  +15\, \lambda_3 \Big)\mbox{Tr}\Big({Y_{d}^{\dagger} Y_u Y_{u}^{\dagger} Y_d }\Big)-\frac{9}{2}\, \lambda_3 \mbox{Tr}\Big({Y_e  Y_{e}^{\dagger}  Y_e  Y_{e}^{\dagger}}\Big) \nonumber \\ 
	&+\Big(-\frac{12}{5}\, g_{1}^{2} +5\, \lambda_3 \Big)\mbox{Tr}\Big({Y_{e}^{\dagger} Y_\nu Y_{\nu}^{\dagger}  Y_e    }\Big) -\frac{27}{2} \,\lambda_3 \mbox{Tr}\Big({Y_u  Y_{u}^{\dagger}  Y_u  Y_{u}^{\dagger}}\Big) -\frac{9}{2}\, \lambda_3 \mbox{Tr}\Big({Y_\nu  Y_{\nu}^{\dagger}  Y_\nu  Y_{\nu}^{\dagger}}\Big) \nonumber \\ 
	&+12\, \mbox{Tr}\Big({Y_{d}^{\dagger} Y_u Y_{u}^{\dagger} Y_d  Y_{d}^{\dagger} Y_d     }\Big) +24\, \mbox{Tr}\Big({Y_{d}^{\dagger}Y_d  Y_{d}^{\dagger} Y_u  Y_{u}^{\dagger}  Y_d    }\Big) +36\, \mbox{Tr}\Big({Y_{d}^{\dagger} Y_u  Y_{u}^{\dagger}  Y_u Y_{u}^{\dagger}   Y_d      }\Big) \nonumber \\ 
	&+4\, \mbox{Tr}\Big({Y_{e}^{\dagger} Y_\nu Y_{\nu}^{\dagger}  Y_e   Y_{e}^{\dagger}  Y_e  }\Big) +8\, \mbox{Tr}\Big({Y_{e}^{\dagger} Y_e Y_{e}^{\dagger}Y_\nu  Y_{\nu}^{\dagger} Y_e         }\Big) +12\, \mbox{Tr}\Big({Y_{e}^{\dagger} Y_\nu Y_{\nu}^{\dagger}  Y_\nu Y_{\nu}^{\dagger}  Y_e   }\Big) \\
	\beta_{\lambda_4}^{(1)} & =  
	+\frac{9}{5}\, g_{1}^{2} g_{2}^{2} -\frac{9}{5}\, g_{1}^{2} \lambda_4 -9\, g_{2}^{2} \lambda_4 +2\, \lambda_1 \lambda_4 +2 \,\lambda_2 \lambda_4 +8\, \lambda_3 \lambda_4 +4\, \lambda_{4}^{2}  +6\, \lambda_4 \mbox{Tr}\Big({Y_d  Y_{d}^{\dagger}}\Big) \nonumber \\ 
	&+2\, \lambda_4 \mbox{Tr}\Big({Y_e  Y_{e}^{\dagger}}\Big) +6 \,\lambda_4 \mbox{Tr}\Big({Y_u  Y_{u}^{\dagger}}\Big) +2\, \lambda_4 \mbox{Tr}\Big({Y_\nu  Y_{\nu}^{\dagger}}\Big) +12\, \mbox{Tr}\Big({Y_{d}^{\dagger}  Y_u Y_{u}^{\dagger} Y_d     }\Big) \nonumber\\
	&+4 \mbox{Tr}\Big({Y_{e}^{\dagger} Y_\nu Y_{\nu}^{\dagger}  Y_e   }\Big) \\ 
	\beta_{\lambda_4}^{(2)} & =  
	-\frac{657}{50}\, g_{1}^{4} g_{2}^{2} -\frac{42}{5}\, g_{1}^{2} g_{2}^{4} +3 \,g_{1}^{2} g_{2}^{2} \lambda_1 +3\, g_{1}^{2} g_{2}^{2} \lambda_2 +\frac{6}{5}\, g_{1}^{2} g_{2}^{2} \lambda_3 +\frac{1413}{200}\, g_{1}^{4} \lambda_4 +\frac{153}{20}\, g_{1}^{2} g_{2}^{2} \lambda_4 \nonumber \\ 
	&-\frac{231}{8} \,g_{2}^{4} \lambda_4 +\frac{12}{5}\, g_{1}^{2} \lambda_1 \lambda_4 -7 \,\lambda_{1}^{2} \lambda_4 +\frac{12}{5}\, g_{1}^{2} \lambda_2 \lambda_4 -7\, \lambda_{2}^{2} \lambda_4 +\frac{12}{5}\, g_{1}^{2} \lambda_3 \lambda_4 +36\, g_{2}^{2} \lambda_3 \lambda_4 \nonumber \\ 
	&-40\, \lambda_1 \lambda_3 \lambda_4 -40\, \lambda_2 \lambda_3 \lambda_4 -28\, \lambda_{3}^{2} \lambda_4 +\frac{24}{5}\, g_{1}^{2} \lambda_{4}^{2} +18\, g_{2}^{2} \lambda_{4}^{2} -20 \,\lambda_1 \lambda_{4}^{2} -20 \,\lambda_2 \lambda_{4}^{2} -28 \,\lambda_3 \lambda_{4}^{2} \nonumber \\ 
	&+\frac{1}{20} \,\Big(225 \,g_{2}^{2} \lambda_4 -240\, \Big( \lambda_1 \lambda_4  + 2 \,\lambda_3 \lambda_4  + \lambda_{4}^{2}\Big) + 800 \,g_{3}^{2} \lambda_4  + g_{1}^{2} \Big(108\, g_{2}^{2}  + 25\, \lambda_4 \Big)\Big)\mbox{Tr}\Big({Y_d  Y_{d}^{\dagger}}\Big) \nonumber \\ 
	&+\frac{1}{20}\, \Big(3\, g_{1}^{2} \Big(25\, \lambda_4  + 44\, g_{2}^{2} \Big) + 75\, g_{2}^{2} \lambda_4  -80\, \Big( \lambda_1 \lambda_4  + 2\, \lambda_3 \lambda_4   + \lambda_{4}^{2}\Big)\Big)\mbox{Tr}\Big({Y_e  Y_{e}^{\dagger}}\Big) \nonumber \\ 
	&+\Big(\frac{63}{5}\, g_{1}^{2} g_{2}^{2}  +\frac{17}{4}\, g_{1}^{2} \lambda_4 +\frac{45}{4}\, g_{2}^{2} \lambda_4 +40 \,g_{3}^{2} \lambda_4 -12\, \lambda_2 \lambda_4  -24\, \lambda_3 \lambda_4  -12\, \lambda_{4}^{2}  \Big)\mbox{Tr}\Big({Y_u  Y_{u}^{\dagger}}\Big) \nonumber\\
	&+\Big(-\frac{3}{5} \,g_{1}^{2} g_{2}^{2}  +\frac{3}{4} \,g_{1}^{2} \lambda_4  +\frac{15}{4} \,g_{2}^{2} \lambda_4  -8 \,\lambda_2 \lambda_4 -8\, \lambda_3 \lambda_4 -4\, \lambda_{4}^{2} \Big)\mbox{Tr}\Big({Y_\nu  Y_{\nu}^{\dagger}}\Big)\nonumber\\
	& -\frac{27}{2} \,\lambda_4 \mbox{Tr}\Big({Y_d  Y_{d}^{\dagger}  Y_d  Y_{d}^{\dagger}}\Big)+\Big(\frac{4}{5} \,g_{1}^{2} +64\, g_{3}^{2}  -24\, \lambda_3 -33\, \lambda_4 \Big)\mbox{Tr}\Big({Y_d  Y_{u}^{\dagger}  Y_u  Y_{d}^{\dagger}}\Big) \nonumber\\
	&-\frac{9}{2}\, \lambda_4 \mbox{Tr}\Big({Y_e  Y_{e}^{\dagger}  Y_e  Y_{e}^{\dagger}}\Big) +\Big(\frac{12}{5}\, g_{1}^{2} -8 \,\lambda_3  -11 \,\lambda_4 \Big)\mbox{Tr}\Big({Y_e  Y_{\nu}^{\dagger}  Y_\nu  Y_{e}^{\dagger}}\Big) \nonumber\\
	&-\frac{27}{2}\, \lambda_4 \mbox{Tr}\Big({Y_u  Y_{u}^{\dagger}  Y_u  Y_{u}^{\dagger}}\Big) -\frac{9}{2}\, \lambda_4 \mbox{Tr}\Big({Y_\nu  Y_{\nu}^{\dagger}  Y_\nu  Y_{\nu}^{\dagger}}\Big) -12\, \mbox{Tr}\Big({Y_d  Y_{d}^{\dagger}  Y_d  Y_{u}^{\dagger}  Y_u  Y_{d}^{\dagger}}\Big) \nonumber\\
	&-12\, \mbox{Tr}\Big({Y_d  Y_{u}^{\dagger}  Y_u  Y_{d}^{\dagger}  Y_d  Y_{d}^{\dagger}}\Big) -24\, \mbox{Tr}\Big({Y_d  Y_{u}^{\dagger}  Y_u  Y_{u}^{\dagger}  Y_u  Y_{d}^{\dagger}}\Big) -4 \,\mbox{Tr}\Big({Y_e  Y_{e}^{\dagger}  Y_e  Y_{\nu}^{\dagger}  Y_\nu  Y_{e}^{\dagger}}\Big)\nonumber\\
	& -4\, \mbox{Tr}\Big({Y_e  Y_{\nu}^{\dagger}  Y_\nu  Y_{e}^{\dagger}  Y_e  Y_{e}^{\dagger}}\Big) -8\, \mbox{Tr}\Big({Y_e  Y_{\nu}^{\dagger}  Y_\nu  Y_{\nu}^{\dagger}  Y_\nu  Y_{e}^{\dagger}}\Big)  
	\end{align}} 

\section{One loop threshold corrections at $M_S$}
\label{Appendix:threshold}
In this appendix, we list the one-loop threshold corrections to the scalar quartic couplings considered in our analysis. These corrections arise from the box and triangle diagrams in MSSM at the scale $M_S$ in $\overline {\rm MS}$ scheme \cite{Haber:1993an,Lee:2015uza}. Note that the following expressions are obtained for a particular choice of the SUSY spectrum given in Eq. (\ref{susy_spectrum}). 
\begin{eqnarray}
\Delta \lambda_1&=&\frac{1}{16\pi^2}\,\Big(\frac{\mu}{M_S}\Big)^4\,\left[-\frac{1}{2}\,\mbox{Tr}\Big({Y_u  Y_{u}^{\dagger}  Y_u  Y_{u}^{\dagger}}\Big)-\frac{1}{6}\,\mbox{Tr}\Big({Y_\nu  Y_{\nu}^{\dagger}  Y_\nu  Y_{\nu}^{\dagger}}\Big)\right]\nonumber\\
&+&\frac{1}{16\pi^2}\,\Big(\frac{\mu}{M_S}\Big)^2\,\frac{g_2^2+g_Y^2}{4}\,\left[3\,\mbox{Tr}\Big({Y_u  Y_{u}^{\dagger}}\Big)+\mbox{Tr}\Big({Y_\nu  Y_{\nu}^{\dagger}}\Big)\right]\\
\Delta \lambda_2&=&\frac{1}{16\pi^2}\,\Big(\frac{\mu}{M_S}\Big)^4\,\left[-\frac{1}{2}\,\mbox{Tr}\Big({Y_d  Y_{d}^{\dagger}  Y_d  Y_{d}^{\dagger}}\Big)-\frac{1}{6}\,\mbox{Tr}\Big({Y_e  Y_{e}^{\dagger}  Y_e Y_{e}^{\dagger}}\Big)\right]\nonumber\\
&+&\frac{1}{16\pi^2}\,\Big(\frac{\mu}{M_S}\Big)^2\,\frac{g_2^2+g_Y^2}{4}\,\left[3\,\mbox{Tr}\Big({Y_d  Y_{d}^{\dagger}}\Big)+\mbox{Tr}\Big({Y_e  Y_{e}^{\dagger}}\Big)\right]\\
\Delta \lambda_3&=&\frac{1}{16\pi^2}\,\Big(\frac{\mu}{M_S}\Big)^2\,\left[\frac{3}{2}\,\mbox{Tr}\Big({Y_u  Y_{u}^{\dagger}  Y_u  Y_{u}^{\dagger}}\Big)+\frac{3}{2}\,\mbox{Tr}\Big({Y_d  Y_{d}^{\dagger}  Y_d  Y_{d}^{\dagger}}\Big)+\frac{1}{2}\,\mbox{Tr}\Big({Y_\nu  Y_{\nu}^{\dagger}  Y_\nu  Y_{\nu}^{\dagger}}\Big)+\frac{1}{2}\,\mbox{Tr}\Big({Y_e  Y_{e}^{\dagger}  Y_e Y_{e}^{\dagger}}\Big)\right]\nonumber\\
&+&\frac{1}{16\pi^2}\,\Big(\frac{\mu}{M_S}\Big)^2\,\Big(1+\Big(\frac{\mu}{M_S}\Big)^2\Big)\,\left[-\frac{1}{2}\,\mbox{Tr}\Big({Y_{d}^{\dagger} Y_u  Y_{u}^{\dagger} Y_d  }\Big)-\frac{1}{6}\,\mbox{Tr}\Big({Y_{e}^{\dagger}Y_\nu Y_{\nu}^{\dagger} Y_e    }\Big)\right]\nonumber\\
&+&\frac{1}{16\pi^2}\,\Big(\frac{\mu}{M_S}\Big)^2\,\frac{g_2^2-g_Y^2}{8}\,\left[3\,\mbox{Tr}\Big({Y_u  Y_{u}^{\dagger}}\Big)+\mbox{Tr}\Big({Y_\nu  Y_{\nu}^{\dagger}}\Big)+3\,\mbox{Tr}\Big({Y_d  Y_{d}^{\dagger}}\Big)+\mbox{Tr}\Big({Y_e  Y_{e}^{\dagger}}\Big)\right]\\
\Delta \lambda_4&=&\frac{1}{16\pi^2}\,\Big(\frac{\mu}{M_S}\Big)^2\,\left[\frac{3}{2}\,\mbox{Tr}\Big({Y_u  Y_{u}^{\dagger}  Y_u  Y_{u}^{\dagger}}\Big)+\frac{3}{2}\,\mbox{Tr}\Big({Y_d  Y_{d}^{\dagger}  Y_d  Y_{d}^{\dagger}}\Big)+\frac{1}{2}\,\mbox{Tr}\Big({Y_\nu  Y_{\nu}^{\dagger}  Y_\nu  Y_{\nu}^{\dagger}}\Big)+\frac{1}{2}\,\mbox{Tr}\Big({Y_e  Y_{e}^{\dagger}  Y_e Y_{e}^{\dagger}}\Big)\right]\nonumber\\
&+&\frac{1}{16\pi^2}\,\Big(\frac{\mu}{M_S}\Big)^2\,\Big(1+\Big(\frac{\mu}{M_S}\Big)^2\Big)\,\left[\frac{1}{2}\,\mbox{Tr}\Big({Y_{d}^{\dagger} Y_u  Y_{u}^{\dagger} Y_d  }\Big)+\frac{1}{6}\,\mbox{Tr}\Big({Y_{e}^{\dagger}Y_\nu Y_{\nu}^{\dagger} Y_e    }\Big)\right]\nonumber\\
&-&\frac{1}{16\pi^2}\,\Big(\frac{\mu}{M_S}\Big)^2\,\frac{g_2^2}{4}\,\left[3\,\mbox{Tr}\Big({Y_u  Y_{u}^{\dagger}}\Big)+\mbox{Tr}\Big({Y_\nu  Y_{\nu}^{\dagger}}\Big)+3\,\mbox{Tr}\Big({Y_d  Y_{d}^{\dagger}}\Big)+\mbox{Tr}\Big({Y_e  Y_{e}^{\dagger}}\Big)\right]
\end{eqnarray}
We assume $\mu = 0.1 M_S$ in evaluating the threshold corrections throughout our analysis.

\section{Details of the gauge and Yukawa couplings at $M_t$}
\label{Appendix:inputs}
In this appendix, we briefly describe our procedure to evaluate the input parameters at the scale $M_t$.  We use experimental values of masses of the quarks and leptons from PDG \cite{Patrignani:2016xqp}. The masses of light quarks, namely $u$, $d$ and $s$, are given at 2 GeV while those of heavy quarks $c$, $b$ and $t$ are determined at their respective pole masses. To calculate the running quark masses at $M_t$ we consider 3-loop running of gauge couplings and mass parameters in effective  QCD theory and perform appropriate matching at the intermediate thresholds with 2-loop threshold corrections. We closely follow the notations and procedure given in \cite{Chetyrkin:2000yt,Xing:2007fb}. 

In a given effective QCD$^{(n_f)}$ theory, where $n_f$ is number of quarks flavours lighter than the renormalization scale $\mu$, the running strong coupling constant and quark masses can be calculated by solving the following 3-loop RG equations.
\begin{eqnarray}
\mu^2 \frac{d}{d\mu^2} \frac{\alpha_s^{(n_f)}(\mu)}{\pi}&=&-\sum_{k=0}^2\beta_k^{(n_f)}\,\left(\frac{\alpha_s^{(n_f)}(\mu)}{\pi}\right)^{k+2}\,,\\
\mu^2 \frac{d}{d\mu^2} m_q^{(n_f)}(\mu)&=& - m_q^{(n_f)}(\mu)\,  \sum_{k=0}^2 \gamma_k^{(n_f)}\, \left(\frac{\alpha_s^{(n_f)}(\mu)}{\pi}\right)^{k+1}\,.
\end{eqnarray}
The explicit forms of coefficients $\beta_i$ and $\gamma_i$ are given in Eq. (2) and Eq. (7) in \cite{Chetyrkin:2000yt}. At the given mass threshold the 2-loop corrected matching between the strong coupling constant and quark masses is performed using
\begin{eqnarray}\label{alphas}
\alpha_s(\mu)^{(n_f-1)} &=& \xi_g\,\alpha_s(\mu)^{(n_f)}\,, \nonumber \\
m_q(\mu)^{(n_f-1)} &=& \xi_m\,m_q(\mu)^{(n_f)}\,
\end{eqnarray}
where $\xi_g$, $\xi_m$ are matching factors in $\overline{\rm MS}$ scheme and we use their expressions as given in Eqs. (20,26) in \cite{Chetyrkin:2000yt}.

We use 2-loop QED and 3-loop QCD corrections while evaluating the effect of charged leptons and the strong coupling constant $\alpha_s$ on the running of electromagnetic coupling constant $\alpha_{\rm em}$. The pole masses of charged leptons obtained from PDG \cite{Patrignani:2016xqp} are converted into $\overline{\rm MS}$ running masses using the following relations
\begin{eqnarray}\label{alphaem}
\mu^2 \frac{d\alpha_{em}(\mu)}{d\mu^2}&=&-\frac{\alpha_{\rm em}^2}{\pi}\left(\beta_0+\beta_1\,\frac{\alpha_{em}}{\pi}+\sum_{i=1,3}\,\rho_i\, \left(\frac{\alpha_s}{\pi}\right)^i\right)\,, 
\nonumber \\
m_l(\mu) &=& M_l\, \left(1-\frac{\alpha_{\rm em}(\mu)}{\pi}\,\left(1+\frac{3}{2} \ln\left(\frac{\mu}{m_l(\mu)}\right) \right) \right)\,,
\end{eqnarray}
where $l=e,\mu,\tau$ and $M_l$ is the pole mass of the corresponding charged lepton.  Detailed expressions for $\beta_i$ and $\rho_i$ can be found in Eqs. (16,17) in \cite{Xing:2007fb}. 

The gauge couplings at $M_t$ are obtained using
\begin{eqnarray}
g_1(M_t)&=&\sqrt{\frac{5}{3}}\frac{\sqrt{4\pi\, \alpha_{\rm em}(M_t)}}{\cos \theta_W}\\
g_2(M_t)&=&\frac{\sqrt{4\pi\, \alpha_{\rm em}(M_t)}} {\sin \theta_W}\\
g_3(M_t)&=&\sqrt{4\pi\, \alpha_{s}(M_t)}
\end{eqnarray} 
where $\theta_W$ is the weak mixing angle in $\overline{\rm MS}$ scheme. In the above expression $\alpha_{\rm em}(M_t)$ and $\alpha_{s}(M_t)$ are calculated using Eqs. (\ref{alphaem}) and (\ref{alphas}) respectively. The values of the gauge couplings and fermion masses obtained at $M_t$ in this way are listed in Table \ref{tab:inputs} in section \ref{sec:rge}.

\section{Effects of uncertainty in the measurement of $M_t$ on the results}
\label{Appendix:mt_var}
\begin{figure}[!ht]
\centering
\subfigure{\includegraphics[width=0.43\textwidth]{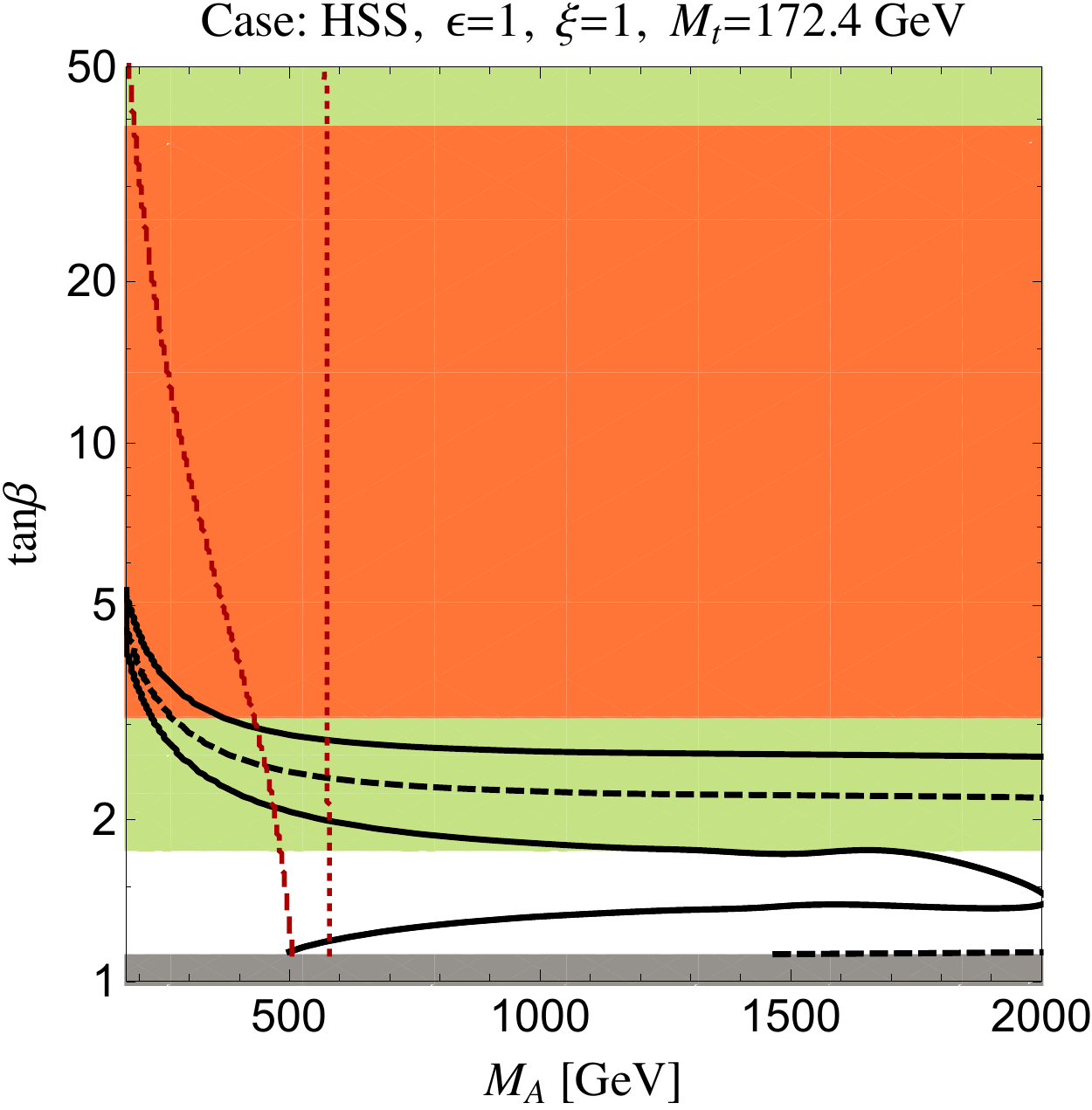}} \hspace*{0.1cm}
\subfigure{\includegraphics[width=0.43\textwidth]{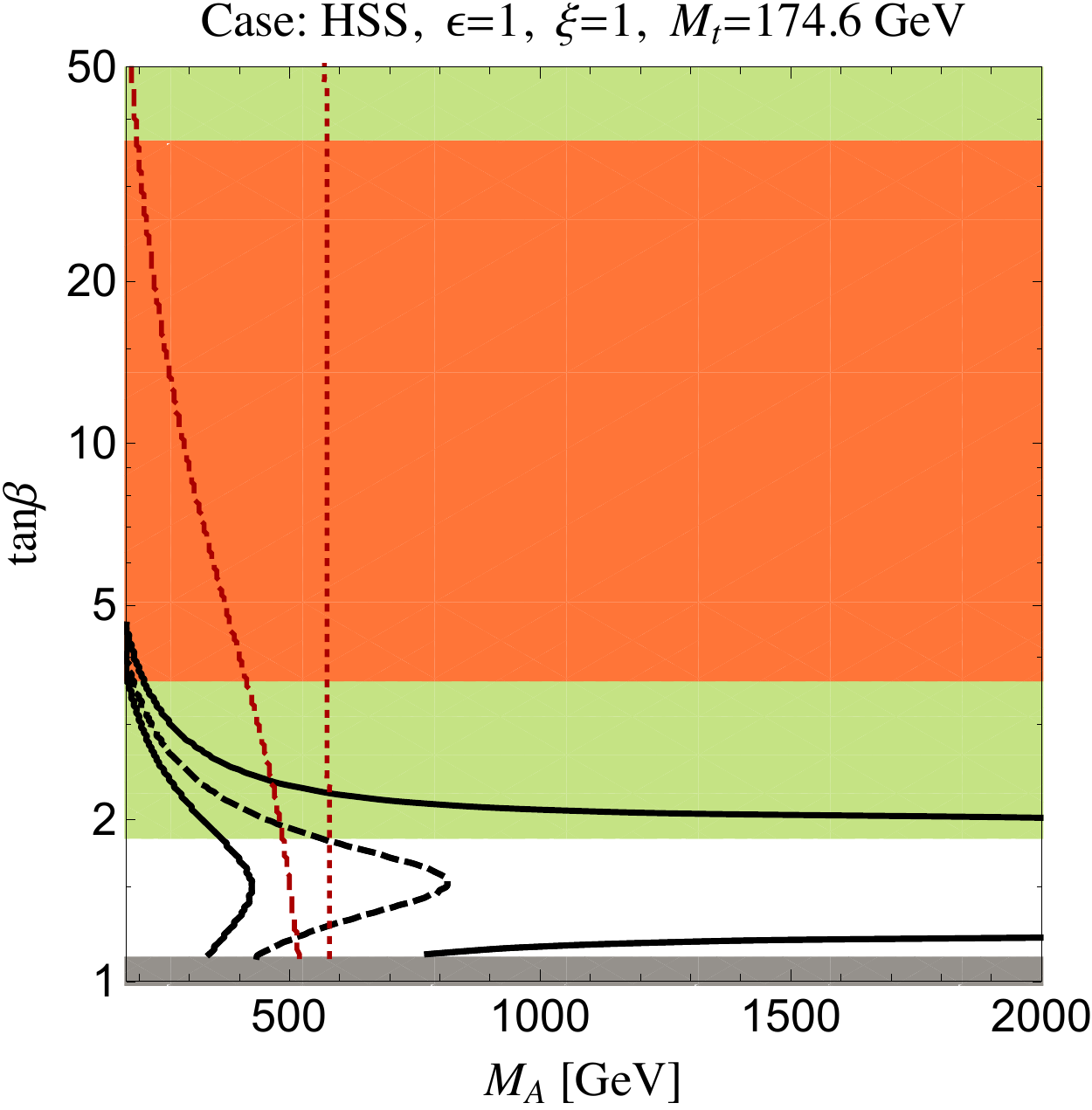}}
\subfigure{\includegraphics[width=0.43\textwidth]{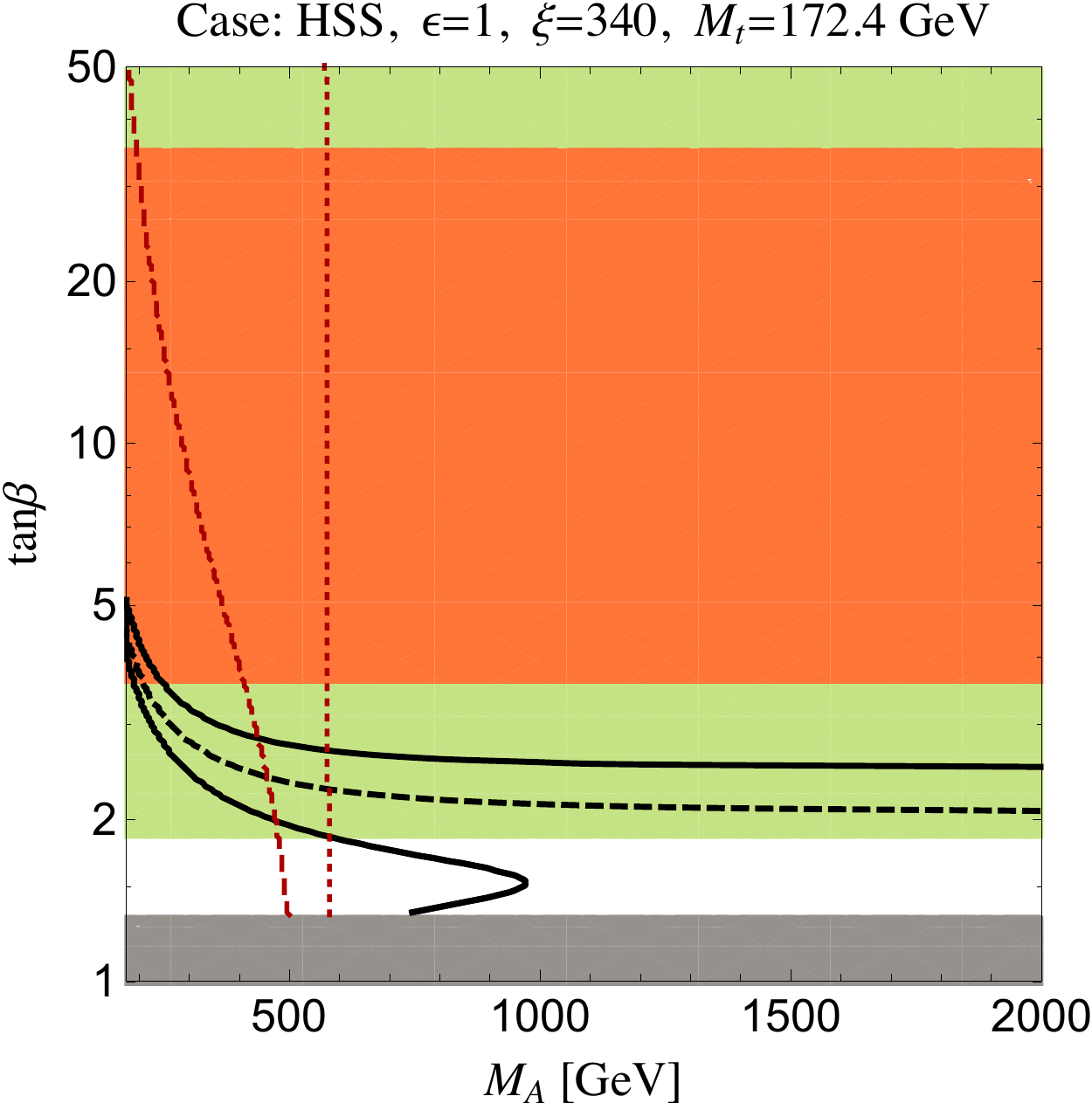}} \hspace*{0.1cm}
\subfigure{\includegraphics[width=0.43\textwidth]{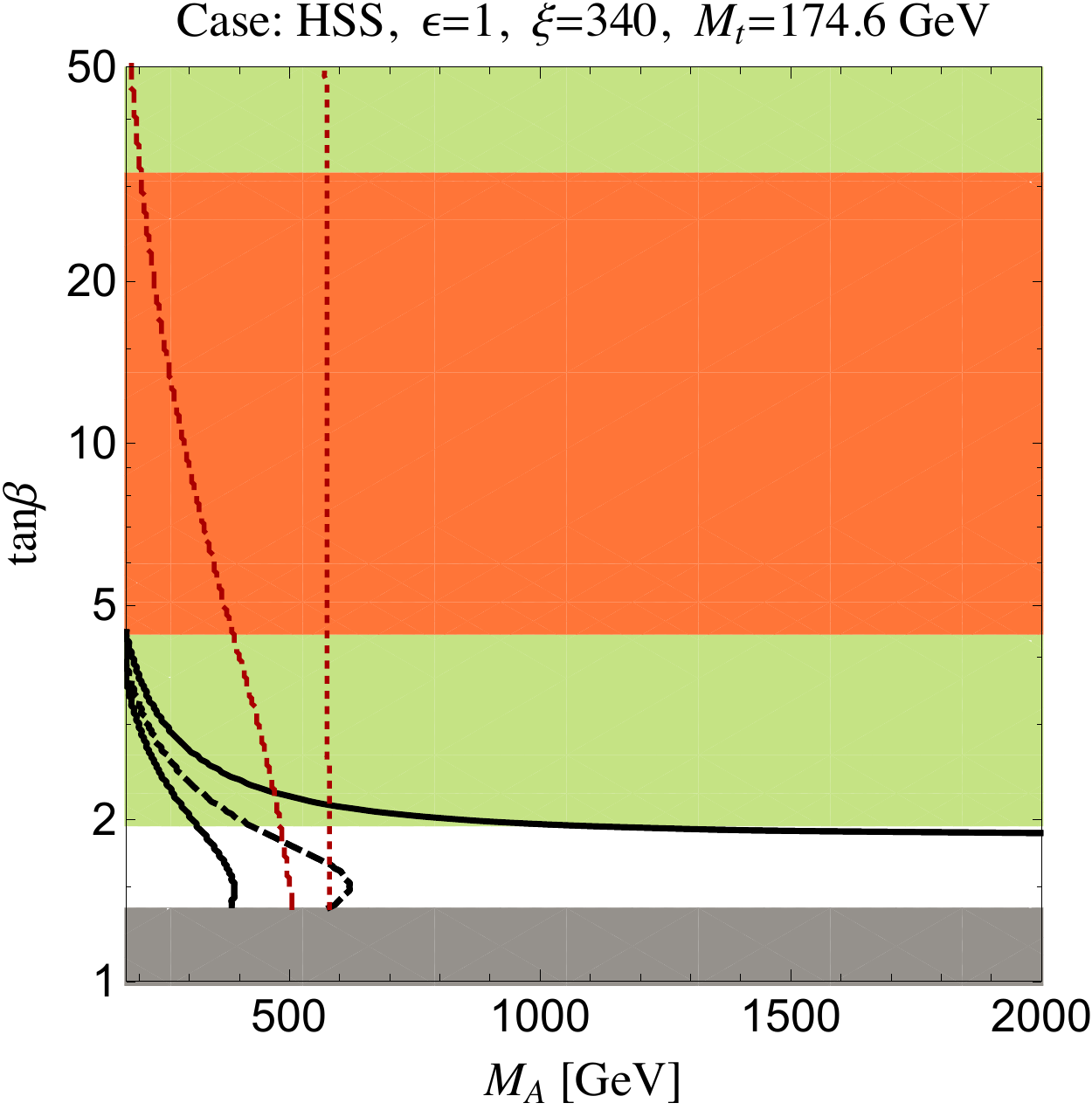}}
\subfigure{\includegraphics[width=0.43\textwidth]{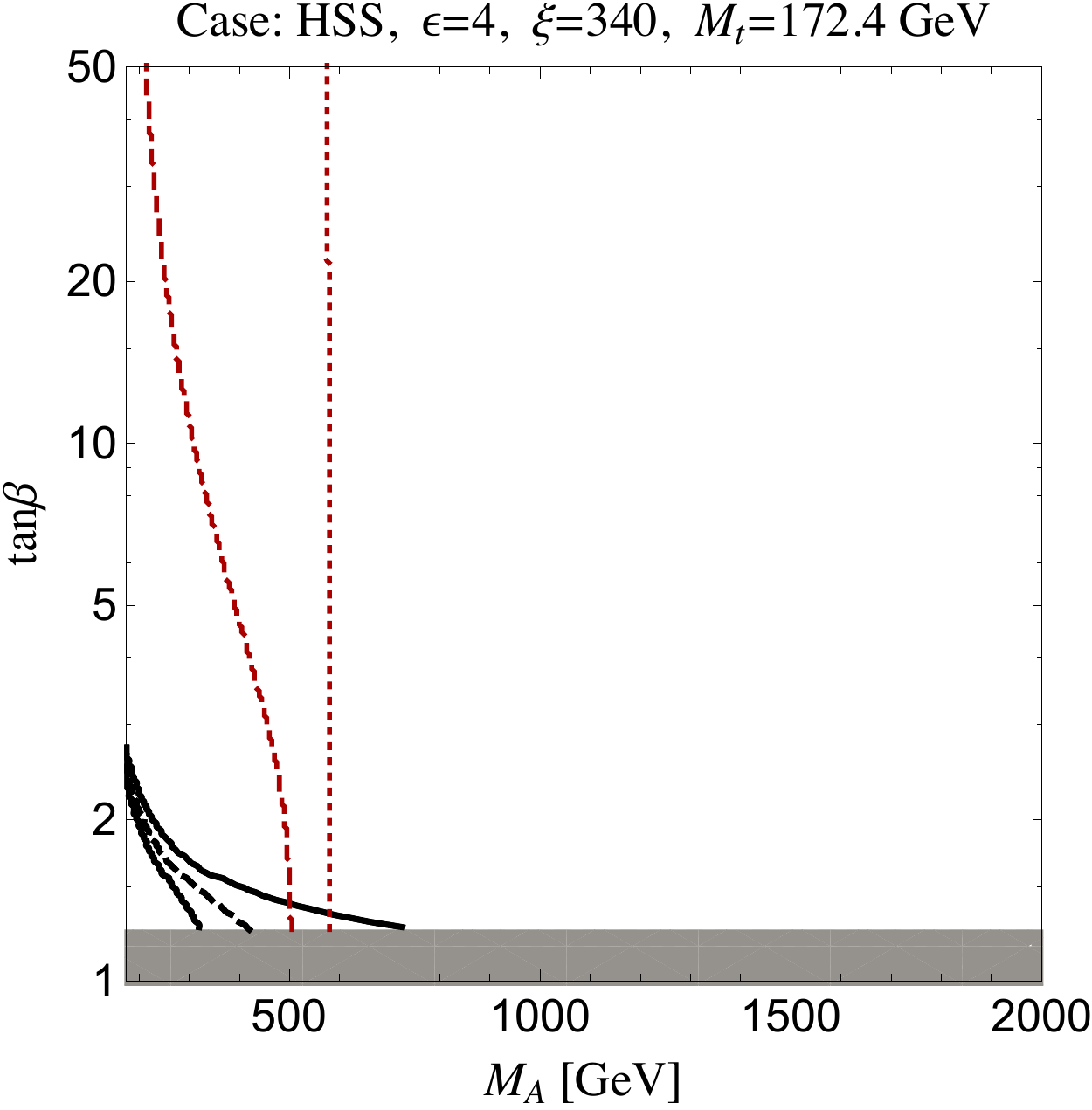}} \hspace*{0.1cm}
\subfigure{\includegraphics[width=0.43\textwidth]{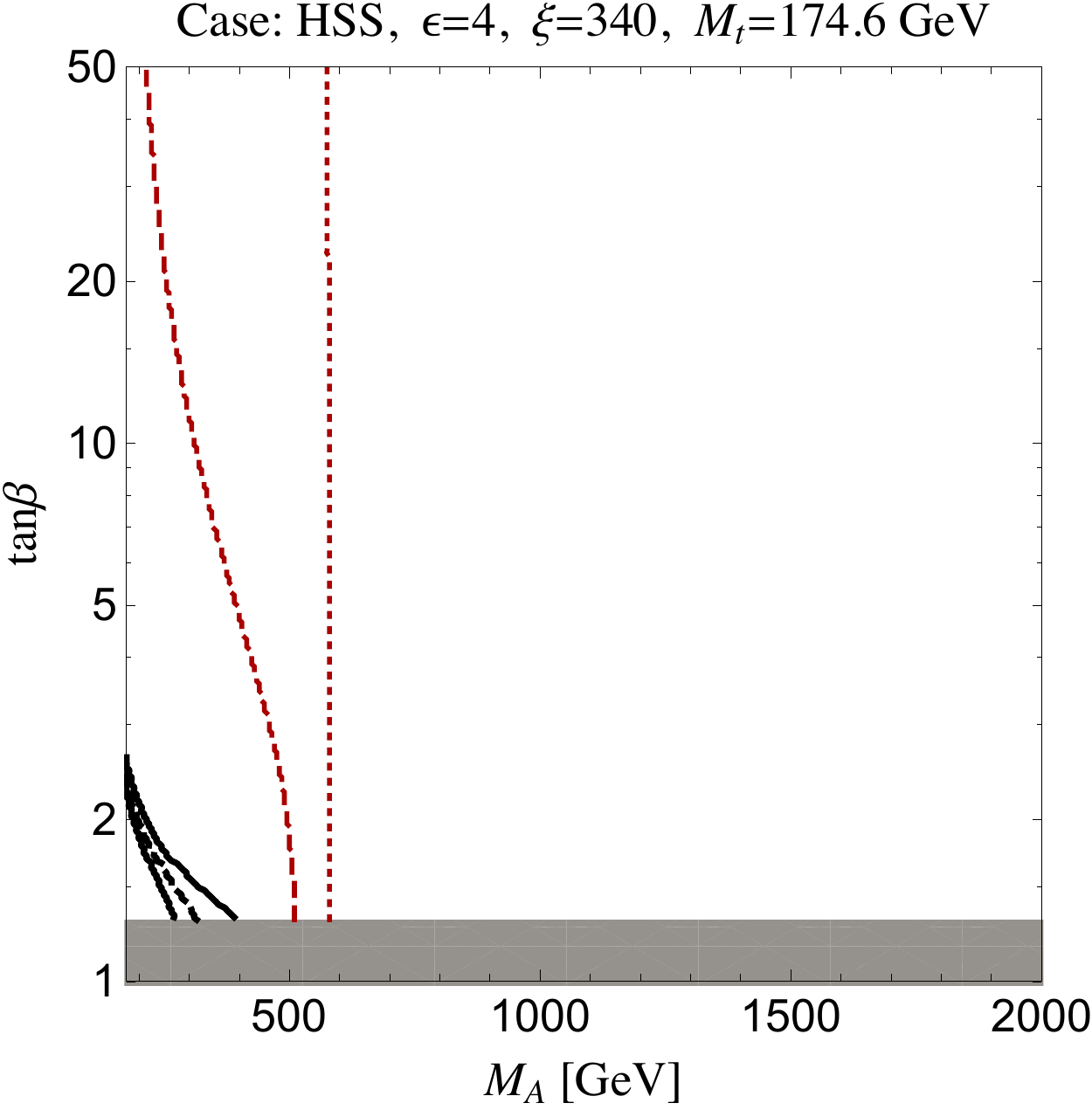}}
\caption{The details are same as given in the caption of Fig. \ref{fig1}. For all the cases, $N=2.07$.}
\label{AP_Fig1}
\end{figure}

\begin{figure}[!ht]
\centering
\subfigure{\includegraphics[width=0.43\textwidth]{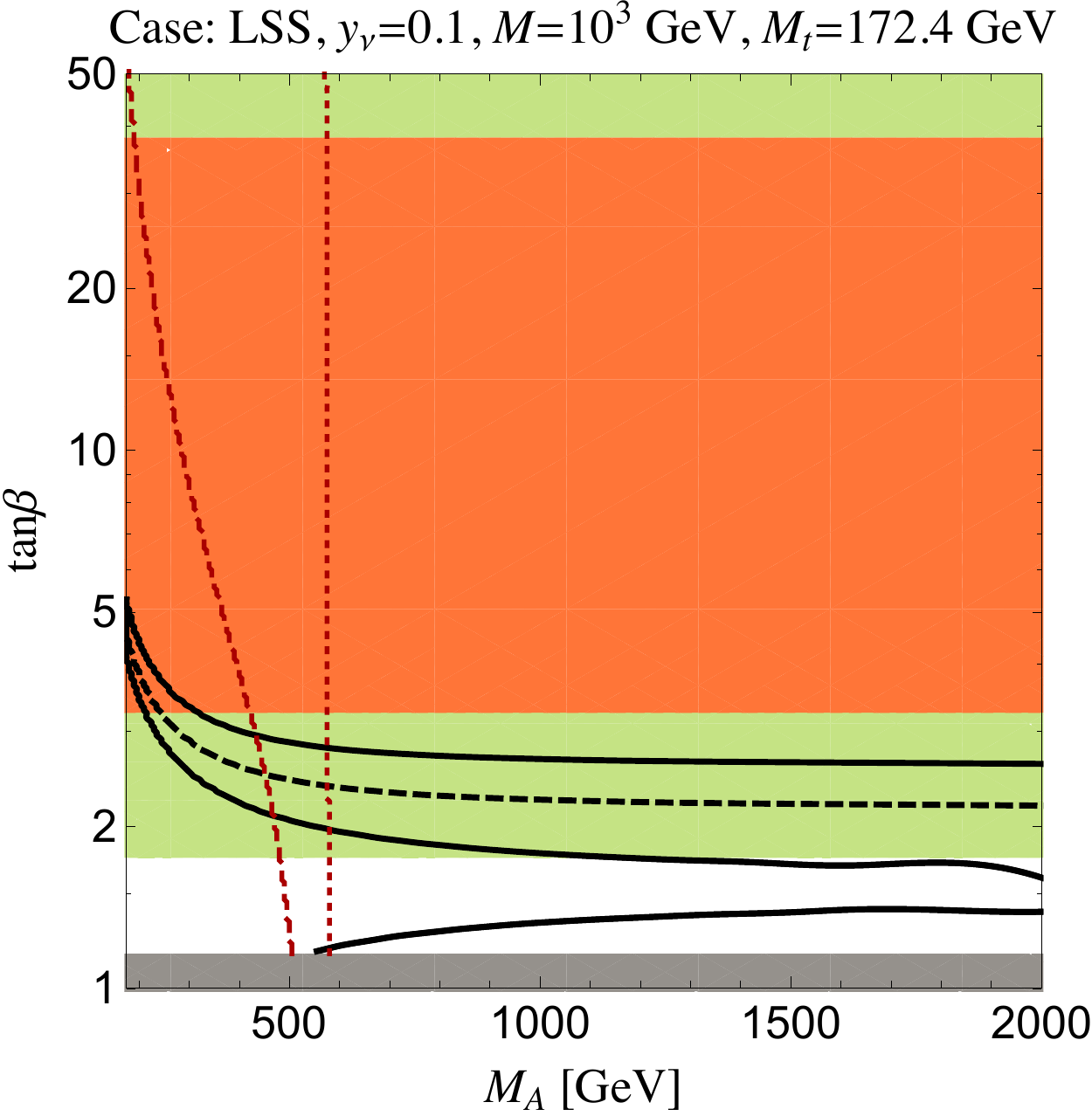}} \hspace*{0.1cm}
\subfigure{\includegraphics[width=0.43\textwidth]{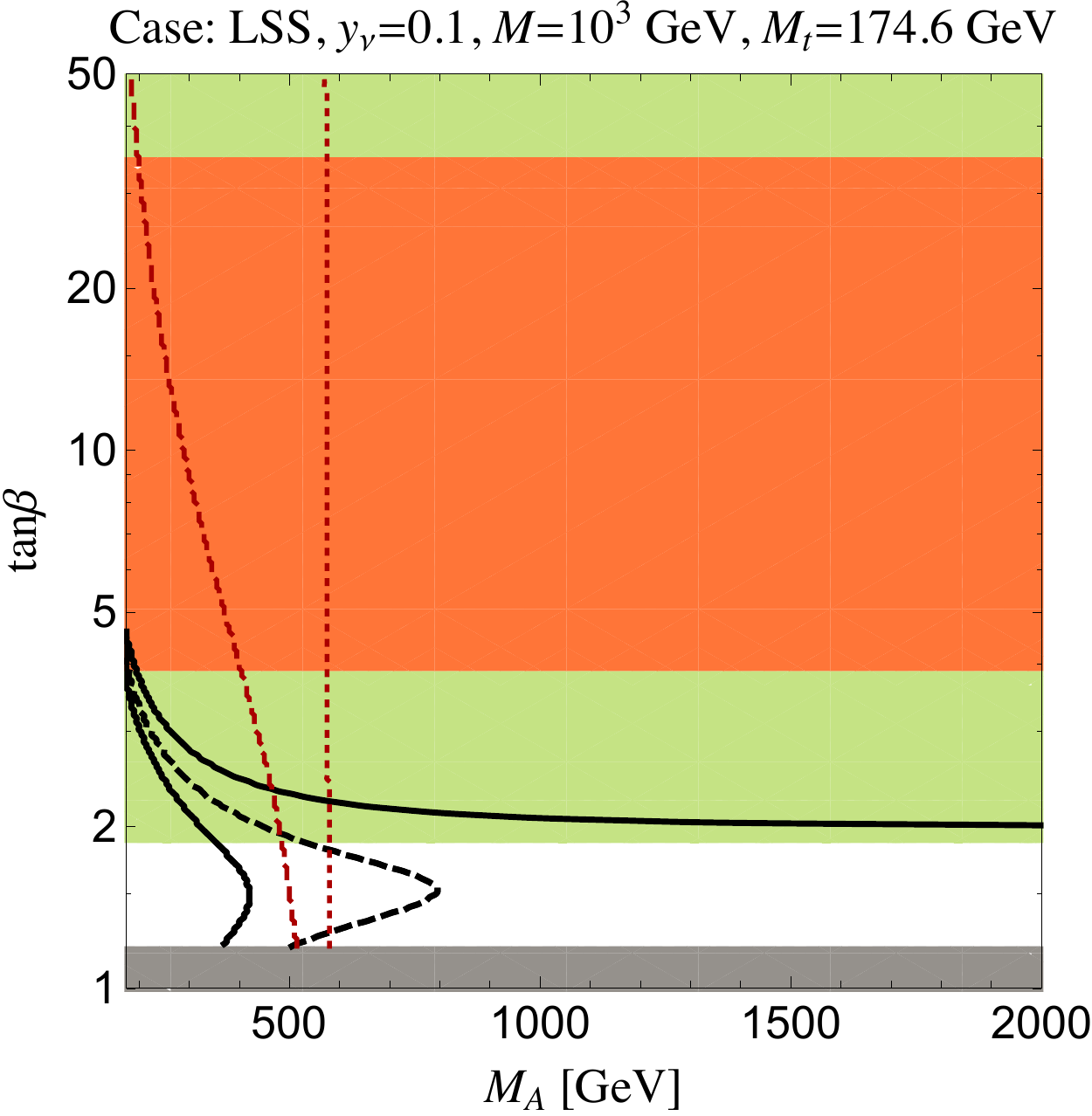}}
\subfigure{\includegraphics[width=0.43\textwidth]{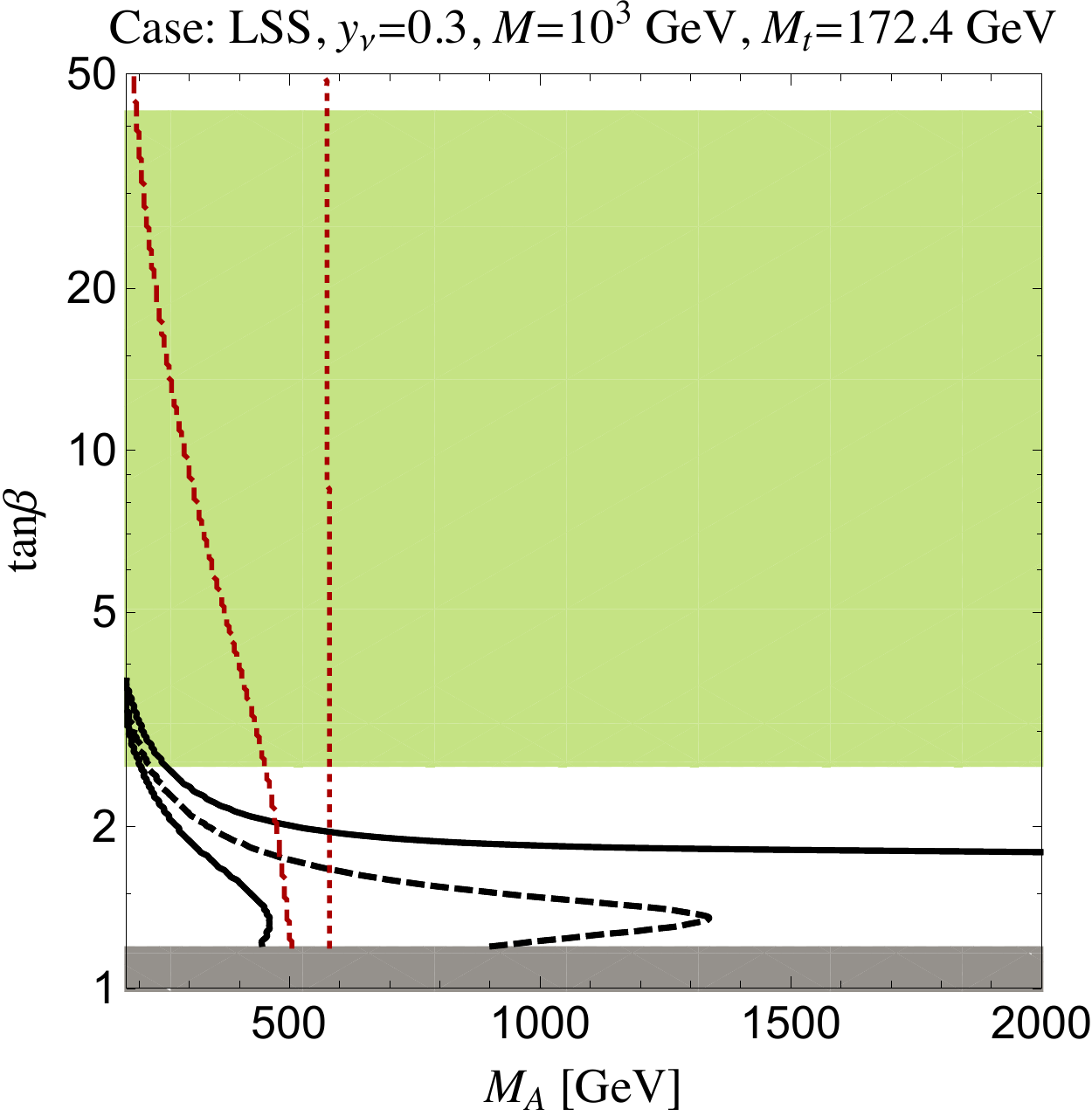}} \hspace*{0.1cm}
\subfigure{\includegraphics[width=0.43\textwidth]{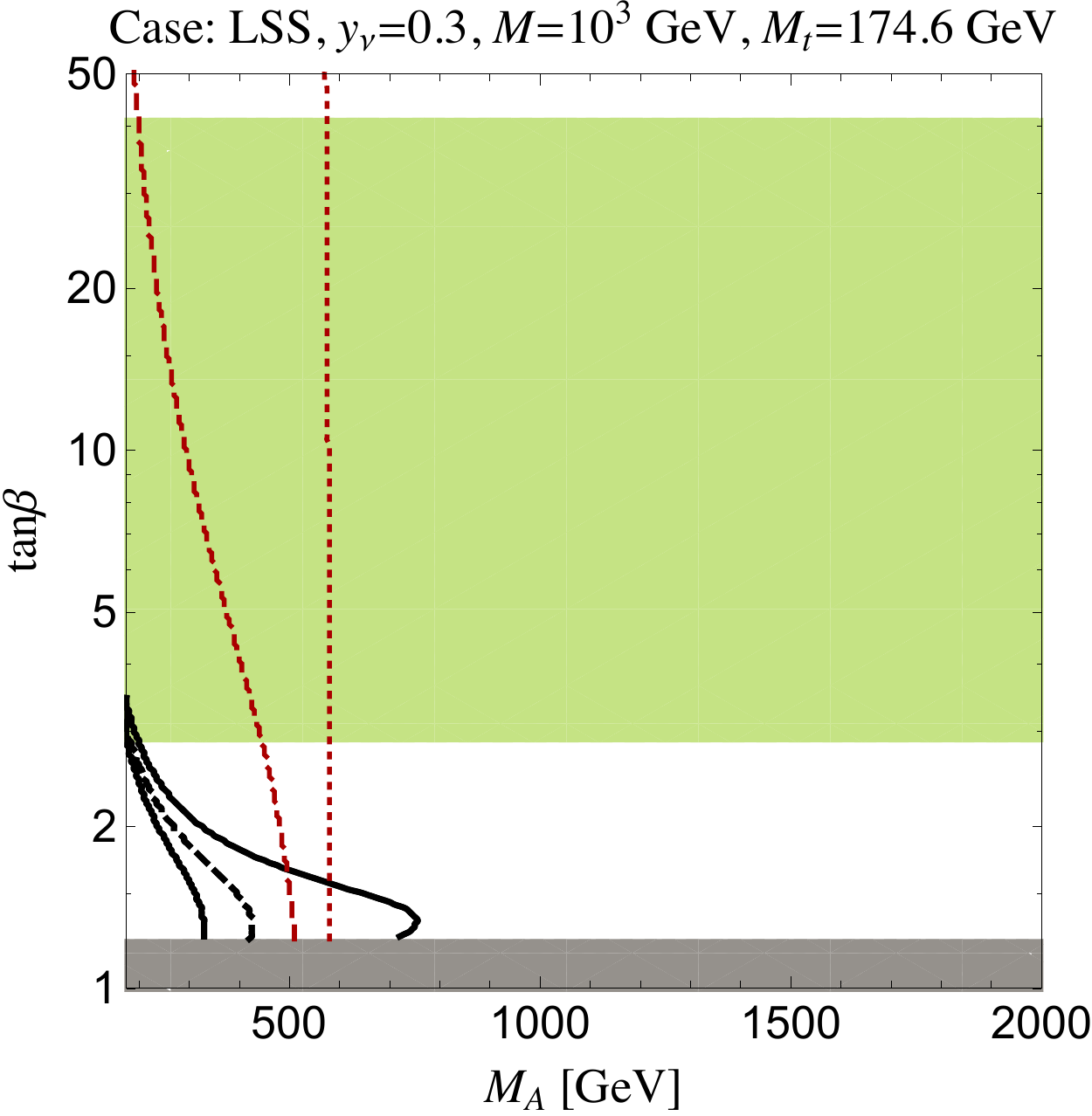}}
\subfigure{\includegraphics[width=0.43\textwidth]{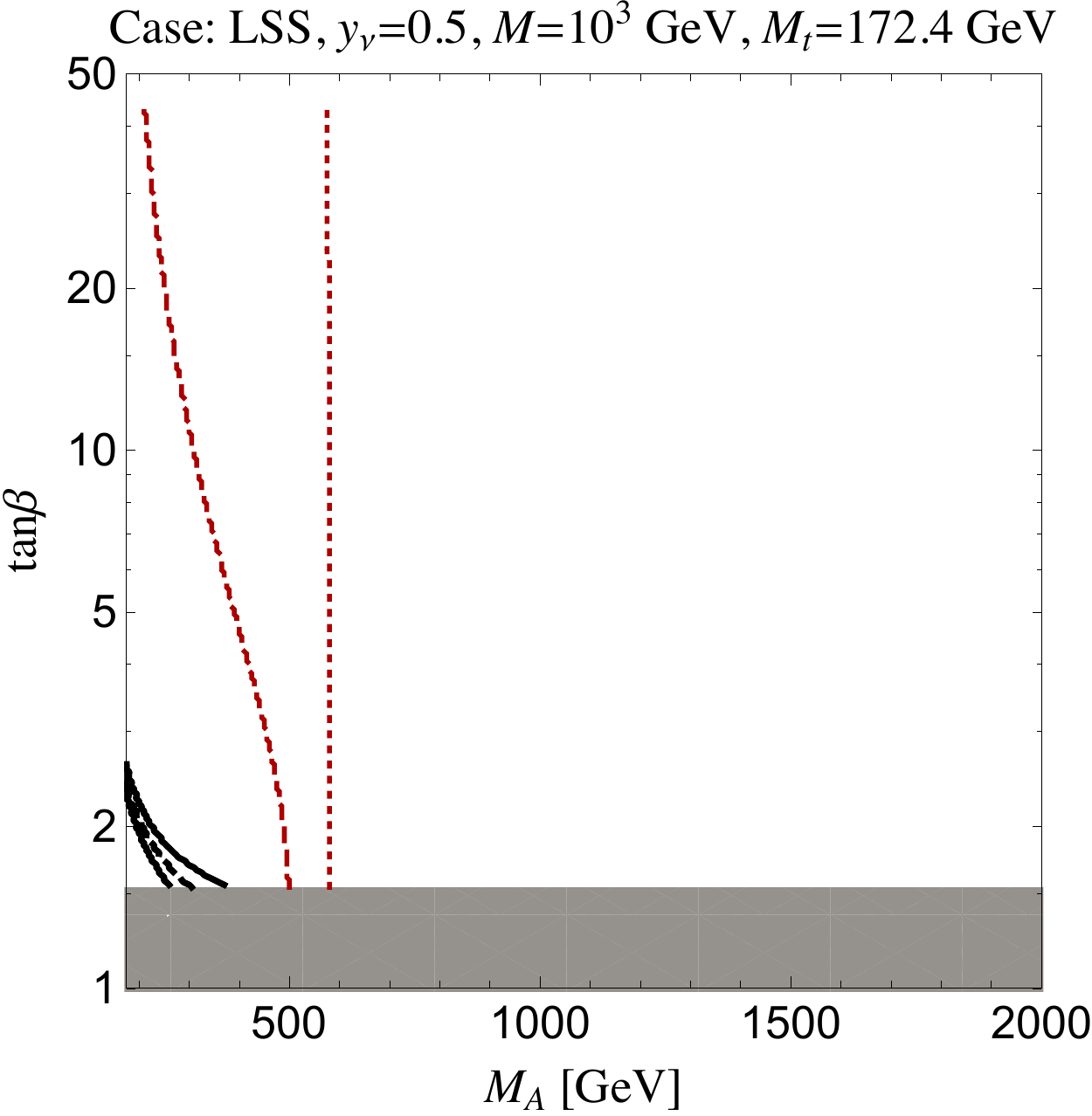}} \hspace*{0.1cm}
\subfigure{\includegraphics[width=0.43\textwidth]{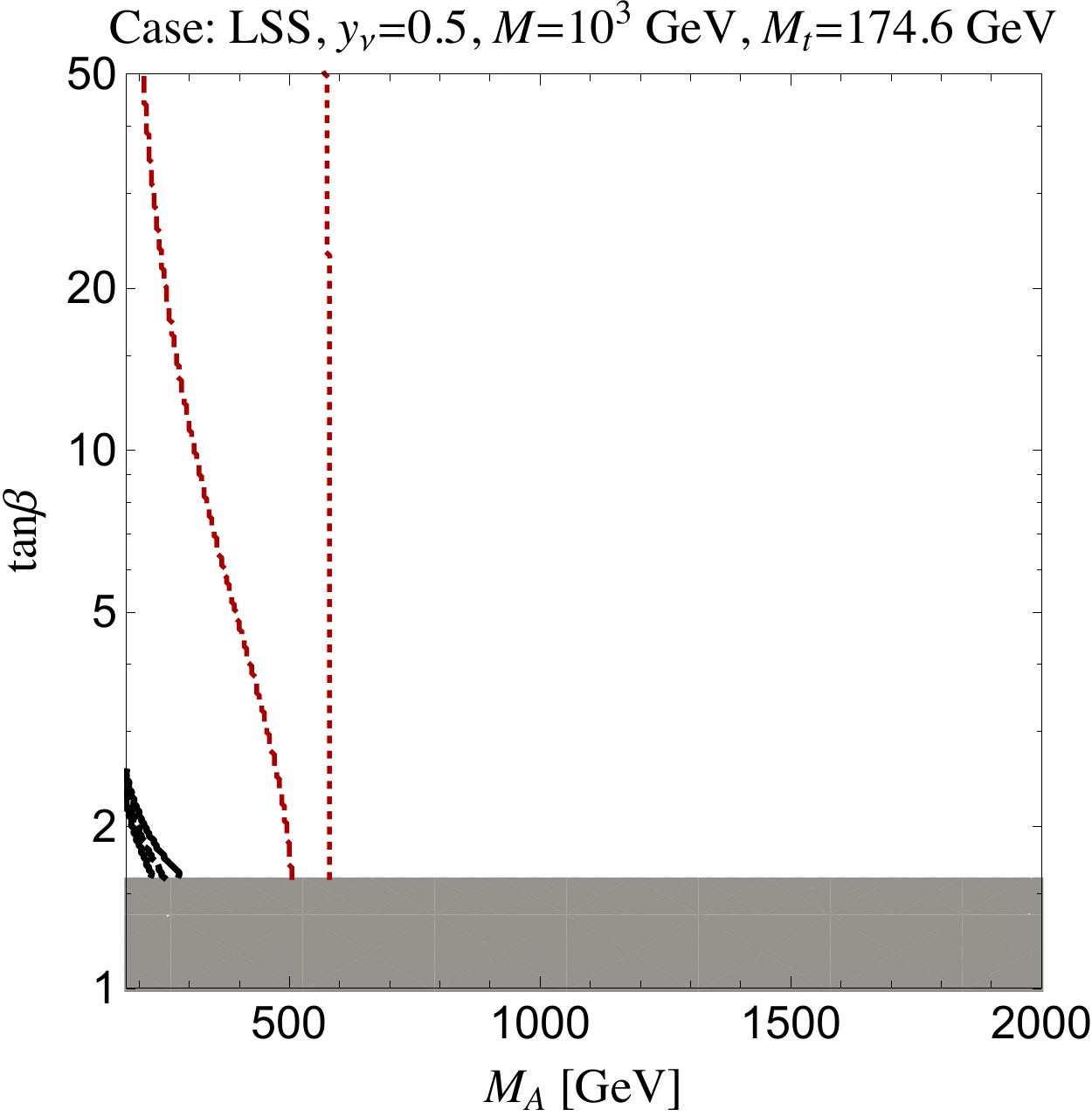}}
\caption{The details are same as given in the caption of Fig. \ref{fig1}.}
\label{AP_Fig2}
\end{figure}

In the following we show the effects of uncertainty in the measurements of $M_t$ on the results obtained by us in the case of high and low scale seesaw displayed in Figs. \ref{fig3} and \ref{fig5}
 respectively. We carry out the similar analysis for $M_t = 172.4$ GeV and $M_t = 174.6$ GeV which are values of top quark threshold at  $-1\sigma$ and $1\sigma$ respectively. The constraints on $\tan\beta-M_A$ plane obtained for various cases and for the above values of $M_t$ are displayed in Figs. \ref{AP_Fig1} and \ref{AP_Fig2}.

\bibliography{references}

\begin{thebibliography}{54}%
\makeatletter
\providecommand \@ifxundefined [1]{%
 \@ifx{#1\undefined}
}%
\providecommand \@ifnum [1]{%
 \ifnum #1\expandafter \@firstoftwo
 \else \expandafter \@secondoftwo
 \fi
}%
\providecommand \@ifx [1]{%
 \ifx #1\expandafter \@firstoftwo
 \else \expandafter \@secondoftwo
 \fi
}%
\providecommand \natexlab [1]{#1}%
\providecommand \enquote  [1]{``#1''}%
\providecommand \bibnamefont  [1]{#1}%
\providecommand \bibfnamefont [1]{#1}%
\providecommand \citenamefont [1]{#1}%
\providecommand \href@noop [0]{\@secondoftwo}%
\providecommand \href [0]{\begingroup \@sanitize@url \@href}%
\providecommand \@href[1]{\@@startlink{#1}\@@href}%
\providecommand \@@href[1]{\endgroup#1\@@endlink}%
\providecommand \@sanitize@url [0]{\catcode `\\12\catcode `\$12\catcode
  `\&12\catcode `\#12\catcode `\^12\catcode `\_12\catcode `\%12\relax}%
\providecommand \@@startlink[1]{}%
\providecommand \@@endlink[0]{}%
\providecommand \url  [0]{\begingroup\@sanitize@url \@url }%
\providecommand \@url [1]{\endgroup\@href {#1}{\urlprefix }}%
\providecommand \urlprefix  [0]{URL }%
\providecommand \Eprint [0]{\href }%
\providecommand \doibase [0]{http://dx.doi.org/}%
\providecommand \selectlanguage [0]{\@gobble}%
\providecommand \bibinfo  [0]{\@secondoftwo}%
\providecommand \bibfield  [0]{\@secondoftwo}%
\providecommand \translation [1]{[#1]}%
\providecommand \BibitemOpen [0]{}%
\providecommand \bibitemStop [0]{}%
\providecommand \bibitemNoStop [0]{.\EOS\space}%
\providecommand \EOS [0]{\spacefactor3000\relax}%
\providecommand \BibitemShut  [1]{\csname bibitem#1\endcsname}%
\let\auto@bib@innerbib\@empty
\bibitem [{\citenamefont {Giudice}\ and\ \citenamefont
  {Romanino}(2004)}]{Giudice:2004tc}%
  \BibitemOpen
  \bibfield  {author} {\bibinfo {author} {\bibfnamefont {G.~F.}\ \bibnamefont
  {Giudice}}\ and\ \bibinfo {author} {\bibfnamefont {A.}~\bibnamefont
  {Romanino}},\ }\bibfield  {title} {\enquote {\bibinfo {title} {{Split
  supersymmetry}},}\ }\href {\doibase 10.1016/j.nuclphysb.2004.11.048,
  10.1016/j.nuclphysb.2004.08.001} {\bibfield  {journal} {\bibinfo  {journal}
  {Nucl. Phys.}\ }\textbf {\bibinfo {volume} {B699}},\ \bibinfo {pages}
  {65--89} (\bibinfo {year} {2004})},\ \bibinfo {note} {[Erratum: Nucl.
  Phys.B706,487(2005)]},\ \Eprint {http://arxiv.org/abs/hep-ph/0406088}
  {arXiv:hep-ph/0406088 [hep-ph]} \BibitemShut {NoStop}%
\bibitem [{\citenamefont {Arkani-Hamed}\ and\ \citenamefont
  {Dimopoulos}(2005)}]{ArkaniHamed:2004fb}%
  \BibitemOpen
  \bibfield  {author} {\bibinfo {author} {\bibfnamefont {Nima}\ \bibnamefont
  {Arkani-Hamed}}\ and\ \bibinfo {author} {\bibfnamefont {Savas}\ \bibnamefont
  {Dimopoulos}},\ }\bibfield  {title} {\enquote {\bibinfo {title}
  {{Supersymmetric unification without low energy supersymmetry and signatures
  for fine-tuning at the LHC}},}\ }\href {\doibase
  10.1088/1126-6708/2005/06/073} {\bibfield  {journal} {\bibinfo  {journal}
  {JHEP}\ }\textbf {\bibinfo {volume} {06}},\ \bibinfo {pages} {073} (\bibinfo
  {year} {2005})},\ \Eprint {http://arxiv.org/abs/hep-th/0405159}
  {arXiv:hep-th/0405159 [hep-th]} \BibitemShut {NoStop}%
\bibitem [{\citenamefont {Green}\ \emph {et~al.}(1988)\citenamefont {Green},
  \citenamefont {Schwarz},\ and\ \citenamefont {Witten}}]{Green:1987sp}%
  \BibitemOpen
  \bibfield  {author} {\bibinfo {author} {\bibfnamefont {Michael~B.}\
  \bibnamefont {Green}}, \bibinfo {author} {\bibfnamefont {J.~H.}\ \bibnamefont
  {Schwarz}}, \ and\ \bibinfo {author} {\bibfnamefont {Edward}\ \bibnamefont
  {Witten}},\ }\href
  {http://www.cambridge.org/us/academic/subjects/physics/theoretical-physics-and-mathematical-physics/superstring-theory-volume-1}
  {\emph {\bibinfo {title} {{SUPERSTRING THEORY. VOL. 1: INTRODUCTION}}}},\
  Cambridge Monographs on Mathematical Physics\ (\bibinfo {year}
  {1988})\BibitemShut {NoStop}%
\bibitem [{\citenamefont {Kitano}\ and\ \citenamefont
  {Li}(2003)}]{Kitano:2003cn}%
  \BibitemOpen
  \bibfield  {author} {\bibinfo {author} {\bibfnamefont {Ryuichiro}\
  \bibnamefont {Kitano}}\ and\ \bibinfo {author} {\bibfnamefont {Tian-jun}\
  \bibnamefont {Li}},\ }\bibfield  {title} {\enquote {\bibinfo {title} {{Flavor
  hierarchy in SO(10) grand unified theories via five-dimensional wave function
  localization}},}\ }\href {\doibase 10.1103/PhysRevD.67.116004} {\bibfield
  {journal} {\bibinfo  {journal} {Phys. Rev.}\ }\textbf {\bibinfo {volume}
  {D67}},\ \bibinfo {pages} {116004} (\bibinfo {year} {2003})},\ \Eprint
  {http://arxiv.org/abs/hep-ph/0302073} {arXiv:hep-ph/0302073 [hep-ph]}
  \BibitemShut {NoStop}%
\bibitem [{\citenamefont {Buchmuller}\ \emph {et~al.}(2015)\citenamefont
  {Buchmuller}, \citenamefont {Dierigl}, \citenamefont {Ruehle},\ and\
  \citenamefont {Schweizer}}]{Buchmuller:2015jna}%
  \BibitemOpen
  \bibfield  {author} {\bibinfo {author} {\bibfnamefont {Wilfried}\
  \bibnamefont {Buchmuller}}, \bibinfo {author} {\bibfnamefont {Markus}\
  \bibnamefont {Dierigl}}, \bibinfo {author} {\bibfnamefont {Fabian}\
  \bibnamefont {Ruehle}}, \ and\ \bibinfo {author} {\bibfnamefont {Julian}\
  \bibnamefont {Schweizer}},\ }\bibfield  {title} {\enquote {\bibinfo {title}
  {{Split symmetries}},}\ }\href {\doibase 10.1016/j.physletb.2015.09.069}
  {\bibfield  {journal} {\bibinfo  {journal} {Phys. Lett.}\ }\textbf {\bibinfo
  {volume} {B750}},\ \bibinfo {pages} {615--619} (\bibinfo {year} {2015})},\
  \Eprint {http://arxiv.org/abs/1507.06819} {arXiv:1507.06819 [hep-th]}
  \BibitemShut {NoStop}%
\bibitem [{\citenamefont {Aaboud}\ \emph {et~al.}(2017)\citenamefont {Aaboud}
  \emph {et~al.}}]{Aaboud:2017bac}%
  \BibitemOpen
  \bibfield  {author} {\bibinfo {author} {\bibfnamefont {Morad}\ \bibnamefont
  {Aaboud}} \emph {et~al.} (\bibinfo {collaboration} {ATLAS}),\ }\bibfield
  {title} {\enquote {\bibinfo {title} {{Search for squarks and gluinos in
  events with an isolated lepton, jets, and missing transverse momentum at
  $\sqrt{s}=13$ TeV with the ATLAS detector}},}\ }\href {\doibase
  10.1103/PhysRevD.96.112010} {\bibfield  {journal} {\bibinfo  {journal} {Phys.
  Rev.}\ }\textbf {\bibinfo {volume} {D96}},\ \bibinfo {pages} {112010}
  (\bibinfo {year} {2017})},\ \Eprint {http://arxiv.org/abs/1708.08232}
  {arXiv:1708.08232 [hep-ex]} \BibitemShut {NoStop}%
\bibitem [{\citenamefont {Sirunyan}\ \emph {et~al.}(2017)\citenamefont
  {Sirunyan} \emph {et~al.}}]{Sirunyan:2017cwe}%
  \BibitemOpen
  \bibfield  {author} {\bibinfo {author} {\bibfnamefont {Albert~M}\
  \bibnamefont {Sirunyan}} \emph {et~al.} (\bibinfo {collaboration} {CMS}),\
  }\bibfield  {title} {\enquote {\bibinfo {title} {{Search for supersymmetry in
  multijet events with missing transverse momentum in proton-proton collisions
  at 13 TeV}},}\ }\href {\doibase 10.1103/PhysRevD.96.032003} {\bibfield
  {journal} {\bibinfo  {journal} {Phys. Rev.}\ }\textbf {\bibinfo {volume}
  {D96}},\ \bibinfo {pages} {032003} (\bibinfo {year} {2017})},\ \Eprint
  {http://arxiv.org/abs/1704.07781} {arXiv:1704.07781 [hep-ex]} \BibitemShut
  {NoStop}%
\bibitem [{\citenamefont {Giudice}\ and\ \citenamefont
  {Strumia}(2012)}]{Giudice:2011cg}%
  \BibitemOpen
  \bibfield  {author} {\bibinfo {author} {\bibfnamefont {Gian~F.}\ \bibnamefont
  {Giudice}}\ and\ \bibinfo {author} {\bibfnamefont {Alessandro}\ \bibnamefont
  {Strumia}},\ }\bibfield  {title} {\enquote {\bibinfo {title} {{Probing
  High-Scale and Split Supersymmetry with Higgs Mass Measurements}},}\ }\href
  {\doibase 10.1016/j.nuclphysb.2012.01.001} {\bibfield  {journal} {\bibinfo
  {journal} {Nucl. Phys.}\ }\textbf {\bibinfo {volume} {B858}},\ \bibinfo
  {pages} {63--83} (\bibinfo {year} {2012})},\ \Eprint
  {http://arxiv.org/abs/1108.6077} {arXiv:1108.6077 [hep-ph]} \BibitemShut
  {NoStop}%
\bibitem [{\citenamefont {Elias-Miro}\ \emph {et~al.}(2012)\citenamefont
  {Elias-Miro}, \citenamefont {Espinosa}, \citenamefont {Giudice},
  \citenamefont {Isidori}, \citenamefont {Riotto},\ and\ \citenamefont
  {Strumia}}]{EliasMiro:2011aa}%
  \BibitemOpen
  \bibfield  {author} {\bibinfo {author} {\bibfnamefont {Joan}\ \bibnamefont
  {Elias-Miro}}, \bibinfo {author} {\bibfnamefont {Jose~R.}\ \bibnamefont
  {Espinosa}}, \bibinfo {author} {\bibfnamefont {Gian~F.}\ \bibnamefont
  {Giudice}}, \bibinfo {author} {\bibfnamefont {Gino}\ \bibnamefont {Isidori}},
  \bibinfo {author} {\bibfnamefont {Antonio}\ \bibnamefont {Riotto}}, \ and\
  \bibinfo {author} {\bibfnamefont {Alessandro}\ \bibnamefont {Strumia}},\
  }\bibfield  {title} {\enquote {\bibinfo {title} {{Higgs mass implications on
  the stability of the electroweak vacuum}},}\ }\href {\doibase
  10.1016/j.physletb.2012.02.013} {\bibfield  {journal} {\bibinfo  {journal}
  {Phys. Lett.}\ }\textbf {\bibinfo {volume} {B709}},\ \bibinfo {pages}
  {222--228} (\bibinfo {year} {2012})},\ \Eprint
  {http://arxiv.org/abs/1112.3022} {arXiv:1112.3022 [hep-ph]} \BibitemShut
  {NoStop}%
\bibitem [{\citenamefont {Draper}\ \emph {et~al.}(2014)\citenamefont {Draper},
  \citenamefont {Lee},\ and\ \citenamefont {Wagner}}]{Draper:2013oza}%
  \BibitemOpen
  \bibfield  {author} {\bibinfo {author} {\bibfnamefont {Patrick}\ \bibnamefont
  {Draper}}, \bibinfo {author} {\bibfnamefont {Gabriel}\ \bibnamefont {Lee}}, \
  and\ \bibinfo {author} {\bibfnamefont {Carlos E.~M.}\ \bibnamefont
  {Wagner}},\ }\bibfield  {title} {\enquote {\bibinfo {title} {{Precise
  estimates of the Higgs mass in heavy supersymmetry}},}\ }\href {\doibase
  10.1103/PhysRevD.89.055023} {\bibfield  {journal} {\bibinfo  {journal} {Phys.
  Rev.}\ }\textbf {\bibinfo {volume} {D89}},\ \bibinfo {pages} {055023}
  (\bibinfo {year} {2014})},\ \Eprint {http://arxiv.org/abs/1312.5743}
  {arXiv:1312.5743 [hep-ph]} \BibitemShut {NoStop}%
\bibitem [{\citenamefont {Ellis}\ and\ \citenamefont
  {Wells}(2017)}]{Ellis:2017erg}%
  \BibitemOpen
  \bibfield  {author} {\bibinfo {author} {\bibfnamefont {Sebastian A.~R.}\
  \bibnamefont {Ellis}}\ and\ \bibinfo {author} {\bibfnamefont {James~D.}\
  \bibnamefont {Wells}},\ }\bibfield  {title} {\enquote {\bibinfo {title}
  {{High-scale supersymmetry, the Higgs boson mass, and gauge unification}},}\
  }\href {\doibase 10.1103/PhysRevD.96.055024} {\bibfield  {journal} {\bibinfo
  {journal} {Phys. Rev.}\ }\textbf {\bibinfo {volume} {D96}},\ \bibinfo {pages}
  {055024} (\bibinfo {year} {2017})},\ \Eprint
  {http://arxiv.org/abs/1706.00013} {arXiv:1706.00013 [hep-ph]} \BibitemShut
  {NoStop}%
\bibitem [{\citenamefont {Gorbahn}\ \emph {et~al.}(2011)\citenamefont
  {Gorbahn}, \citenamefont {Jager}, \citenamefont {Nierste},\ and\
  \citenamefont {Trine}}]{Gorbahn:2009pp}%
  \BibitemOpen
  \bibfield  {author} {\bibinfo {author} {\bibfnamefont {Martin}\ \bibnamefont
  {Gorbahn}}, \bibinfo {author} {\bibfnamefont {Sebastian}\ \bibnamefont
  {Jager}}, \bibinfo {author} {\bibfnamefont {Ulrich}\ \bibnamefont {Nierste}},
  \ and\ \bibinfo {author} {\bibfnamefont {Stephanie}\ \bibnamefont {Trine}},\
  }\bibfield  {title} {\enquote {\bibinfo {title} {{The supersymmetric Higgs
  sector and $B-\bar{B}$ mixing for large tan $\beta$}},}\ }\href {\doibase
  10.1103/PhysRevD.84.034030} {\bibfield  {journal} {\bibinfo  {journal} {Phys.
  Rev.}\ }\textbf {\bibinfo {volume} {D84}},\ \bibinfo {pages} {034030}
  (\bibinfo {year} {2011})},\ \Eprint {http://arxiv.org/abs/0901.2065}
  {arXiv:0901.2065 [hep-ph]} \BibitemShut {NoStop}%
\bibitem [{\citenamefont {Lee}\ and\ \citenamefont
  {Wagner}(2015)}]{Lee:2015uza}%
  \BibitemOpen
  \bibfield  {author} {\bibinfo {author} {\bibfnamefont {Gabriel}\ \bibnamefont
  {Lee}}\ and\ \bibinfo {author} {\bibfnamefont {Carlos E.~M.}\ \bibnamefont
  {Wagner}},\ }\bibfield  {title} {\enquote {\bibinfo {title} {{Higgs bosons in
  heavy supersymmetry with an intermediate m$_A$}},}\ }\href {\doibase
  10.1103/PhysRevD.92.075032} {\bibfield  {journal} {\bibinfo  {journal} {Phys.
  Rev.}\ }\textbf {\bibinfo {volume} {D92}},\ \bibinfo {pages} {075032}
  (\bibinfo {year} {2015})},\ \Eprint {http://arxiv.org/abs/1508.00576}
  {arXiv:1508.00576 [hep-ph]} \BibitemShut {NoStop}%
\bibitem [{\citenamefont {Bagnaschi}\ \emph {et~al.}(2014)\citenamefont
  {Bagnaschi}, \citenamefont {Giudice}, \citenamefont {Slavich},\ and\
  \citenamefont {Strumia}}]{Bagnaschi:2014rsa}%
  \BibitemOpen
  \bibfield  {author} {\bibinfo {author} {\bibfnamefont {Emanuele}\
  \bibnamefont {Bagnaschi}}, \bibinfo {author} {\bibfnamefont {Gian~F.}\
  \bibnamefont {Giudice}}, \bibinfo {author} {\bibfnamefont {Pietro}\
  \bibnamefont {Slavich}}, \ and\ \bibinfo {author} {\bibfnamefont
  {Alessandro}\ \bibnamefont {Strumia}},\ }\bibfield  {title} {\enquote
  {\bibinfo {title} {{Higgs Mass and Unnatural Supersymmetry}},}\ }\href
  {\doibase 10.1007/JHEP09(2014)092} {\bibfield  {journal} {\bibinfo  {journal}
  {JHEP}\ }\textbf {\bibinfo {volume} {09}},\ \bibinfo {pages} {092} (\bibinfo
  {year} {2014})},\ \Eprint {http://arxiv.org/abs/1407.4081} {arXiv:1407.4081
  [hep-ph]} \BibitemShut {NoStop}%
\bibitem [{\citenamefont {Branco}\ \emph {et~al.}(2012)\citenamefont {Branco},
  \citenamefont {Ferreira}, \citenamefont {Lavoura}, \citenamefont {Rebelo},
  \citenamefont {Sher},\ and\ \citenamefont {Silva}}]{Branco:2011iw}%
  \BibitemOpen
  \bibfield  {author} {\bibinfo {author} {\bibfnamefont {G.~C.}\ \bibnamefont
  {Branco}}, \bibinfo {author} {\bibfnamefont {P.~M.}\ \bibnamefont
  {Ferreira}}, \bibinfo {author} {\bibfnamefont {L.}~\bibnamefont {Lavoura}},
  \bibinfo {author} {\bibfnamefont {M.~N.}\ \bibnamefont {Rebelo}}, \bibinfo
  {author} {\bibfnamefont {Marc}\ \bibnamefont {Sher}}, \ and\ \bibinfo
  {author} {\bibfnamefont {Joao~P.}\ \bibnamefont {Silva}},\ }\bibfield
  {title} {\enquote {\bibinfo {title} {{Theory and phenomenology of
  two-Higgs-doublet models}},}\ }\href {\doibase 10.1016/j.physrep.2012.02.002}
  {\bibfield  {journal} {\bibinfo  {journal} {Phys. Rept.}\ }\textbf {\bibinfo
  {volume} {516}},\ \bibinfo {pages} {1--102} (\bibinfo {year} {2012})},\
  \Eprint {http://arxiv.org/abs/1106.0034} {arXiv:1106.0034 [hep-ph]}
  \BibitemShut {NoStop}%
\bibitem [{\citenamefont {Bagnaschi}\ \emph {et~al.}(2016)\citenamefont
  {Bagnaschi}, \citenamefont {Brummer}, \citenamefont {Buchmuller},
  \citenamefont {Voigt},\ and\ \citenamefont {Weiglein}}]{Bagnaschi:2015pwa}%
  \BibitemOpen
  \bibfield  {author} {\bibinfo {author} {\bibfnamefont {Emanuele}\
  \bibnamefont {Bagnaschi}}, \bibinfo {author} {\bibfnamefont {Felix}\
  \bibnamefont {Brummer}}, \bibinfo {author} {\bibfnamefont {Wilfried}\
  \bibnamefont {Buchmuller}}, \bibinfo {author} {\bibfnamefont {Alexander}\
  \bibnamefont {Voigt}}, \ and\ \bibinfo {author} {\bibfnamefont {Georg}\
  \bibnamefont {Weiglein}},\ }\bibfield  {title} {\enquote {\bibinfo {title}
  {{Vacuum stability and supersymmetry at high scales with two Higgs
  doublets}},}\ }\href {\doibase 10.1007/JHEP03(2016)158} {\bibfield  {journal}
  {\bibinfo  {journal} {JHEP}\ }\textbf {\bibinfo {volume} {03}},\ \bibinfo
  {pages} {158} (\bibinfo {year} {2016})},\ \Eprint
  {http://arxiv.org/abs/1512.07761} {arXiv:1512.07761 [hep-ph]} \BibitemShut
  {NoStop}%
\bibitem [{\citenamefont {Bhattacharyya}\ \emph {et~al.}(2017)\citenamefont
  {Bhattacharyya}, \citenamefont {Das}, \citenamefont {Pérez}, \citenamefont
  {Saha}, \citenamefont {Santamaria},\ and\ \citenamefont
  {Vives}}]{Bhattacharyya:2017ksj}%
  \BibitemOpen
  \bibfield  {author} {\bibinfo {author} {\bibfnamefont {Gautam}\ \bibnamefont
  {Bhattacharyya}}, \bibinfo {author} {\bibfnamefont {Dipankar}\ \bibnamefont
  {Das}}, \bibinfo {author} {\bibfnamefont {M.~Jay}\ \bibnamefont {Pérez}},
  \bibinfo {author} {\bibfnamefont {Ipsita}\ \bibnamefont {Saha}}, \bibinfo
  {author} {\bibfnamefont {Arcadi}\ \bibnamefont {Santamaria}}, \ and\ \bibinfo
  {author} {\bibfnamefont {Oscar}\ \bibnamefont {Vives}},\ }\bibfield  {title}
  {\enquote {\bibinfo {title} {{Can measurements of 2HDM parameters provide
  hints for high scale supersymmetry?}}}\ }\href@noop {} {\  (\bibinfo {year}
  {2017})},\ \Eprint {http://arxiv.org/abs/1712.00791} {arXiv:1712.00791
  [hep-ph]} \BibitemShut {NoStop}%
\bibitem [{\citenamefont {Minkowski}(1977)}]{Minkowski:1977sc}%
  \BibitemOpen
  \bibfield  {author} {\bibinfo {author} {\bibfnamefont {Peter}\ \bibnamefont
  {Minkowski}},\ }\bibfield  {title} {\enquote {\bibinfo {title} {{$\mu \to e
  \gamma$ at a Rate of One Out of 1-Billion Muon Decays?}}}\ }\href {\doibase
  10.1016/0370-2693(77)90435-X} {\bibfield  {journal} {\bibinfo  {journal}
  {Phys.Lett.}\ }\textbf {\bibinfo {volume} {B67}},\ \bibinfo {pages} {421}
  (\bibinfo {year} {1977})}\BibitemShut {NoStop}%
\bibitem [{\citenamefont {Yanagida}(1979)}]{Yanagida:1979as}%
  \BibitemOpen
  \bibfield  {author} {\bibinfo {author} {\bibfnamefont {Tsutomu}\ \bibnamefont
  {Yanagida}},\ }\bibfield  {title} {\enquote {\bibinfo {title} {{HORIZONTAL
  SYMMETRY AND MASSES OF NEUTRINOS}},}\ }\href@noop {} {\bibfield  {journal}
  {\bibinfo  {journal} {Conf.Proc.}\ }\textbf {\bibinfo {volume} {C7902131}},\
  \bibinfo {pages} {95--99} (\bibinfo {year} {1979})}\BibitemShut {NoStop}%
\bibitem [{\citenamefont {Gell-Mann}\ \emph {et~al.}(1979)\citenamefont
  {Gell-Mann}, \citenamefont {Ramond},\ and\ \citenamefont
  {Slansky}}]{GellMann:1980vs}%
  \BibitemOpen
  \bibfield  {author} {\bibinfo {author} {\bibfnamefont {Murray}\ \bibnamefont
  {Gell-Mann}}, \bibinfo {author} {\bibfnamefont {Pierre}\ \bibnamefont
  {Ramond}}, \ and\ \bibinfo {author} {\bibfnamefont {Richard}\ \bibnamefont
  {Slansky}},\ }\bibfield  {title} {\enquote {\bibinfo {title} {{Complex
  Spinors and Unified Theories}},}\ }\href@noop {} {\bibfield  {journal}
  {\bibinfo  {journal} {Conf.Proc.}\ }\textbf {\bibinfo {volume} {C790927}},\
  \bibinfo {pages} {315--321} (\bibinfo {year} {1979})},\ \Eprint
  {http://arxiv.org/abs/1306.4669} {arXiv:1306.4669 [hep-th]} \BibitemShut
  {NoStop}%
\bibitem [{\citenamefont {Glashow}(1980)}]{Glashow:1979nm}%
  \BibitemOpen
  \bibfield  {author} {\bibinfo {author} {\bibfnamefont {S.L.}\ \bibnamefont
  {Glashow}},\ }\bibfield  {title} {\enquote {\bibinfo {title} {{The Future of
  Elementary Particle Physics}},}\ }\href@noop {} {\bibfield  {journal}
  {\bibinfo  {journal} {NATO Adv.Study Inst.Ser.B Phys.}\ }\textbf {\bibinfo
  {volume} {59}},\ \bibinfo {pages} {687} (\bibinfo {year} {1980})}\BibitemShut
  {NoStop}%
\bibitem [{\citenamefont {Lazarides}\ \emph {et~al.}(1981)\citenamefont
  {Lazarides}, \citenamefont {Shafi},\ and\ \citenamefont
  {Wetterich}}]{Lazarides:1980nt}%
  \BibitemOpen
  \bibfield  {author} {\bibinfo {author} {\bibfnamefont {George}\ \bibnamefont
  {Lazarides}}, \bibinfo {author} {\bibfnamefont {Q.}~\bibnamefont {Shafi}}, \
  and\ \bibinfo {author} {\bibfnamefont {C.}~\bibnamefont {Wetterich}},\
  }\bibfield  {title} {\enquote {\bibinfo {title} {{Proton Lifetime and Fermion
  Masses in an SO(10) Model}},}\ }\href {\doibase 10.1016/0550-3213(81)90354-0}
  {\bibfield  {journal} {\bibinfo  {journal} {Nucl.Phys.}\ }\textbf {\bibinfo
  {volume} {B181}},\ \bibinfo {pages} {287--300} (\bibinfo {year}
  {1981})}\BibitemShut {NoStop}%
\bibitem [{\citenamefont {Schechter}\ and\ \citenamefont
  {Valle}(1980)}]{Schechter:1980gr}%
  \BibitemOpen
  \bibfield  {author} {\bibinfo {author} {\bibfnamefont {J.}~\bibnamefont
  {Schechter}}\ and\ \bibinfo {author} {\bibfnamefont {J.W.F.}\ \bibnamefont
  {Valle}},\ }\bibfield  {title} {\enquote {\bibinfo {title} {{Neutrino Masses
  in SU(2) x U(1) Theories}},}\ }\href {\doibase 10.1103/PhysRevD.22.2227}
  {\bibfield  {journal} {\bibinfo  {journal} {Phys.Rev.}\ }\textbf {\bibinfo
  {volume} {D22}},\ \bibinfo {pages} {2227} (\bibinfo {year}
  {1980})}\BibitemShut {NoStop}%
\bibitem [{\citenamefont {Mohapatra}\ and\ \citenamefont
  {Senjanovic}(1981)}]{Mohapatra:1980yp}%
  \BibitemOpen
  \bibfield  {author} {\bibinfo {author} {\bibfnamefont {Rabindra~N.}\
  \bibnamefont {Mohapatra}}\ and\ \bibinfo {author} {\bibfnamefont {Goran}\
  \bibnamefont {Senjanovic}},\ }\bibfield  {title} {\enquote {\bibinfo {title}
  {{Neutrino Masses and Mixings in Gauge Models with Spontaneous Parity
  Violation}},}\ }\href {\doibase 10.1103/PhysRevD.23.165} {\bibfield
  {journal} {\bibinfo  {journal} {Phys.Rev.}\ }\textbf {\bibinfo {volume}
  {D23}},\ \bibinfo {pages} {165} (\bibinfo {year} {1981})}\BibitemShut
  {NoStop}%
\bibitem [{\citenamefont {Buchmuller}\ and\ \citenamefont
  {Schweizer}(2017)}]{Buchmuller:2017vho}%
  \BibitemOpen
  \bibfield  {author} {\bibinfo {author} {\bibfnamefont {Wilfried}\
  \bibnamefont {Buchmuller}}\ and\ \bibinfo {author} {\bibfnamefont {Julian}\
  \bibnamefont {Schweizer}},\ }\bibfield  {title} {\enquote {\bibinfo {title}
  {{Flavor mixings in flux compactifications}},}\ }\href {\doibase
  10.1103/PhysRevD.95.075024} {\bibfield  {journal} {\bibinfo  {journal} {Phys.
  Rev.}\ }\textbf {\bibinfo {volume} {D95}},\ \bibinfo {pages} {075024}
  (\bibinfo {year} {2017})},\ \Eprint {http://arxiv.org/abs/1701.06935}
  {arXiv:1701.06935 [hep-ph]} \BibitemShut {NoStop}%
\bibitem [{\citenamefont {Buchmuller}\ and\ \citenamefont
  {Patel}(2018)}]{Buchmuller:2017vut}%
  \BibitemOpen
  \bibfield  {author} {\bibinfo {author} {\bibfnamefont {Wilfried}\
  \bibnamefont {Buchmuller}}\ and\ \bibinfo {author} {\bibfnamefont {Ketan~M.}\
  \bibnamefont {Patel}},\ }\bibfield  {title} {\enquote {\bibinfo {title}
  {{Flavour physics without flavour symmetries}},}\ }\href {\doibase
  10.1103/PhysRevD.97.075019} {\bibfield  {journal} {\bibinfo  {journal} {Phys.
  Rev.}\ }\textbf {\bibinfo {volume} {D97}},\ \bibinfo {pages} {075019}
  (\bibinfo {year} {2018})},\ \Eprint {http://arxiv.org/abs/1712.06862}
  {arXiv:1712.06862 [hep-ph]} \BibitemShut {NoStop}%
\bibitem [{\citenamefont {Haber}\ and\ \citenamefont
  {Hempfling}(1993)}]{Haber:1993an}%
  \BibitemOpen
  \bibfield  {author} {\bibinfo {author} {\bibfnamefont {Howard~E.}\
  \bibnamefont {Haber}}\ and\ \bibinfo {author} {\bibfnamefont {Ralf}\
  \bibnamefont {Hempfling}},\ }\bibfield  {title} {\enquote {\bibinfo {title}
  {{The Renormalization group improved Higgs sector of the minimal
  supersymmetric model}},}\ }\href {\doibase 10.1103/PhysRevD.48.4280}
  {\bibfield  {journal} {\bibinfo  {journal} {Phys. Rev.}\ }\textbf {\bibinfo
  {volume} {D48}},\ \bibinfo {pages} {4280--4309} (\bibinfo {year} {1993})},\
  \Eprint {http://arxiv.org/abs/hep-ph/9307201} {arXiv:hep-ph/9307201 [hep-ph]}
  \BibitemShut {NoStop}%
\bibitem [{\citenamefont {Casas}\ and\ \citenamefont
  {Ibarra}(2001)}]{Casas:2001sr}%
  \BibitemOpen
  \bibfield  {author} {\bibinfo {author} {\bibfnamefont {J.~A.}\ \bibnamefont
  {Casas}}\ and\ \bibinfo {author} {\bibfnamefont {A.}~\bibnamefont {Ibarra}},\
  }\bibfield  {title} {\enquote {\bibinfo {title} {{Oscillating neutrinos and
  muon ---> e, gamma}},}\ }\href {\doibase 10.1016/S0550-3213(01)00475-8}
  {\bibfield  {journal} {\bibinfo  {journal} {Nucl. Phys.}\ }\textbf {\bibinfo
  {volume} {B618}},\ \bibinfo {pages} {171--204} (\bibinfo {year} {2001})},\
  \Eprint {http://arxiv.org/abs/hep-ph/0103065} {arXiv:hep-ph/0103065 [hep-ph]}
  \BibitemShut {NoStop}%
\bibitem [{\citenamefont {Fritzsch}\ and\ \citenamefont
  {Minkowski}(1975)}]{Fritzsch:1974nn}%
  \BibitemOpen
  \bibfield  {author} {\bibinfo {author} {\bibfnamefont {Harald}\ \bibnamefont
  {Fritzsch}}\ and\ \bibinfo {author} {\bibfnamefont {Peter}\ \bibnamefont
  {Minkowski}},\ }\bibfield  {title} {\enquote {\bibinfo {title} {{Unified
  Interactions of Leptons and Hadrons}},}\ }\href {\doibase
  10.1016/0003-4916(75)90211-0} {\bibfield  {journal} {\bibinfo  {journal}
  {Annals Phys.}\ }\textbf {\bibinfo {volume} {93}},\ \bibinfo {pages}
  {193--266} (\bibinfo {year} {1975})}\BibitemShut {NoStop}%
\bibitem [{\citenamefont {Babu}\ and\ \citenamefont
  {Mohapatra}(1993)}]{Babu:1992ia}%
  \BibitemOpen
  \bibfield  {author} {\bibinfo {author} {\bibfnamefont {K.~S.}\ \bibnamefont
  {Babu}}\ and\ \bibinfo {author} {\bibfnamefont {R.~N.}\ \bibnamefont
  {Mohapatra}},\ }\bibfield  {title} {\enquote {\bibinfo {title} {{Predictive
  neutrino spectrum in minimal SO(10) grand unification}},}\ }\href {\doibase
  10.1103/PhysRevLett.70.2845} {\bibfield  {journal} {\bibinfo  {journal}
  {Phys. Rev. Lett.}\ }\textbf {\bibinfo {volume} {70}},\ \bibinfo {pages}
  {2845--2848} (\bibinfo {year} {1993})},\ \Eprint
  {http://arxiv.org/abs/hep-ph/9209215} {arXiv:hep-ph/9209215 [hep-ph]}
  \BibitemShut {NoStop}%
\bibitem [{\citenamefont {Joshipura}\ and\ \citenamefont
  {Patel}(2011)}]{Joshipura:2011nn}%
  \BibitemOpen
  \bibfield  {author} {\bibinfo {author} {\bibfnamefont {Anjan~S.}\
  \bibnamefont {Joshipura}}\ and\ \bibinfo {author} {\bibfnamefont {Ketan~M.}\
  \bibnamefont {Patel}},\ }\bibfield  {title} {\enquote {\bibinfo {title}
  {{Fermion Masses in SO(10) Models}},}\ }\href {\doibase
  10.1103/PhysRevD.83.095002} {\bibfield  {journal} {\bibinfo  {journal} {Phys.
  Rev.}\ }\textbf {\bibinfo {volume} {D83}},\ \bibinfo {pages} {095002}
  (\bibinfo {year} {2011})},\ \Eprint {http://arxiv.org/abs/1102.5148}
  {arXiv:1102.5148 [hep-ph]} \BibitemShut {NoStop}%
\bibitem [{\citenamefont {Feruglio}\ \emph {et~al.}(2014)\citenamefont
  {Feruglio}, \citenamefont {Patel},\ and\ \citenamefont
  {Vicino}}]{Feruglio:2014jla}%
  \BibitemOpen
  \bibfield  {author} {\bibinfo {author} {\bibfnamefont {Ferruccio}\
  \bibnamefont {Feruglio}}, \bibinfo {author} {\bibfnamefont {Ketan~M.}\
  \bibnamefont {Patel}}, \ and\ \bibinfo {author} {\bibfnamefont {Denise}\
  \bibnamefont {Vicino}},\ }\bibfield  {title} {\enquote {\bibinfo {title}
  {{Order and Anarchy hand in hand in 5D SO(10)}},}\ }\href {\doibase
  10.1007/JHEP09(2014)095} {\bibfield  {journal} {\bibinfo  {journal} {JHEP}\
  }\textbf {\bibinfo {volume} {09}},\ \bibinfo {pages} {095} (\bibinfo {year}
  {2014})},\ \Eprint {http://arxiv.org/abs/1407.2913} {arXiv:1407.2913
  [hep-ph]} \BibitemShut {NoStop}%
\bibitem [{\citenamefont {Feruglio}\ \emph {et~al.}(2015)\citenamefont
  {Feruglio}, \citenamefont {Patel},\ and\ \citenamefont
  {Vicino}}]{Feruglio:2015iua}%
  \BibitemOpen
  \bibfield  {author} {\bibinfo {author} {\bibfnamefont {Ferruccio}\
  \bibnamefont {Feruglio}}, \bibinfo {author} {\bibfnamefont {Ketan~M.}\
  \bibnamefont {Patel}}, \ and\ \bibinfo {author} {\bibfnamefont {Denise}\
  \bibnamefont {Vicino}},\ }\bibfield  {title} {\enquote {\bibinfo {title} {{A
  realistic pattern of fermion masses from a five-dimensional SO(10) model}},}\
  }\href {\doibase 10.1007/JHEP09(2015)040} {\bibfield  {journal} {\bibinfo
  {journal} {JHEP}\ }\textbf {\bibinfo {volume} {09}},\ \bibinfo {pages} {040}
  (\bibinfo {year} {2015})},\ \Eprint {http://arxiv.org/abs/1507.00669}
  {arXiv:1507.00669 [hep-ph]} \BibitemShut {NoStop}%
\bibitem [{\citenamefont {Kersten}\ and\ \citenamefont
  {Smirnov}(2007)}]{Kersten:2007vk}%
  \BibitemOpen
  \bibfield  {author} {\bibinfo {author} {\bibfnamefont {Jorn}\ \bibnamefont
  {Kersten}}\ and\ \bibinfo {author} {\bibfnamefont {Alexei~{\relax Yu}.}\
  \bibnamefont {Smirnov}},\ }\bibfield  {title} {\enquote {\bibinfo {title}
  {{Right-Handed Neutrinos at CERN LHC and the Mechanism of Neutrino Mass
  Generation}},}\ }\href {\doibase 10.1103/PhysRevD.76.073005} {\bibfield
  {journal} {\bibinfo  {journal} {Phys. Rev.}\ }\textbf {\bibinfo {volume}
  {D76}},\ \bibinfo {pages} {073005} (\bibinfo {year} {2007})},\ \Eprint
  {http://arxiv.org/abs/0705.3221} {arXiv:0705.3221 [hep-ph]} \BibitemShut
  {NoStop}%
\bibitem [{\citenamefont {Lee}\ \emph {et~al.}(2013)\citenamefont {Lee},
  \citenamefont {Bhupal~Dev},\ and\ \citenamefont {Mohapatra}}]{Dev:2013oxa}%
  \BibitemOpen
  \bibfield  {author} {\bibinfo {author} {\bibfnamefont {Chang-Hun}\
  \bibnamefont {Lee}}, \bibinfo {author} {\bibfnamefont {P.~S.}\ \bibnamefont
  {Bhupal~Dev}}, \ and\ \bibinfo {author} {\bibfnamefont {R.~N.}\ \bibnamefont
  {Mohapatra}},\ }\bibfield  {title} {\enquote {\bibinfo {title} {{Natural
  TeV-scale left-right seesaw mechanism for neutrinos and experimental
  tests}},}\ }\href {\doibase 10.1103/PhysRevD.88.093010} {\bibfield  {journal}
  {\bibinfo  {journal} {Phys. Rev.}\ }\textbf {\bibinfo {volume} {D88}},\
  \bibinfo {pages} {093010} (\bibinfo {year} {2013})},\ \Eprint
  {http://arxiv.org/abs/1309.0774} {arXiv:1309.0774 [hep-ph]} \BibitemShut
  {NoStop}%
\bibitem [{\citenamefont {Chattopadhyay}\ and\ \citenamefont
  {Patel}(2017)}]{Chattopadhyay:2017zvs}%
  \BibitemOpen
  \bibfield  {author} {\bibinfo {author} {\bibfnamefont {Pratik}\ \bibnamefont
  {Chattopadhyay}}\ and\ \bibinfo {author} {\bibfnamefont {Ketan~M.}\
  \bibnamefont {Patel}},\ }\bibfield  {title} {\enquote {\bibinfo {title}
  {{Discrete symmetries for electroweak natural type-I seesaw mechanism}},}\
  }\href {\doibase 10.1016/j.nuclphysb.2017.06.008} {\bibfield  {journal}
  {\bibinfo  {journal} {Nucl. Phys.}\ }\textbf {\bibinfo {volume} {B921}},\
  \bibinfo {pages} {487--506} (\bibinfo {year} {2017})},\ \Eprint
  {http://arxiv.org/abs/1703.09541} {arXiv:1703.09541 [hep-ph]} \BibitemShut
  {NoStop}%
\bibitem [{\citenamefont {Staub}(2014)}]{Staub:2013tta}%
  \BibitemOpen
  \bibfield  {author} {\bibinfo {author} {\bibfnamefont {Florian}\ \bibnamefont
  {Staub}},\ }\bibfield  {title} {\enquote {\bibinfo {title} {{SARAH 4 : A tool
  for (not only SUSY) model builders}},}\ }\href {\doibase
  10.1016/j.cpc.2014.02.018} {\bibfield  {journal} {\bibinfo  {journal}
  {Comput. Phys. Commun.}\ }\textbf {\bibinfo {volume} {185}},\ \bibinfo
  {pages} {1773--1790} (\bibinfo {year} {2014})},\ \Eprint
  {http://arxiv.org/abs/1309.7223} {arXiv:1309.7223 [hep-ph]} \BibitemShut
  {NoStop}%
\bibitem [{\citenamefont {Gunion}\ and\ \citenamefont
  {Haber}(2003)}]{Gunion:2002zf}%
  \BibitemOpen
  \bibfield  {author} {\bibinfo {author} {\bibfnamefont {John~F.}\ \bibnamefont
  {Gunion}}\ and\ \bibinfo {author} {\bibfnamefont {Howard~E.}\ \bibnamefont
  {Haber}},\ }\bibfield  {title} {\enquote {\bibinfo {title} {{The CP
  conserving two Higgs doublet model: The Approach to the decoupling limit}},}\
  }\href {\doibase 10.1103/PhysRevD.67.075019} {\bibfield  {journal} {\bibinfo
  {journal} {Phys. Rev.}\ }\textbf {\bibinfo {volume} {D67}},\ \bibinfo {pages}
  {075019} (\bibinfo {year} {2003})},\ \Eprint
  {http://arxiv.org/abs/hep-ph/0207010} {arXiv:hep-ph/0207010 [hep-ph]}
  \BibitemShut {NoStop}%
\bibitem [{\citenamefont {Isidori}\ \emph {et~al.}(2001)\citenamefont
  {Isidori}, \citenamefont {Ridolfi},\ and\ \citenamefont
  {Strumia}}]{Isidori:2001bm}%
  \BibitemOpen
  \bibfield  {author} {\bibinfo {author} {\bibfnamefont {Gino}\ \bibnamefont
  {Isidori}}, \bibinfo {author} {\bibfnamefont {Giovanni}\ \bibnamefont
  {Ridolfi}}, \ and\ \bibinfo {author} {\bibfnamefont {Alessandro}\
  \bibnamefont {Strumia}},\ }\bibfield  {title} {\enquote {\bibinfo {title}
  {{On the metastability of the standard model vacuum}},}\ }\href {\doibase
  10.1016/S0550-3213(01)00302-9} {\bibfield  {journal} {\bibinfo  {journal}
  {Nucl. Phys.}\ }\textbf {\bibinfo {volume} {B609}},\ \bibinfo {pages}
  {387--409} (\bibinfo {year} {2001})},\ \Eprint
  {http://arxiv.org/abs/hep-ph/0104016} {arXiv:hep-ph/0104016 [hep-ph]}
  \BibitemShut {NoStop}%
\bibitem [{\citenamefont {Aad}\ \emph {et~al.}(2015)\citenamefont {Aad} \emph
  {et~al.}}]{Aad:2015zhl}%
  \BibitemOpen
  \bibfield  {author} {\bibinfo {author} {\bibfnamefont {Georges}\ \bibnamefont
  {Aad}} \emph {et~al.} (\bibinfo {collaboration} {ATLAS, CMS}),\ }\bibfield
  {title} {\enquote {\bibinfo {title} {{Combined Measurement of the Higgs Boson
  Mass in $pp$ Collisions at $\sqrt{s}=7$ and 8 TeV with the ATLAS and CMS
  Experiments}},}\ }\href {\doibase 10.1103/PhysRevLett.114.191803} {\bibfield
  {journal} {\bibinfo  {journal} {Phys. Rev. Lett.}\ }\textbf {\bibinfo
  {volume} {114}},\ \bibinfo {pages} {191803} (\bibinfo {year} {2015})},\
  \Eprint {http://arxiv.org/abs/1503.07589} {arXiv:1503.07589 [hep-ex]}
  \BibitemShut {NoStop}%
\bibitem [{\citenamefont {Misiak}\ and\ \citenamefont
  {Steinhauser}(2017)}]{Misiak:2017bgg}%
  \BibitemOpen
  \bibfield  {author} {\bibinfo {author} {\bibfnamefont {Mikolaj}\ \bibnamefont
  {Misiak}}\ and\ \bibinfo {author} {\bibfnamefont {Matthias}\ \bibnamefont
  {Steinhauser}},\ }\bibfield  {title} {\enquote {\bibinfo {title} {{Weak
  radiative decays of the B meson and bounds on $M_{H^\pm }$ in the
  Two-Higgs-Doublet Model}},}\ }\href {\doibase 10.1140/epjc/s10052-017-4776-y}
  {\bibfield  {journal} {\bibinfo  {journal} {Eur. Phys. J.}\ }\textbf
  {\bibinfo {volume} {C77}},\ \bibinfo {pages} {201} (\bibinfo {year}
  {2017})},\ \Eprint {http://arxiv.org/abs/1702.04571} {arXiv:1702.04571
  [hep-ph]} \BibitemShut {NoStop}%
\bibitem [{\citenamefont {Chowdhury}\ and\ \citenamefont
  {Eberhardt}(2017)}]{Chowdhury:2017aav}%
  \BibitemOpen
  \bibfield  {author} {\bibinfo {author} {\bibfnamefont {Debtosh}\ \bibnamefont
  {Chowdhury}}\ and\ \bibinfo {author} {\bibfnamefont {Otto}\ \bibnamefont
  {Eberhardt}},\ }\bibfield  {title} {\enquote {\bibinfo {title} {{Update of
  Global Two-Higgs-Doublet Model Fits}},}\ }\href@noop {} {\  (\bibinfo {year}
  {2017})},\ \Eprint {http://arxiv.org/abs/1711.02095} {arXiv:1711.02095
  [hep-ph]} \BibitemShut {NoStop}%
\bibitem [{\citenamefont {Broggio}\ \emph {et~al.}(2014)\citenamefont
  {Broggio}, \citenamefont {Chun}, \citenamefont {Passera}, \citenamefont
  {Patel},\ and\ \citenamefont {Vempati}}]{Broggio:2014mna}%
  \BibitemOpen
  \bibfield  {author} {\bibinfo {author} {\bibfnamefont {Alessandro}\
  \bibnamefont {Broggio}}, \bibinfo {author} {\bibfnamefont {Eung~Jin}\
  \bibnamefont {Chun}}, \bibinfo {author} {\bibfnamefont {Massimo}\
  \bibnamefont {Passera}}, \bibinfo {author} {\bibfnamefont {Ketan~M.}\
  \bibnamefont {Patel}}, \ and\ \bibinfo {author} {\bibfnamefont {Sudhir~K.}\
  \bibnamefont {Vempati}},\ }\bibfield  {title} {\enquote {\bibinfo {title}
  {{Limiting two-Higgs-doublet models}},}\ }\href {\doibase
  10.1007/JHEP11(2014)058} {\bibfield  {journal} {\bibinfo  {journal} {JHEP}\
  }\textbf {\bibinfo {volume} {11}},\ \bibinfo {pages} {058} (\bibinfo {year}
  {2014})},\ \Eprint {http://arxiv.org/abs/1409.3199} {arXiv:1409.3199
  [hep-ph]} \BibitemShut {NoStop}%
\bibitem [{\citenamefont {Patrignani}\ \emph {et~al.}(2016)\citenamefont
  {Patrignani} \emph {et~al.}}]{Patrignani:2016xqp}%
  \BibitemOpen
  \bibfield  {author} {\bibinfo {author} {\bibfnamefont {C.}~\bibnamefont
  {Patrignani}} \emph {et~al.} (\bibinfo {collaboration} {Particle Data
  Group}),\ }\bibfield  {title} {\enquote {\bibinfo {title} {{Review of
  Particle Physics}},}\ }\href {\doibase 10.1088/1674-1137/40/10/100001}
  {\bibfield  {journal} {\bibinfo  {journal} {Chin. Phys.}\ }\textbf {\bibinfo
  {volume} {C40}},\ \bibinfo {pages} {100001} (\bibinfo {year}
  {2016})}\BibitemShut {NoStop}%
\bibitem [{\citenamefont {Esteban}\ \emph {et~al.}(2017)\citenamefont
  {Esteban}, \citenamefont {Gonzalez-Garcia}, \citenamefont {Maltoni},
  \citenamefont {Martinez-Soler},\ and\ \citenamefont
  {Schwetz}}]{Esteban:2016qun}%
  \BibitemOpen
  \bibfield  {author} {\bibinfo {author} {\bibfnamefont {Ivan}\ \bibnamefont
  {Esteban}}, \bibinfo {author} {\bibfnamefont {M.~C.}\ \bibnamefont
  {Gonzalez-Garcia}}, \bibinfo {author} {\bibfnamefont {Michele}\ \bibnamefont
  {Maltoni}}, \bibinfo {author} {\bibfnamefont {Ivan}\ \bibnamefont
  {Martinez-Soler}}, \ and\ \bibinfo {author} {\bibfnamefont {Thomas}\
  \bibnamefont {Schwetz}},\ }\bibfield  {title} {\enquote {\bibinfo {title}
  {{Updated fit to three neutrino mixing: exploring the accelerator-reactor
  complementarity (NuFIT 3.2 (2018), www.nu-fit.org)}},}\ }\href {\doibase
  10.1007/JHEP01(2017)087} {\bibfield  {journal} {\bibinfo  {journal} {JHEP}\
  }\textbf {\bibinfo {volume} {01}},\ \bibinfo {pages} {087} (\bibinfo {year}
  {2017})},\ \Eprint {http://arxiv.org/abs/1611.01514} {arXiv:1611.01514
  [hep-ph]} \BibitemShut {NoStop}%
\bibitem [{\citenamefont {Rodejohann}\ and\ \citenamefont
  {Zhang}(2012)}]{Rodejohann:2012px}%
  \BibitemOpen
  \bibfield  {author} {\bibinfo {author} {\bibfnamefont {Werner}\ \bibnamefont
  {Rodejohann}}\ and\ \bibinfo {author} {\bibfnamefont {He}~\bibnamefont
  {Zhang}},\ }\bibfield  {title} {\enquote {\bibinfo {title} {{Impact of
  massive neutrinos on the Higgs self-coupling and electroweak vacuum
  stability}},}\ }\href {\doibase 10.1007/JHEP06(2012)022} {\bibfield
  {journal} {\bibinfo  {journal} {JHEP}\ }\textbf {\bibinfo {volume} {06}},\
  \bibinfo {pages} {022} (\bibinfo {year} {2012})},\ \Eprint
  {http://arxiv.org/abs/1203.3825} {arXiv:1203.3825 [hep-ph]} \BibitemShut
  {NoStop}%
\bibitem [{\citenamefont {Khan}\ \emph {et~al.}(2014)\citenamefont {Khan},
  \citenamefont {Goswami},\ and\ \citenamefont {Roy}}]{Khan:2012zw}%
  \BibitemOpen
  \bibfield  {author} {\bibinfo {author} {\bibfnamefont {Subrata}\ \bibnamefont
  {Khan}}, \bibinfo {author} {\bibfnamefont {Srubabati}\ \bibnamefont
  {Goswami}}, \ and\ \bibinfo {author} {\bibfnamefont {Sourov}\ \bibnamefont
  {Roy}},\ }\bibfield  {title} {\enquote {\bibinfo {title} {{Vacuum Stability
  constraints on the minimal singlet TeV Seesaw Model}},}\ }\href {\doibase
  10.1103/PhysRevD.89.073021} {\bibfield  {journal} {\bibinfo  {journal} {Phys.
  Rev.}\ }\textbf {\bibinfo {volume} {D89}},\ \bibinfo {pages} {073021}
  (\bibinfo {year} {2014})},\ \Eprint {http://arxiv.org/abs/1212.3694}
  {arXiv:1212.3694 [hep-ph]} \BibitemShut {NoStop}%
\bibitem [{\citenamefont {Chakrabortty}\ \emph {et~al.}(2013)\citenamefont
  {Chakrabortty}, \citenamefont {Das},\ and\ \citenamefont
  {Mohanty}}]{Chakrabortty:2012np}%
  \BibitemOpen
  \bibfield  {author} {\bibinfo {author} {\bibfnamefont {Joydeep}\ \bibnamefont
  {Chakrabortty}}, \bibinfo {author} {\bibfnamefont {Moumita}\ \bibnamefont
  {Das}}, \ and\ \bibinfo {author} {\bibfnamefont {Subhendra}\ \bibnamefont
  {Mohanty}},\ }\bibfield  {title} {\enquote {\bibinfo {title} {{Constraints on
  TeV scale Majorana neutrino phenomenology from the Vacuum Stability of the
  Higgs}},}\ }\href {\doibase 10.1142/S0217732313500326} {\bibfield  {journal}
  {\bibinfo  {journal} {Mod. Phys. Lett.}\ }\textbf {\bibinfo {volume} {A28}},\
  \bibinfo {pages} {1350032} (\bibinfo {year} {2013})},\ \Eprint
  {http://arxiv.org/abs/1207.2027} {arXiv:1207.2027 [hep-ph]} \BibitemShut
  {NoStop}%
\bibitem [{\citenamefont {Delle~Rose}\ \emph {et~al.}(2015)\citenamefont
  {Delle~Rose}, \citenamefont {Marzo},\ and\ \citenamefont
  {Urbano}}]{Rose:2015fua}%
  \BibitemOpen
  \bibfield  {author} {\bibinfo {author} {\bibfnamefont {Luigi}\ \bibnamefont
  {Delle~Rose}}, \bibinfo {author} {\bibfnamefont {Carlo}\ \bibnamefont
  {Marzo}}, \ and\ \bibinfo {author} {\bibfnamefont {Alfredo}\ \bibnamefont
  {Urbano}},\ }\bibfield  {title} {\enquote {\bibinfo {title} {{On the
  stability of the electroweak vacuum in the presence of low-scale seesaw
  models}},}\ }\href {\doibase 10.1007/JHEP12(2015)050} {\bibfield  {journal}
  {\bibinfo  {journal} {JHEP}\ }\textbf {\bibinfo {volume} {12}},\ \bibinfo
  {pages} {050} (\bibinfo {year} {2015})},\ \Eprint
  {http://arxiv.org/abs/1506.03360} {arXiv:1506.03360 [hep-ph]} \BibitemShut
  {NoStop}%
\bibitem [{\citenamefont {Bambhaniya}\ \emph {et~al.}(2017)\citenamefont
  {Bambhaniya}, \citenamefont {Bhupal~Dev}, \citenamefont {Goswami},
  \citenamefont {Khan},\ and\ \citenamefont {Rodejohann}}]{Bambhaniya:2016rbb}%
  \BibitemOpen
  \bibfield  {author} {\bibinfo {author} {\bibfnamefont {Gulab}\ \bibnamefont
  {Bambhaniya}}, \bibinfo {author} {\bibfnamefont {P.S.}\ \bibnamefont
  {Bhupal~Dev}}, \bibinfo {author} {\bibfnamefont {Srubabati}\ \bibnamefont
  {Goswami}}, \bibinfo {author} {\bibfnamefont {Subrata}\ \bibnamefont {Khan}},
  \ and\ \bibinfo {author} {\bibfnamefont {Werner}\ \bibnamefont
  {Rodejohann}},\ }\bibfield  {title} {\enquote {\bibinfo {title}
  {{Naturalness, Vacuum Stability and Leptogenesis in the Minimal Seesaw
  Model}},}\ }\href {\doibase 10.1103/PhysRevD.95.095016} {\bibfield  {journal}
  {\bibinfo  {journal} {Phys. Rev.}\ }\textbf {\bibinfo {volume} {D95}},\
  \bibinfo {pages} {095016} (\bibinfo {year} {2017})},\ \Eprint
  {http://arxiv.org/abs/1611.03827} {arXiv:1611.03827 [hep-ph]} \BibitemShut
  {NoStop}%
\bibitem [{\citenamefont {Coriano}\ \emph {et~al.}(2014)\citenamefont
  {Coriano}, \citenamefont {Delle~Rose},\ and\ \citenamefont
  {Marzo}}]{Coriano:2014mpa}%
  \BibitemOpen
  \bibfield  {author} {\bibinfo {author} {\bibfnamefont {Claudio}\ \bibnamefont
  {Coriano}}, \bibinfo {author} {\bibfnamefont {Luigi}\ \bibnamefont
  {Delle~Rose}}, \ and\ \bibinfo {author} {\bibfnamefont {Carlo}\ \bibnamefont
  {Marzo}},\ }\bibfield  {title} {\enquote {\bibinfo {title} {{Vacuum Stability
  in U(1)-Prime Extensions of the Standard Model with TeV Scale Right Handed
  Neutrinos}},}\ }\href {\doibase 10.1016/j.physletb.2014.09.001} {\bibfield
  {journal} {\bibinfo  {journal} {Phys. Lett.}\ }\textbf {\bibinfo {volume}
  {B738}},\ \bibinfo {pages} {13--19} (\bibinfo {year} {2014})},\ \Eprint
  {http://arxiv.org/abs/1407.8539} {arXiv:1407.8539 [hep-ph]} \BibitemShut
  {NoStop}%
\bibitem [{\citenamefont {Coriano}\ \emph {et~al.}(2016)\citenamefont
  {Coriano}, \citenamefont {Delle~Rose},\ and\ \citenamefont
  {Marzo}}]{Coriano:2015sea}%
  \BibitemOpen
  \bibfield  {author} {\bibinfo {author} {\bibfnamefont {Claudio}\ \bibnamefont
  {Coriano}}, \bibinfo {author} {\bibfnamefont {Luigi}\ \bibnamefont
  {Delle~Rose}}, \ and\ \bibinfo {author} {\bibfnamefont {Carlo}\ \bibnamefont
  {Marzo}},\ }\bibfield  {title} {\enquote {\bibinfo {title} {{Constraints on
  abelian extensions of the Standard Model from two-loop vacuum stability and
  $U(1)_{B-L}$}},}\ }\href {\doibase 10.1007/JHEP02(2016)135} {\bibfield
  {journal} {\bibinfo  {journal} {JHEP}\ }\textbf {\bibinfo {volume} {02}},\
  \bibinfo {pages} {135} (\bibinfo {year} {2016})},\ \Eprint
  {http://arxiv.org/abs/1510.02379} {arXiv:1510.02379 [hep-ph]} \BibitemShut
  {NoStop}%
\bibitem [{\citenamefont {Chetyrkin}\ \emph {et~al.}(2000)\citenamefont
  {Chetyrkin}, \citenamefont {Kuhn},\ and\ \citenamefont
  {Steinhauser}}]{Chetyrkin:2000yt}%
  \BibitemOpen
  \bibfield  {author} {\bibinfo {author} {\bibfnamefont {K.~G.}\ \bibnamefont
  {Chetyrkin}}, \bibinfo {author} {\bibfnamefont {Johann~H.}\ \bibnamefont
  {Kuhn}}, \ and\ \bibinfo {author} {\bibfnamefont {M.}~\bibnamefont
  {Steinhauser}},\ }\bibfield  {title} {\enquote {\bibinfo {title} {{RunDec: A
  Mathematica package for running and decoupling of the strong coupling and
  quark masses}},}\ }\href {\doibase 10.1016/S0010-4655(00)00155-7} {\bibfield
  {journal} {\bibinfo  {journal} {Comput. Phys. Commun.}\ }\textbf {\bibinfo
  {volume} {133}},\ \bibinfo {pages} {43--65} (\bibinfo {year} {2000})},\
  \Eprint {http://arxiv.org/abs/hep-ph/0004189} {arXiv:hep-ph/0004189 [hep-ph]}
  \BibitemShut {NoStop}%
\bibitem [{\citenamefont {Xing}\ \emph {et~al.}(2008)\citenamefont {Xing},
  \citenamefont {Zhang},\ and\ \citenamefont {Zhou}}]{Xing:2007fb}%
  \BibitemOpen
  \bibfield  {author} {\bibinfo {author} {\bibfnamefont {Zhi-zhong}\
  \bibnamefont {Xing}}, \bibinfo {author} {\bibfnamefont {He}~\bibnamefont
  {Zhang}}, \ and\ \bibinfo {author} {\bibfnamefont {Shun}\ \bibnamefont
  {Zhou}},\ }\bibfield  {title} {\enquote {\bibinfo {title} {{Updated Values of
  Running Quark and Lepton Masses}},}\ }\href {\doibase
  10.1103/PhysRevD.77.113016} {\bibfield  {journal} {\bibinfo  {journal} {Phys.
  Rev.}\ }\textbf {\bibinfo {volume} {D77}},\ \bibinfo {pages} {113016}
  (\bibinfo {year} {2008})},\ \Eprint {http://arxiv.org/abs/0712.1419}
  {arXiv:0712.1419 [hep-ph]} \BibitemShut {NoStop}%
\end{thebibliography}%
\end{document}